\documentclass[11pt,a4paper]{article}

\usepackage{myjheppub}
\makeatletter
\gdef\@fpheader{}
\makeatother

\hypersetup{colorlinks = true, linkcolor = blue, urlcolor = blue, citecolor = blue, anchorcolor = blue}

\usepackage[T1]{fontenc}
\usepackage[latin9]{inputenc}
\usepackage{babel}
\usepackage{nccmath}
\usepackage{orcidlink}
\usepackage{placeins}
\usepackage{subcaption}
\usepackage{tikz}
\usetikzlibrary{calc}
\usepackage{tikz-feynman}
\usetikzlibrary{decorations.pathmorphing}
\tikzfeynmanset{
  Bprop/.style={
    draw,
    decorate,
    decoration={snake, amplitude=1.2mm, segment length=5mm},
  },
}

\usepackage{amsthm}

\theoremstyle{definition}
\newtheorem{definition}{Definition}[section]

\theoremstyle{remark}

\newcommand{\mathbbm}[1]{\text{\usefont{U}{bbm}{m}{n}#1}} 

\usepackage[bb=boondox]{mathalfa}  

\usepackage{graphicx,amsmath,amssymb,color}
\usepackage[normalem]{ulem}
\usepackage{amsfonts}
\usepackage{url} 
\usepackage{latexsym}
\usepackage{algpseudocode}
\usepackage{mathrsfs}
\usepackage{color,verbatim}
\usepackage{psfrag}

\usepackage{relsize}
\usepackage{etoolbox}
\newcommand{\circled}[2][]{%
  \tikz[baseline=(char.base)]{%
    \node[shape = circle, draw, inner sep = 1pt]
    (char) {\phantom{\ifblank{\smaller \smaller #1}{\smaller \smaller #2}{\smaller \smaller #1}}};%
    \node at (char.center) {\makebox[0pt][c]{\smaller \smaller #2}};}}
\robustify{\circled}

\newcommand{\figref}[1]{Fig.~\ref{#1}}

\newcommand{\bra}[1]{\langle#1 |}
\newcommand{\ket}[1]{|#1 \rangle}
\newcommand{\braket}[2]{\left \langle #1 \mid #2 \right \rangle}

\def\beq{\begin{equation}}
\def\eeq{\end{equation}}
\def\bali{{\begin{align}}}
\def\eali{{\end{align}}}
\def\ie{{\it i.e.~}}
\def\eg{{\it e.g.~}}

\newcommand{\listitem}[1]{\noindent --{\it #1}\\ \vspace{-0.3cm} {} \\
\noindent { }}

\DeclareRobustCommand{\Eq}[1]{Eq.~(\ref{#1})}

\usepackage[labelfont=bf,labelsep=colon]{caption}
\renewcommand{\figurename}{Figure}

\title{When Renormalisation Remembers:\\ UV/IR Mixing as an Entanglement Bridge}

\author{Steven Abel\,\orcidlink{0000-0003-1213-907X}}

\affiliation{Institute for Particle Physics Phenomenology and \\
Department of Mathematical Sciences,\\
Durham University, Durham DH1 3LE, UK}

\emailAdd{s.a.abel@durham.ac.uk}
\newcommand{\preprint}[1]{\begin{flushright}#1\end{flushright}}

\abstract{%
Renormalisation is traditionally understood to be a Wilsonian memoryless process in which ultraviolet (UV) degrees of freedom gradually decouple, leaving an autonomous infrared (IR) description. However this need not be the case: in UV/IR mixed theories correlations between widely separated scales can persist. In this work I recast UV/IR mixing as a Hilbert-space phenomenon, realised as correlations across renormalisation scales. This formulation is implemented using the Born-Reciprocal Tensor Network (BRTN), a new configuration of tensor network that is globally symmetric under phase-space reciprocity. On this network I prepare the vacuum and reproduce the expected radiative corrections. The resulting renormalisation geometry exhibits memory, with a bridge linking reciprocal representations of IR physics, whose cross-bridge entanglement provides a precise criterion for the viability of an effective description. I analyse when this criterion is met, and show that there is a large-volume limit, with the fundamental scale held fixed, in which the obstruction to a local description scales away: Wilsonian behaviour is restored and renormalisation forgets. The BRTN therefore provides a concrete and calculable platform for UV/IR mixing.
}

\begin{document}

\preprint{IPPP/26/48}

\maketitle

\section{Introduction}

In conventional renormalisable quantum field theories, heavy degrees of freedom decouple from low-energy physics~\cite{Appelquist:1974tg}. This decoupling is the bedrock of the effective field theory paradigm. It allows us to discard all the baggage of the UV, and   describe the dynamics at a given energy by ``integrating out'' the modes above that scale~\cite{Wilson:1983xri}. 

However, this Wilsonian picture assumes that a clear separation between light and heavy degrees of freedom can be achieved locally in phase space. Sometimes it cannot. In the presence of UV/IR mixing for example~\cite{Seiberg:1999vs,Minwalla:1999px,Matusis:2000jf}, a definition of ``UV'' and ``IR'' that is purely local in phase space fails to  separate short- and long-distance physics cleanly, leaving residual imprints of ultraviolet structure on the long-distance theory. 

This breakdown of a strictly Wilsonian separation of scales is known to arise in a variety of contexts, including theories that exhibit Regge behaviour~\cite{Donoghue:2009mn}, and more broadly in settings where duality symmetries relate degrees of freedom across scales. 
A particularly important context in which such a breakdown is expected to occur is if UV/IR mixing plays a role in UV-completion or naturalness, a possibility that has attracted growing interest in recent years~\cite{Cohen:1998zx,Dienes:2001se,Lust:2017wrl,Craig:2019zbn,Abel:2021tyt,Arkani-Hamed:2021xlp,Castellano:2021mmx,Craig:2022eqo,Kephart:2022vfr,Abel:2023hkk,Abel:2024twz,Nortier:2025gmc,Cribiori:2025oek,Aoufia:2026bau}. If Nature exhibits such mixing, then the Wilsonian paradigm of strict scale separation will be insufficient to  capture all aspects of renormalisation properly, because the flow can no longer be regarded as a purely local coarse-graining procedure. 

Whether or not an effective low-energy description then remains possible in such theories depends on the global structure of the theory, in particular on the nature of its cross-scale correlations. Classical correlations, such as mean-field backreaction from heavy modes, can be absorbed into renormalised parameters, to yield a viable effective theory. Quantum correlations, by contrast, can entangle degrees of freedom across scales in a non-separable way. The resulting corrections to the theory would be finite and controlled but it might be impossible to absorb them into a local theory. In such theories a Wilsonian description may remain  valid in some regime, or it may simply never give a good description of the physics. 

Distinguishing between these two situations calls for a framework that incorporates cross-scale correlations within the global structure of the theory, rather than eliminating them.
To this end, this paper builds a general framework for UV/IR mixed theories that centres on information flow. 
It is based on the 
Tree Tensor Network (TTN)~\cite{Shi:2006zz} paradigm, which implements a scale-by-scale coarse-graining of Hilbert space, providing an entropic and geometric language for renormalisation. 

The purpose of this work is not  to introduce yet another tensor network algorithm, but to demonstrate how UV/IR mixing can be reformulated in TTN language as an unusual kind of renormalisation geometry that incorporates two competing descriptions of the IR.
This is achieved by extending the TTN to incorporate a global Born-reciprocal mirror, realised at the level of the global Hilbert space structure but not manifest in the low-energy physics once a particular IR description is selected. I will call this doubled TTN the Born-Reciprocal Tensor Network (BRTN).
In this formulation, renormalisation is no longer represented as a one-dimensional flow in scale, but as a global geometric organisation of Hilbert space.

Correlations between the two reciprocal IR descriptions remain encoded in such a renormalisation geometry, forming a geometric bridge that cannot be eliminated by successive local coarse-graining. 
The influence that the two IR descriptions exert on each other through this bridge then becomes the criterion by which the viability of a Wilsonian effective theory can be judged. 
Entanglement across the bridge may render the theory non-separable, while even classical correlations can induce non-locality and the breaking of translational invariance. If such effects were to persist in the large-volume limit then a Wilsonian description would be impossible: they therefore provide a measure of the departure from Wilsonian behaviour. A crucial property of Born-reciprocal networks, demonstrated in this work, is that the cross-bridge entanglement they generate is intensive, and dilutes in the large-volume limit, allowing a Wilsonian picture to re-emerge. This gives the phenomenon of UV/IR mixing a modern information-theoretic interpretation~\cite{Calabrese:2004eu,Calabrese:2005zw,Vidal:2007hda,Casini:2011kv,Casini:2012ei}.

This paper is organised as follows. As the primary audience for this paper may not be familiar with the workings of TTNs, or with their role in renormalisation, Section~\ref{sec:TTNintro}
contains a review of them. It  describes the links between information flow, coarse-graining and renormalisation. This introduces the basic TTN concepts and will serve as a reference for later work; readers familiar with TTNs may wish to skip ahead. 
As Born-reciprocity works with the  continuous-variable degrees of freedom of a genuine scalar field, as opposed to the spin-chain and Ising-type models that have been the primary focus of tensor network studies to date, the first crucial step towards the BRTN is to develop the machinery for normal QFT. This is done in Section~\ref{sec:QFTImp} which implements scalar QFT on the TTN, which will be the working model for this paper. As well as the implementation itself, here we will see how to prepare and verify the QFT vacuum, and demonstrate the measurement of radiative corrections (to the mass-squareds of the scalars). With this technology to hand, the BRTN itself is introduced in Section \ref{sec:BRTN}. In Subsection \ref{subsec:BRTNIntro} we motivate the overall structure of the BRTN as a doubled network which can host Born-reciprocity. Subsection \ref{subsec:FourierBridge}
considers the  properties required of the Hamiltonian that we place upon such a network, in order for the system to be genuinely  Born-reciprocal. 
In Section \ref{sec:BRTNImp} we discuss the practical implementation of the BRTN, and in Section \ref{sec:testimp}
put the framework to the test: we place a regular scalar QFT on the network, discuss the preparation of its vacuum, and measure the expected radiative corrections. 
Section \ref{sec:BRTNObs} then probes  radiative corrections in genuinely Born-reciprocal systems on the BRTN.  
Although these systems are non-perturbative from the perspective of QFT because there is no continuum field theory description for them, leading order radiative corrections can nevertheless be estimated by performing perturbation theory on the quantum mechanical Hilbert space of the network. Our measured radiative corrections reproduce these theoretical expectations with some accuracy. Section~\ref{sec:breaking} discusses the physical ramifications of the non-Wilsonian behaviour, namely nonlocality and breaking of translational invariance, and demonstrates that there is a large volume limit of the 
 Born-reciprocal network in which these effects decouple, allowing standard Wilsonian behaviour to re-emerge. 
Section~\ref{sec:physicalmeaning} discusses the general physical implications of this framework and relates it to other ideas about UV/IR mixing. 

\section{Renormalisation as Entanglement Flow: Review of the TTN Language}
\label{sec:TTNintro}

The modern tensor network  paradigm is the result of several decades of development. It has its origins in the density matrix renormalisation group (DMRG) introduced by White~\cite{White:1992zz,White:1993zz}, which was later cast in the language of Matrix Product States (MPS) and related  frameworks~\cite{Ostlund:1995zz,Fannes:1992zz}.   
MPS and their higher-dimensional generalisation, projected entangled-pair states (PEPS), were subsequently established as a unified framework for representing quantum many-body states~\cite{Verstraete:2004cf,Verstraete:2004uw,PerezGarcia:2006rz,10.21468/SciPostPhysLectNotes.8}. 

TTNs~\cite{Shi:2006zz}, which are the primary setting for the discussion in this work, 
were the first hierarchical structures to appear that were directly motivated by renormalisation. It will become clear as we progress why they are preferred for the framework being developed here. Related approaches include the HaPPY code and the multi-scale entanglement renormalisation ansatz (MERA), all of which became important in the context of holography~\cite{Vidal:2007hda,Vidal:2008zz,Levin:2006jai,Swingle:2009bg,Evenbly:2015uca,Pastawski:2015qua,Harlow:2018fse}. In this work we will not discuss holography directly, although certain echoes of it will emerge along the way.

Entropy provides the natural language for describing information flow, which in quantum systems lends itself naturally to a Hilbert space formulation of renormalisation. A good starting point therefore is to  review the relation between entropy and renormalisation. This will naturally lead us on to the TTN and prepare the ground for the introduction of the BRTN.\\

To begin then, imagine a universe consisting of two qubits described by the bipartite state
\begin{equation}
\label{eq:simp}
    \ket{\Psi}~=~ \sqrt{p}\,\ket{0}_A\ket{0}_B + \sqrt{1-p}\,\ket{1}_A\ket{1}_B ~. 
\end{equation}
Although the global state is pure, it is entangled. 
 Now, as the $B$ state always tracks the $A$ state, we might decide that it is more convenient to treat them as a single block and work within an effective theory that has a reduced basis, $\ket{\delta} \equiv \ket{\delta}_A\ket{\delta}_B$. From a renormalisation perspective this amounts to a coarse-graining. The block states $\ket{\delta}$ ({\it cf.} block spins) form a new basis for that part of the global state that describes our effective theory (\ie the remaining IR quantum degrees of freedom), while we are integrating out the possibility that $\ket{\delta}_B$ might ever be different from $\ket{\delta}_A$.  Mathematically this is equivalent to declaring one of the states, $B$  say, to be redundant and tracing over it. 

A fundamental observation is that the resulting effective theory has a different entropy from that of the global universe. Indeed, as the global state is pure
its density matrix is  $\rho = \ket{\Psi}\bra{\Psi}$, and hence its von Neumann entropy is simply 
\begin{equation}
    S_{\rm global} ~=~ -  {\rm Tr} \,  \rho \log \rho ~=~ 0~.
\end{equation}
By contrast the reduced density matrix after we have traced over the redundant $\ket{\delta}_B$ states is
\begin{equation}
    \rho_{\rm eff} ~=~ p \ket{0}\bra{0} + (1-p) \ket{1}\bra{1}  ~,
\end{equation}
 and consequently the entropy of the effective system is larger than that of the global one: 
\begin{equation}
    S_{\rm eff} ~=~ -  p \log p - (1-p) \log(1-p) ~ >~0~.
\end{equation} 
Thus we can infer an important relationship between entropy and information flow in renormalisation. The entanglement between the 
constituents of the block states is what allows us to go to that conveniently reduced basis, but in doing so we choose to ignore the quantum information that the entanglement contains.
This quantum information is transformed into entropy in the effective system. This  is precisely quantified by the  von Neumann entropy of the reduced density matrix.

To conceptualise this flow of information, there is a convenient network representation that we will be using throughout. Beginning with our simple example, for this trivial case the Wilsonian tracing operation can be represented by a single node as in  \figref{fig:simple}a, in which the box represents what is left after tracing over the redundant factor in the Hilbert space. \figref{fig:simple}b shows the (equally trivial) Tensor Network equivalent, which also contains only a single node, representing the entire state, and two legs, representing the physical indices $\ket{A}$ and $\ket{B}$ that we are considering combining. In the language of TTNs the node here (and indeed the top node of any TTN) is called the {\it root node}. Although the network geometry looks the same, in the TTN we have not yet lost any information from the system: \emph{a priori} the root node simply represents the 
entire global state, and the network diagram indicates a {\it possible} coarse-graining in which the two indices indicated by the legs are combined.

\definecolor{tensorblue}{RGB}{30,110,200}

\begin{figure}[t]
\centering

\begin{minipage}{0.45\columnwidth}
\centering
\begin{tikzpicture}[x=0.5cm, y=0.5cm,
font=\boldmath,
 tensor/.style={
    circle,
    draw=black,
    shade,
ball color=tensorblue,
 fill opacity=0.5,
text opacity=1,
    minimum size=6mm,
    inner sep=0pt,
    line width=0.2pt
  },
  leg/.style={line width=0.8pt},
  highlight/.style={line width=1.4pt, red!80},
  correl/.style={line width=1.4pt, green!60!black}
]

\node (S) at (-1.5,-0.5) {$A$};
\node (E) at (1.5,-0.5) {$B$};

\node[
    rectangle,
    draw=black,
    shading=radial,
    inner color=tensorblue!20,
    outer color=tensorblue!60,
    minimum width=1cm,
    minimum height=0.5cm,
    line width=0.3pt
]   (W) at (0,1.2) {$\rho_\delta$};

\draw (S) -- (W);
\draw (E) -- (W);

\node at (-2,-1.3) {\small retained};
\node at (2,-1.3) {\small traced out};

\end{tikzpicture}

\vspace{0.3em}
(a) Wilsonian tracing 
\end{minipage}
\hspace{-0.8cm}

\begin{minipage}{0.45\columnwidth}
\centering
\begin{tikzpicture}[x=0.5cm, y=0.5cm,
font=\boldmath,
 tensor/.style={
    circle,
    draw=black,
    shade,
ball color=tensorblue,
 fill opacity=0.5,
text opacity=1,
    minimum size=8mm,
    inner sep=0pt,
    line width=0.2pt
  },
  leg/.style={line width=0.8pt},
  highlight/.style={line width=1.4pt, red!80},
  correl/.style={line width=1.4pt, green!60!black}
]
\node (S) at (-1,-0.5) {$A$};
\node (E) at (1,-0.5) {$B$};

\node[tensor] (W) at (0,1.3) {$\ket{\Psi}$};

\draw[leg] (S) -- (W);
\draw[leg] (E) -- (W);

\end{tikzpicture}

\vspace{0.3em}
(b) TTN coarse-graining
\end{minipage}

\caption{Wilsonian integration versus tensor network coarse-graining for the trivial case. The former is a trace over redundant degrees of freedom, which results in local coarse-graining and a mixed state. The TTN retains all degrees of freedom and remains pure.}
\label{fig:simple}

\end{figure}

Let us now generalise the bipartite state. Suppose that the global state is instead of the form 
\begin{equation}
    \ket{\Psi}~=~  \psi_{ij} ~\ket{i}_A\ket{j}_B ~,
\end{equation}
where $\ket{i}_{A,B}$ are dimension $d$ Hilbert spaces, 
where here and throughout repeated indices imply summation, and where the coefficients $\psi_{ij}$ need not be diagonal. One may always bring this state into diagonal form by
singular value decomposition (SVD). That is by SVD on the matrix $\psi_{ij}$, the state can be written
 \begin{align}
    \ket{\Psi}~&=~  U_{i\delta} \lambda_\delta V^\dagger_{\delta j} \ket{i}_A\ket{j}_B ~
\nonumber \\
&=~  \lambda_\delta \ket{\delta}_A\ket{\delta}_B ~,
\end{align}
where 
\begin{align} 
\ket{\delta}_A & = U _{i\delta}\ket{\alpha}_A~~, \ket{\delta}_B = V^\dagger_{\delta j} \ket{\beta}_B~.
\label{eq:svd1}
\end{align}
This diagonal form of a state is its {\it Schmidt decomposition}. 
Once the state is in the Schmidt basis it is again possible  to coarse-grain the system by combining $\ket{\delta}_A \ket{\delta}_B$ into single block states $\ket{\delta}$. 
However as in the previous example we do not actually do this: we simply interpret the diagram in  \figref{fig:simple}b as indicating that this coarse-graining option is available. Thus the eigenvalues $\lambda_\delta$ define what can be thought of as the ``latent entropy'' at that node -- \ie 
they {\it would} give the reduced density matrix should we choose to coarse-grain and integrate out the redundant degrees of freedom, whereupon we would find $\rho_{\rm eff}  = \lambda_\delta^2 \ket{\delta}\bra{\delta}$, and hence von Neumann entropy $- \sum_\delta \lambda_\delta^2 \log\lambda_\delta^2 $. 
As in our previous example, this would represent the entropy of the system if we were to keep only the Hilbert subspace spanned by the block basis states $\ket{\delta}$ corresponding to this bipartition, and discard the orthogonal subspace.
 
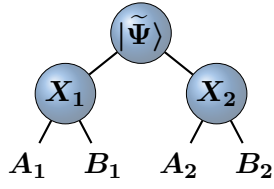
\begin{figure}[t]
\centering

\begin{tikzpicture}[x=0.5cm, y=0.5cm,
font=\boldmath,
 tensor/.style={
    circle,
    draw=black,
    shade,
ball color=tensorblue,
 fill opacity=0.5,
text opacity=1,
    minimum size=8mm,
    inner sep=0pt,
    line width=0.2pt
  },
  leg/.style={line width=0.8pt},
  highlight/.style={line width=1.4pt, red!80},
  correl/.style={line width=1.4pt, green!60!black}
]

\node (A1) at (-3,-0.5) {$A_1$};
\node (B1) at (-1,-0.5) {$B_1$};
\node (A2) at (1,-0.5) {$A_2$};
\node (B2) at (3,-0.5) {$B_2$};

\node[tensor] (W1) at (-2,1.4) {$X_1$};
\node[tensor] (W2) at (2,1.4) {$X_2$};

\node[tensor] (Root) at (0,3) {$\ket{\widetilde\Psi}$};

\draw[leg] (A1) -- (W1);
\draw[leg] (B1) -- (W1);
\draw[leg] (A2) -- (W2);
\draw[leg] (B2) -- (W2);

\draw[leg] (W1) -- (Root);
\draw[leg] (W2) -- (Root);

\end{tikzpicture}

\caption{A two leaf-node TTN (4 physical legs).}
\label{fig:2leaf}

\end{figure}

Thus while Wilsonian coarse-graining is a successive tracing procedure where we cast off information, the TTN always represents the global state. The advantage is of course that other coarse-graining options will become available as we proceed to larger systems, and the TTN will keep track of them all. Continuing in this vein then, consider doubling the size of the system to a four index state, \ie the TTN now grows two {\it leaf nodes}, as in \figref{fig:2leaf}.
In general such a state has the form 
    \begin{equation}
    \label{eq:4indexstate}
    \ket{\Psi}~=~  \psi_{k_1\ell_1 k_2\ell_2} ~\ket{k_1}_{A_1}\ket{\ell_1}_{B_1}\ket{k_2}_{A_2}\ket{\ell_2}_{B_2} ~.
\end{equation}
At first sight this state does not seem to have the factorization of the tree topology in \figref{fig:2leaf} but it may be rendered in that form by repeated application of the same decomposition procedure that we performed on the previous case. 

We can do this in the natural direction of the renormalisation flow by collapsing the bases using SVD. Two  SVD diagonalisations of $\psi_{k_1\ell_1 k_2\ell_2} $ in its $(k_1\ell_1)$ indices and its 
$(k_2\ell_2)$ indices 
yields the state in the Schmidt bases
    \begin{equation}
    \ket{\Psi}~=~  \psi_{i  j} ~\ket{i}_{A_1}\ket{i}_{B_1}\ket{j}_{A_2}\ket{j}_{B_2} ~.
\end{equation}
To coarse-grain, we could again combine them such that 
    \begin{equation}
    \ket{\Psi}~=~  \psi_{i  j} ~\ket{i}_{1}\ket{j}_{2} ~, 
\end{equation}
and now clearly the system is of the same form as the previous bipartite state, and it can be diagonalised with a final SVD in the same way.

How is what we have done isomorphic to the tree in \figref{fig:2leaf}? In order to move from the summation over the physical indices to a summation over the $\ket{i}_1$ Schmidt basis, we need to contract the original coefficients $\psi_{k_1\ell_1  k_2 \ell_2} $ with the three-index tensor 
\begin{equation}
\label{eq:tens}
    X_{1,\, \ell_1 i k_1} ~=~  U^\dagger _{i k_1}   V_{\ell_1 i }~,
\end{equation}
where  as previously the $U$ and $V$ matrices are the SVD matrices. The same is clearly true for the $\ket{j}_2$ basis where we derive an entirely independent three-index tensor $X_{2,\, \ell_2 j k_2} = U^\dagger _{j k_2}   V_{\ell_2 j }$. It is these tensors that can be thought of as occupying the nodes $X_1$ and $X_2$ in the tree.
Proceeding in this way until the root tensor is diagonalised, the entire structure of the state in this form is encoded by the Schmidt spectrum at the root together with the set of three-index tensors that govern how the indices contract by SVD as we flow from the leaves to the root.
In this representation  of the TTN, for which we would say the {\it orthogonality centre} is at the root, we see by inspection of Eq.~\eqref{eq:tens} that the tensors are isometric towards the orthogonality centre (\ie the root in this case). That is $X_{1,\ell_1 i k_1} X^*_{1,\ell_1 i' k_1} = \delta_{ii'}$. When all the tensors in the network are isometric towards the orthogonality centre, the network is said to be in {\it canonical form} about that particular node.

As in the previous examples, the procedure so far can be seen as simply an exact reorganisation of 
our original four index state in Eq.~\eqref{eq:4indexstate}
into a hierarchical form, with the root node again representing the entire global state. 
In this form the state can again be thought of as ``coarse-graining ready''. However, should we decide to commit to a Wilsonian description we now have a choice: we could decide to coarse-grain at either the root node, or at a given depth in the tree. 
To coarse-grain at a certain scale one has to trace out all structure below a cut through the tree. The resulting entropy of the effective theory is then the sum of the latent von Neumann entropies across all the bonds that are severed, each of which is encoded by the singular values  associated with the bipartition of the Hilbert space at that node. 

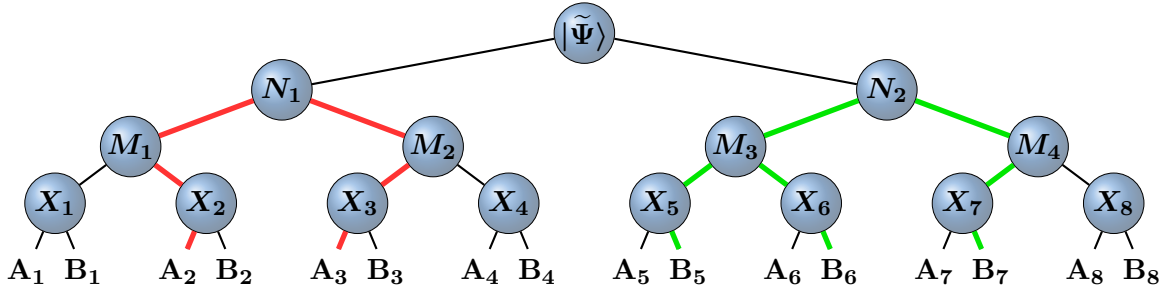
\begin{figure}[t]
\centering
\begin{tikzpicture}[x=0.5cm, y=0.5cm,
font=\boldmath,
 tensor/.style={
    circle,
    draw=black,
    shade,
ball color=tensorblue,
 fill opacity=0.5,
text opacity=1,
    minimum size=8mm,
    inner sep=0pt,
    line width=0.2pt
  },
  leg/.style={line width=0.8pt},
  highlight/.style={line width=2.0pt, red!80},
  correl/.style={line width=2.0pt, green!90!black}
]

\foreach \x/\n in {-14/1,-10/2,-6/3,-2/4,2/5,6/6,10/7,14/8}
{
  \node (A\n) at (\x-0.75,-0.3) {$\bf A_{\n}$};
  \node (B\n) at (\x+0.75,-0.3) {$\bf B_{\n}$};
}

\foreach \x/\n in {-14/1,-10/2,-6/3,-2/4,2/5,6/6,10/7,14/8}
{
  \node[tensor] (L\n) at (\x,1.5) {$X_{\n}$};
  \draw[leg] (A\n) -- (L\n);
  \draw[leg] (B\n) -- (L\n);
}

\node[tensor] (M1) at (-12,3) {$M_1$};
\node[tensor] (M2) at (-4,3)  {$M_2$};
\node[tensor] (M3) at (4,3)   {$M_3$};
\node[tensor] (M4) at (12,3)  {$M_4$};

\draw[leg] (L1) -- (M1);
\draw[leg] (L2) -- (M1);
\draw[leg] (L3) -- (M2);
\draw[leg] (L4) -- (M2);
\draw[leg] (L5) -- (M3);
\draw[leg] (L6) -- (M3);
\draw[leg] (L7) -- (M4);
\draw[leg] (L8) -- (M4);

\node[tensor] (N1) at (-8,4.5) {$N_1$};
\node[tensor] (N2) at (8,4.5)  {$N_2$};

\draw[leg] (M1) -- (N1);
\draw[leg] (M2) -- (N1);
\draw[leg] (M3) -- (N2);
\draw[leg] (M4) -- (N2);

\node[tensor] (Root) at (0,6) {$\ket{\widetilde{\Psi}}$};

\draw[leg] (N1) -- (Root);
\draw[leg] (N2) -- (Root);

\draw[highlight] (A2) -- (L2);
\draw[highlight] (L2) -- (M1);
\draw[highlight] (M1) -- (N1);
\draw[highlight] (N1) -- (M2);
\draw[highlight] (M2) -- (L3);
\draw[highlight] (L3) -- (A3);

\draw[correl] (L5) -- (M3);
\draw[correl] (L5) -- (B5);
\draw[correl] (L6) -- (B6);
\draw[correl] (L6) -- (M3);
\draw[correl] (M3) -- (N2);
\draw[correl] (L7) -- (B7);
\draw[correl] (L7) -- (M4);
\draw[correl] (M4) -- (N2);

\end{tikzpicture}

\caption{An eight leaf node TTN with two sets of physical indices $A_n$ and $B_n$. This is a prototypical set-up for a one dimensional QFT with two scalar fields $\phi_A$ and $\phi_B$  discretized over eight space points $X_{1\ldots 8}$ as the quadrature variables of the local Hilbert spaces, $\phi_A\equiv \hat q_{A,n}$ and  $\phi_B\equiv \hat q_{B,n}$ respectively. Highlighted in red is the path connecting points $A_2$ and $A_3$ through their lowest common ancestor (LCA) at $N_1$, which has to be traversed in order to implement the  $e^{i\delta t \, \hat q_{A,2} \, \hat q_{A,3}} $ link gate. The cluster of bonds in green shows the nontrivial contractions within the entanglement wedge of the $\ell = 3$ contiguous region ${\phi_B(X_5)\phi_B(X_6)\phi_B(X_7)} $. This roughly determines how correlation lengths translate into tree-depth.}
\label{fig:8leaf}
\end{figure}

To make this process feasible for larger trees, for example the eight leaf node tree in \figref{fig:8leaf}, we now introduce some additional tools. 
To handle larger trees it becomes necessary to approximate the full state using {\it truncation}.
Typically one imposes a truncation on both the minimum allowed singular value (the {\it cut-off}), and also the allowed number of singular values (the {\it maximum bond dimension}). The latter naturally tends to grow as bonds from lower down the tree feed into fewer bonds.
This  loss of information in the TTN through truncation gives it  an irreversibility that it did not previously have, and we accordingly  write the root node in \ref{fig:2leaf} with a tilde, as  $\ket{\widetilde \Psi }$.
Note this does not change the important structural difference between Wilsonian renormalisation and the TTN, namely that the former generally yields mixed states, while in a TTN the truncated state $\ket{\widetilde \Psi}$ remains pure: in this sense it is always trying to approximate the global state. 

Truncation, and indeed any local operation on the tree such as finding the entropy associated with a particular bond, requires the ability to work at a specific node. This is made possible by the fact that the orthogonality centre can be moved to any node by SVD, with the network taking  canonical form at that location. That is one may write the state around a node as
\begin{equation}
    \ket{\Psi}~=~ T_{ik\ell}\, \ket{P_i}\, \ket{C_{L,k}}\,\ket{C_{R,\ell}}~,
\end{equation}
where $T_{ijk}$ is the tensor at the node, and where the node's parent $\ket{P_i}$ and its left- and right-children $\ket{C_{L,k}}$ and $\ket{C_{R,
\ell}}$ form orthonormal bases for the subtrees.  
As long as one keeps track of the orthogonality centre (which is typically started off at the root node), it is possible to move it to any desired node.
In this canonical form all the tensors below the orthogonality centre are isometric towards it, and the Schmidt decomposition and truncation can be  performed locally by SVD as 
\begin{equation}
   T_{ik\ell}~=~ U_{i\delta } \lambda_\delta V^\dagger _{\delta,k\ell}
\end{equation} 
where the $k\ell$  indices of the child nodes have been flattened into a single index that appears on the single diagonalisation matrix $V$. As before, the singular values  are the Schmidt coefficients, $\lambda_\delta$, now associated with the bipartition of the Hilbert space at this particular node. They define the entropy that {\it would} be incurred if the parent bond were severed. The truncation is then systematically performed by retaining only the largest of these Schmidt eigenvalues feeding into the parent bond. 

This ability to systematically  truncate TTNs is crucial for making them tractable. Without it, carrying out the above decomposition  in a generic system would generate bond dimensions and hence tensors whose size grows exponentially with the system. For example, two child bonds of dimension $d$ lead to $d^2 $ Schmidt eigenvalues. Without truncation at tree depth $n$ we would be dealing with bond dimensions of order $\chi_{\rm max} \sim d^{2^n}$, which would soon become computationally intractable for a network of any reasonable size. Indeed, with no truncation we would gain precisely nothing by expressing the state as a TTN: the number of physical indices on a tree is $N=2^n$, and therefore without truncation the largest bond dimension is $d^N$, the same as the dimensionality of the entire state. 

By only retaining a truncated  spectrum of Schmidt values at each node we retain the most important contributions to long range entanglement.
Extending the topology as in \figref{fig:8leaf} to $N\gg 1$ leaf nodes makes it possible to encode a one-dimensional field theory of two scalar fields, with the separate leaf nodes $X_{n=1\ldots N}$ corresponding to the $N$-site discretized space-time, and the hope is that the TTN will be able to keep track of long-range entanglement even in such a large system. 
The power of TTNs lies in the fact that  one can now adopt a variational procedure. The tree is initialised with trivial internal bonds  (\ie bonds of dimension one) with only the leaf nodes carrying non-trivial structure. A physical condition is then imposed on the TTN, for example that the state collectively approximates a quantum field theory (QFT) vacuum. Entanglement is thereby allowed to grow dynamically along the tree towards the root, up to the limits imposed by the chosen cut-off and maximum bond dimension. In this way the entanglement in the entire tree encodes the response of the state to the physical condition. 

This then is the structure of the TTN. But how does it tie in with the entropic scaling in renormalisation which was the original motivation for this discussion? Compared to the linearly arranged MPS encoding of field theory the great advantage of the TTN is that it has a more natural and efficient encoding of information flow, because entanglement structure is organised into subtrees. Thus correlations across a given distance in space naturally correspond to depth in the network. For example in \figref{fig:8leaf} the green lines show the subtree for a contiguous block of nodes of length $\ell =3$ which can be thought of as its ``entanglement wedge''. Correlations between any pair of leaves in this block only have to traverse the path passing through the {\it lowest common ancestor} (LCA) of the two leaf nodes. In the block shown for example $N_2$ is the LCA of $X_6$ and $X_7$ while $M_3$ is the LCA of $X_5$ and $X_6$. 
In general the number of bonds along the path connecting two leaf nodes a distance $\ell$ apart scales as $\log_2 \ell$, as opposed to the linear scaling of an MPS. This is in turn reflected in the entropies as follows. In a block of size $\ell$ there are of order $\log_2\ell$ bonds to cut to detach the subtree from the TTN. Thus the von Neumann entropy $S(\ell)$ contained in a contiguous block of size $\ell$ scales as $S(\ell ) \sim  (\log_2\ell) \log\chi $, where $\chi$ is the bond dimension (since a flat distribution of Schmidt values at a single bond would have $\lambda_i^2 = 1/\chi$ and hence entropy $-\sum_i \lambda_i ^2 \log \lambda_i^2 = \log\chi$ across that bond). This allows a natural identification of renormalisation coarse-graining scale with tree-depth. By contrast an MPS has entropy $S(\ell ) \sim  \log\chi$ regardless of the size $\ell$ of a contiguous block, and thus the bond dimensions themselves must grow with the correlation length in order to encode long range entanglement, which can make it technically more challenging to model. This appealing geometric TTN  picture of coarse-graining will be central to our discussion.

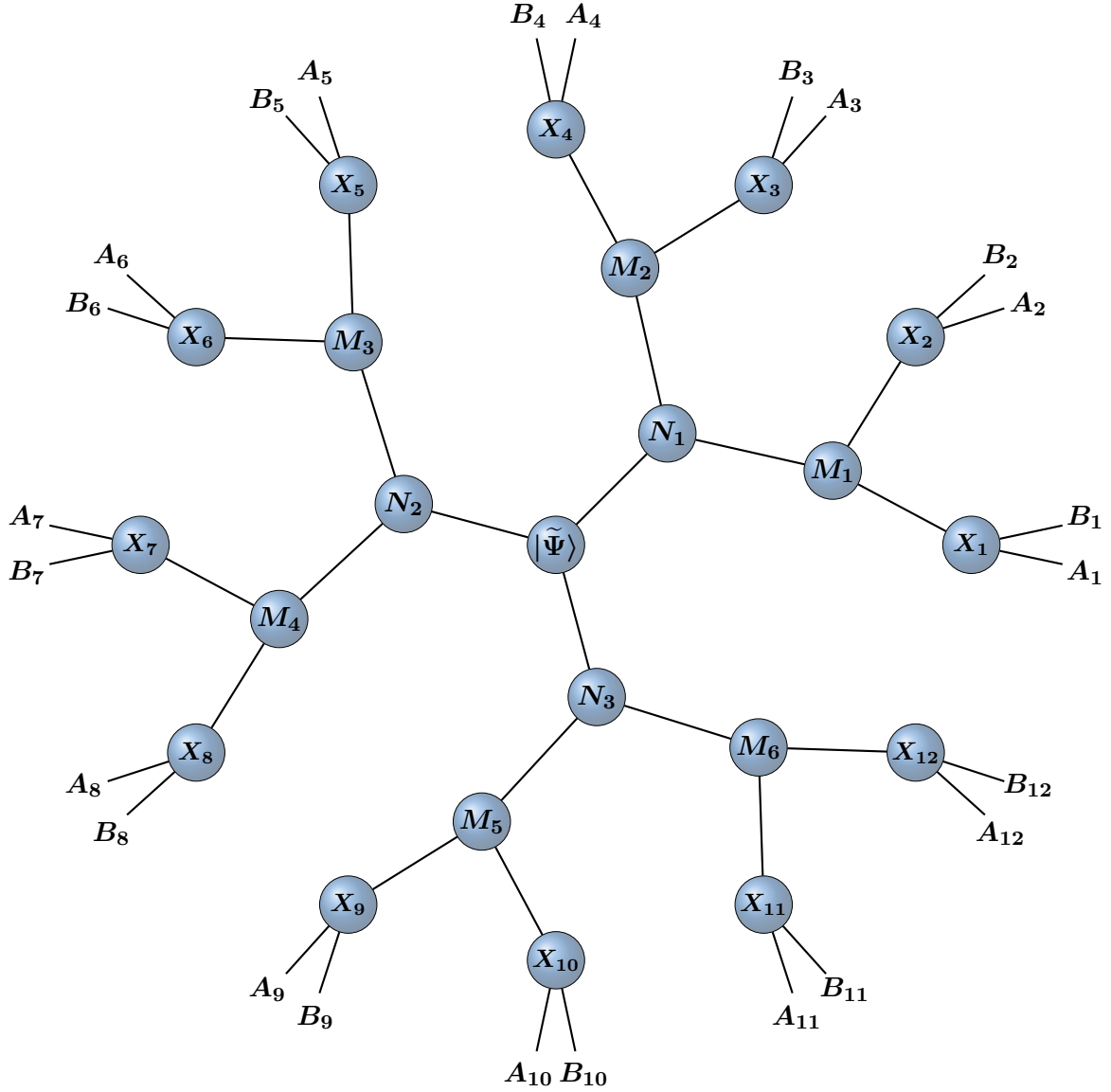
\begin{figure}[t]
\centering
\begin{tikzpicture}[
  font=\boldmath,
  tensor/.style={
    circle,
    draw=black,
    shade,
    ball color=tensorblue,
    fill opacity=0.5,
    text opacity=1,
    minimum size=8mm,
    inner sep=0pt,
    line width=0.2pt
  },
  leg/.style={line width=0.8pt},
]

\def\RN{2.2}
\def\RM{4.0}
\def\RL{5.8}


\node[tensor] (Root) at (0,0) {$\ket{\widetilde{\Psi}}$};

\node[tensor] (N1) at (45:\RN)  {$N_1$};
\node[tensor] (N2) at (165:\RN) {$N_2$};
\node[tensor] (N3) at (285:\RN) {$N_3$};
\draw[leg] (Root) -- (N1);
\draw[leg] (Root) -- (N2);
\draw[leg] (Root) -- (N3);

\node[tensor] (M1) at (15:\RM)  {$M_1$};
\node[tensor] (M2) at (75:\RM)  {$M_2$};
\node[tensor] (M3) at (135:\RM) {$M_3$};
\node[tensor] (M4) at (195:\RM) {$M_4$};
\node[tensor] (M5) at (255:\RM) {$M_5$};
\node[tensor] (M6) at (315:\RM) {$M_6$};
\draw[leg] (N1) -- (M1);
\draw[leg] (N1) -- (M2);
\draw[leg] (N2) -- (M3);
\draw[leg] (N2) -- (M4);
\draw[leg] (N3) -- (M5);
\draw[leg] (N3) -- (M6);

\node[tensor] (L1)  at (0:\RL)   {\small $X_1$};
\node[tensor] (L2)  at (30:\RL)  {\small $X_2$};
\node[tensor] (L3)  at (60:\RL)  {\small $X_3$};
\node[tensor] (L4)  at (90:\RL)  {\small $X_4$};
\node[tensor] (L5)  at (120:\RL) {\small $X_5$};
\node[tensor] (L6)  at (150:\RL) {\small $X_6$};
\node[tensor] (L7)  at (180:\RL) {\small $X_7$};
\node[tensor] (L8)  at (210:\RL) {\small $X_8$};
\node[tensor] (L9)  at (240:\RL) {\small $X_9$};
\node[tensor] (L10) at (270:\RL) {\small $X_{10}$};
\node[tensor] (L11) at (300:\RL) {\small $X_{11}$};
\node[tensor] (L12) at (330:\RL) {\small $X_{12}$};

\draw[leg] (M1) -- (L1);
\draw[leg] (M1) -- (L2);
\draw[leg] (M2) -- (L3);
\draw[leg] (M2) -- (L4);
\draw[leg] (M3) -- (L5);
\draw[leg] (M3) -- (L6);
\draw[leg] (M4) -- (L7);
\draw[leg] (M4) -- (L8);
\draw[leg] (M5) -- (L9);
\draw[leg] (M5) -- (L10);
\draw[leg] (M6) -- (L11);
\draw[leg] (M6) -- (L12);


\foreach \ang/\n in {0/1, 30/2, 60/3, 90/4, 120/5, 150/6,
                     180/7, 210/8, 240/9, 270/10, 300/11, 330/12} {
  \draw[leg] (L\n) -- ++(\ang-12:1.3);
  \draw[leg] (L\n) -- ++(\ang+12:1.3);
  \node[font=\boldmath] at (\ang-3.0:7.4) {$A_{\n}$};
  \node[font=\boldmath] at (\ang+3.0:7.4) {$B_{\n}$};
}

\end{tikzpicture}
\caption{A more symmetrical version of a TTN in which every node, including the root node, carries three indices. The bulk encodes the renormalisation geometry while the physical degrees of freedom live on the boundary. In the gapped system that we consider here we do not need to impose hyperbolic tiling: if one does so then in principle the root node no longer occupies a special position. }
\label{fig:12leaf_circular}
\end{figure}

As a final remark we note that other network topologies now present themselves as interesting  alternatives, and this is an opportunity to briefly comment on the links to holographic networks. Consider for example the network in \figref{fig:12leaf_circular}. This network has the same physical interpretation as the standard TTN, with the UV leaves now living on the circular boundary, however now the root at the centre has the same rank as all the other tensors. This representation is suggestive and is as we will see more natural for systems with periodic boundary conditions. However the root node still constitutes a special location in this network, as it marks the endpoint of renormalisation flow. By contrast holographic networks such as those in the HaPPY code \cite{Pastawski:2015qua} implement a hyperbolic tiling structure within the 
network. At infinite $N$ in such a network there is no preferred bulk point and the root node becomes simply a choice of coordinate. This provides a nice geometric understanding of the special properties of holographic systems. Since the systems we will consider here are gapped, there is no particular need to reproduce hyperbolic tiling, and the root node will remain special for us. 

\section{Implementing Scalar QFT on the TTN}
\label{sec:QFTImp} 

Thus far I have  described the general framework for encoding renormalisation in a Hilbert space built from a lattice of leaf nodes on a standard TTN. We now need to build the foundations for the BRTN within this framework. Importantly the BRTN  requires the continuous-variable degrees of freedom of a genuine field -- as distinct from the spin-chain and Ising-type models that have been the main focus of tensor network studies to date. Therefore, as a first step we need to develop the machinery for encoding a QFT on the TTN, which is the purpose of this section. 

Developing this machinery will elucidate two important aspects that  will carry forwards directly to the BRTN. First we will see how to prepare the QFT vacuum by so-called imaginary-time evolution using {\it time-evolving block decimation} (TEBD). Second we will demonstrate how entanglement spreads throughout the tree during this process, by computing the average bond entanglement at each layer. This provides a benchmark against which the Born-Reciprocal Tensor Network (BRTN) construction will be tested in the sections that follow.
  
As this particular study is concerned with the BRTN, we will not proceed to consider explicit real time Hamiltonian evolution to perform for example scattering. However, once the QFT vacuum has been prepared, we will have built all the machinery required to do so. Future work will look at these physical applications~\cite{tocome}. 

 The presentation will be specifically aimed at TTNs that can model a QFT of a pair of interacting scalars, which will allow a more explicit and self-contained discussion: that is all the tensors except the root node will be three-index, the internal bond dimensions will as mentioned be allowed to grow arbitrarily up to a defined maximum, and the leaf nodes will be defined with a large dimension local Hilbert space in order to approximate continuous field values.  Throughout, numerical tensor-network calculations will be performed using the {\tt ITensor} library in {\tt Julia}~\cite{itensor,itensor-r0.3} (see also \cite{10.21468/SciPostPhysLectNotes.8} for a compendium of algorithms and terminology). This software provides index-labelled tensor objects, contraction routines and automated SVD with controlled truncation, with the TTN geometries themselves being defined explicitly. 

As mentioned, here we will be considering the TEBD method for time evolution. This aligns closely with the truncation picture of renormalisation that was presented in the introduction. It is worth mentioning an alternative method that could have been used, which is the  
 time-dependent variational principle (TDVP)~\cite{Haegeman:2011,Lubich:2015,Paeckel:2019yjf} which was extended to tensor network geometries in Ref.~\cite{Holtz:2012,Lubich:2013,Haegeman:2016}, and which 
 has been used for real-time applications such as scattering in for example Refs.~\cite{Rigobello:2021fxw,  PhysRevResearch.6.033057,PRXQuantum.5.037001}. In TDVP the Schr\"odinger evolution is projected onto the TTN manifold, and entanglement growth is constrained explicitly by geometry rather than by successive truncation. Thus there is less of a `state-counting' interpretation of the information flow. Nevertheless, it could be an interesting alternative approach. 

\subsection{The set-up for scalar field theory}
 
 For this initial study it will as mentioned above be possible to consider a two-scalar QFT, with the nodes $X_n$ corresponding to space points in a 1+1D field theory. The dimensionful space coordinate we define as $x_n = a X_n $ where $a$ is a length scale. 
  As there are two subspaces at each leaf node of the TTN, $A$ and $B$, it will be these legs that encode the two scalar fields, $\phi_A$ and $\phi_B$. 

In order to simulate QFT there are essentially two main tasks. The first is to encode the fields $\phi_A$ and $\phi_B$ themselves in terms of the local Hilbert spaces of the lattice. This gives us a representation of the entire QFT at some initial point in time as a huge quantum mechanical state $\ket{\Psi(0)}$. The second task is to then encode its Schr\"odinger evolution:
\begin{equation}
    \ket{\Psi(t)} ~=~ e^{-i H t}\ket{\Psi(0)}~,
\end{equation}
for which we will need the Hamiltonian.

We first describe how the scalar fields that live on the leaves can be built up from the local Hilbert space. The physical lattice interpretation is conventional from a field theory perspective, however, we may think of it as a universal, platform-agnostic prescription for computation. This viewpoint was developed in Refs.~\cite{Abel:2024kuv,Abel:2025zxb,Abel:2025pxa}, where the corresponding structure was introduced as a ``qumode lattice'', making the framework naturally portable to implementations of scalar QFTs on other platforms such as photonic systems, analogous MPS architectures or, as here, TTNs. 

The implementation is as follows. At each of the $N$ leaf nodes of the TTN there is a local Hilbert space, and we build the QFT by coupling together a lattice of these local quantum mechanical Hilbert spaces, regarded as individual oscillators with quasi-continuous quadrature variables $\hat q$ and $\hat p$. The field operator $\phi(x)$ and its corresponding conjugate momentum operator $\pi(x)$ can be directly equated with the quadrature variables of the oscillators (written as a  density) at that position on the lattice, such that one can identify  
\begin{equation}\label{eqn:latticeFields}
\hat q_n(t) ~\leftrightarrow ~  \phi(x_n, t)~, \qquad \hat p_n (t) ~\leftrightarrow ~  \pi( x_n , t)~.
\end{equation}
For the discussion in this section the index $A$ and $B$ corresponding to $\phi_A$ and $\phi_B$ will be omitted: the procedure that we will describe is trivially repeated for each physical index. 

To make the above identification, we first express the spatial derivatives by finite difference,
\begin{equation}\label{eqn:finiteDifference}
\partial_x \phi(x_n, t) ~\leftrightarrow~ \frac{{\hat q}_{n+1}(t) - {\hat q}_n(t)}{a}~.
\end{equation}
There is then a configuration of nearest-neighbour couplings of the oscillators that reproduces the QFT Hamiltonian in the large $N$ limit, which is found by summing the contributions from the coupled individual qumodes over the entire lattice. If the 1+1D field theory potential is $\mathscr{V}(\phi)$, then we may write the Hamiltonian in terms of the dimensionless potential 
\begin{equation}
    V (\hat q ) ~=~ a^2 \mathscr{V} (\hat q )
\end{equation}
as follows:
\begin{equation}
\label{eq:disc_ham}
H ~=~ {a^{-1}} \sum_{n=1}^N\left[  \frac{ {\hat p}_n(t)^2}{2} + 
\frac{(\hat q_{\,{n+1}}(t) - \hat q_n(t))^2}{2} + {V}(\hat q_n) \right]~,
\end{equation}
for arbitrary potential  ${V}(\phi)$, where the conjugate momentum is $\hat p_n \equiv a \, \partial _t \hat q _n $, and  
with variables normalised such that 
\beq 
[ \hat q_n ,\,\hat p_m]~=~ i\delta_{nm} ~.
\eeq
For example the dimensionless potential in a free field theory of mass $m_{\rm phys}$ is $V(\hat q_n ) =\frac{ m^2}{2}\hat q_n ^2$, where $m=m_{\rm phys} a$ is the dimensionless mass. More generally we may regard $V$ as being the field theory potential with all its dimensionful parameters expressed in units of $a$.
The overall $a$ prefactor in the original Hamiltonian ensures that it scales extensively with the size of the lattice, ${\rm Vol}=aN$, corresponding to the space integral in the continuum field theory. For numerical work we may then absorb the remaining overall $a^{-1}$ factor by defining it to be our unit of time.

The only platform-specific question then becomes how to encode the local Hilbert space. On tensor networks it must be discretised, and this can be done in several ways. For example one can simply discretise the quadrature $q_n$ into a $d$-dimensional space $q_n^{j=1\ldots d}$, where
we consider eigenvalues $q_n^j \in [-L,L]$ with 
\begin{equation}
    q_n^j ~=~ -L + j\times  \delta q 
\end{equation}
where $\delta q = 2L/d$. This $d$ dimensional space would correspond to a $d$-dimensional bond on each leg of the leaf node.
Thus the quantum mechanical wavefunction is given in the $\ket{q_n^i}\equiv \ket{\phi(x_n)}$ basis in the obvious way: 
\begin{equation}
    \ket{\psi_n} ~=~ \sum_{i=1}^{d}   \ket{q^i_n} \braket{q^i_n}{\psi_n}~.
\end{equation}
In this basis the operators $\hat q_n $ and $\hat p_n^2 $ act as matrices on the local Hilbert space of site $n$ that can be deduced by finite difference: 
\begin{align}
\hat{q}_n~&=~{\rm diag}
\left(
q^1_n  \ldots q_n^d\right)~~,~~~~
    \hat{p}_n^2  ~=~- \frac{\hbar^2}{\delta q^2} \begin{pmatrix}
 2      & -1 & ~ & ~~ & \ldots \\
-1 & 2 & -1 &~ & \ldots\\
~ & -1 & 2 & ~ & \ldots\\
\vdots & ~ & ~ & \ddots & -1 \\
\vdots & ~ & ~& ~~-1 & 2   
\label{eq:qandp}
\end{pmatrix}~.
\end{align}
This basis is convenient because $\hat q_n$ is diagonal so it is trivial to construct $V(\hat q_n)$. However in practice the dimension $d$ of the local Hilbert space is restricted by resource limitations to be  relatively small (we shall take $d=6$). It is then more efficient to use an improved basis, the so-called discrete variable representation (DVR), in which $\hat q_n$ is also diagonal~\cite{10.1063/1.462100}.

For completeness let us briefly recap how it is constructed. In the DVR basis, rather than approximating operators by finite difference, one introduces a truncated harmonic oscillator basis of dimension $d$ and then diagonalises the position operator in that subspace. Concretely, one begins with the first $d$ harmonic oscillator eigenstates
$\{\ket{m}\}_{m=0}^{d-1}$ and forms the matrix elements of $\hat q$ as a $d\times d$ matrix,
\begin{equation}
    (q_n)_{mm'} ~=~ \bra{m} \hat q_n \ket{m'} .
\end{equation}
Then diagonalising {\it this} matrix with a unitary matrix $U$, $
U^\dagger q_n U ~=~ 
    \mathrm{diag}(\tilde q_n^{\,1},\ldots,\tilde q_n^{\,d}),
$ yields the DVR basis states as
\begin{equation}
    \ket{\tilde q_n^{\,j}} ~=~ \sum_{m=0}^{d-1} U_{mj}\ket{m}.
\end{equation}
For example with $d=6$ the position operator
expressed in the truncated Fock basis is given by 
\begin{equation}
\hat q_n ~=~
\frac{1}{\sqrt{2}}
\begin{pmatrix}
0 & 1 & 0 & 0 & 0 & 0 \\
1 & 0 & \sqrt{2} & 0 & 0 & 0 \\
0 & \sqrt{2} & 0 & \sqrt{3} & 0 & 0 \\
0 & 0 & \sqrt{3} & 0 & 2 & 0 \\
0 & 0 & 0 & 2 & 0 & \sqrt{5} \\
0 & 0 & 0 & 0 & \sqrt{5} & 0
\end{pmatrix}~.
\label{eq:q_HO_truncated}
\end{equation}
The eigenvalues of $\hat q_n$ are the zeros of the Hermite polynomial $H_6(x)$, hence in the DVR basis
with frequency $\omega_0$ we have  
\begin{equation}
\hat{q}_n ~=~
\frac{1}{\sqrt{\omega_0}}\, {\rm diag}\left(-2.35,
 -1.34,
 -0.44,\,
 0.44,\,
 1.34,\,
 2.35\right)~.
 \label{eq:qDVR}
\end{equation}
The operator $\hat p_n^2$ in the DVR basis is then found by rotating it from the truncated Fock basis with the same unitary transformation $U_{mj}$; this yields a dense matrix with entries of the form
\begin{equation}
    \left({\hat p}_n^2\right)_{ij} ~=~ \omega_0 \frac{(-1)^{i+j}}{(\tilde q_n^i-\tilde q_n ^j )^2 }~. \label{eq:p2DVR}
\end{equation}

These operators can now be employed to  perform  time evolution on the state using  Suzuki-Trotter decomposition~\cite{pjm/1103039709, 10.1063/1.526596}, where the full time-evolution operator,
\begin{equation}
    U(t) ~=~ e^{-i H t}~,
\end{equation}
is approximated with 
\begin{equation}
    {U} (t) ~=~ \left[ \prod_i e^ {- i H_i \delta t} \right]^{t/\delta t}~,
\end{equation}
where the Hamiltonian has been split up into a sum of non-commuting parts, $H = \sum_i H_i$, and where the total evolution time is \textit{Trotterised} by discretising $t$ into small steps of $\delta t$. 

The implementation of the $U(t)$ operator on the TTN can be divided into two pieces, one that is relatively straightforward and one that is considerably more challenging. The former consists of the local onsite Hamiltonian which can be read off Eq.~\eqref{eq:disc_ham}:
\begin{equation}
\label{eq:qumode_ham}
H_{{\rm loc}} ~=~ \sum_{n=1}^N \, \frac{1}{2} {\hat p}_n^2 + 
 {\hat q}_{n}^2 + V(\hat q_n) ~,
\end{equation}
where we will henceforth set the lattice-spacing to be $a=1$. 
These operators act locally on both physical indices of the leaves. They do not cause entanglement and can be applied simply using the DVR matrix version of $\hat q$ and $\hat p^2_n$ from Eqs.~\eqref{eq:qDVR} and \eqref{eq:p2DVR} respectively. For future reference note that in a free field theory of mass $m$ the effective angular frequency of the local oscillators is given by $\omega_0^2 = 2+ m^2$ (because of the diagonal contribution from the gradient terms), and $\omega_0$ will be the relevant frequency to use for the DVR basis. Thus the diagonal gate that needs to be applied is
\begin{equation}
U_{{\rm loc},n}(\delta t) ~=~ e^{- i {\delta t}  \,\left( \frac{1}{2} {\hat p}_n^2 + 
 {\hat q}_{n}^2 +V(\hat q_n) \right)  } ~.
\end{equation}
This gate can simply be determined as a single dense matrix that can be applied to each site.

The more troublesome contribution is the hopping term in Eq.~\eqref{eq:disc_ham}, namely 
\begin{equation}
H_{{\rm hop}} ~=~ \sum_{n=1}^{N-1} \, 
 - \, {\hat q}_{n}{\hat q}_{n+1} ~.
\end{equation}
For a single Trotter step we must  apply the following link-gate on each of the sites (for each of the fields $\phi_A $ and $\phi_B$):
\begin{equation}
U_{{\rm hop},n}(\delta t) ~=~ e^{i \delta t  \, 
 {\hat q}_{n}{\hat q}_{n+1} } ~.
\end{equation}
These gates are responsible for the spread of entanglement through the tree and each involves a distinct contraction path within the network.
\figref{fig:8leaf} shows in red the path that must be navigated to implement the  $U_{{\rm hop},2}$ gate for $\phi_A$. As the gate has operators from both leaves, one must perform the following procedure for each gate. First we locate the LCA of the two leaves, \ie  node $N_1$ in \figref{fig:8leaf}. Even though the coupling is  nearest neighbour, the LCA may occur at any height in the tree; \eg for the operator $U_{{\rm hop},N/2}$ on an $N$ leaf-node TTN, the LCA is the root itself. Next we must compute a replacement for the train of tensors along the path, which we refer to as the {\it path-product}. Initially the path-product takes the form 
\begin{align}
    P_{n,n+1}~=~ X^{i_n,j_n}_n \cdot M^{a_1}_1 \ldots M^{a_\ell}_\ell \cdot  {\rm LCA}^\alpha \cdot 
    M^{a_{\ell+1}}_{\ell+1}\ldots M^{a_{2\ell}}_{2\ell}\cdot X^{i_{n+1},j_{n+1}}_{n+1}~,
\end{align}
where the intermediate tensors are labelled generically as $M^{a_k}_{k=1\ldots \ell}$, and where the LCA is at level $\ell+2$. Here the contractions between the tensors are over the parent/child bonds along the path, while we show explicitly the indices $a_k$ that connect to the adjacent subtrees, as well as the parent bond index $\alpha$ of the LCA (which is present if it is not the root), and the physical indices of the leaves. 
For example in \figref{fig:8leaf} we have (now suppressing indices) $P_{2,3} ~=~ X_2 \cdot M_1 \cdot N_1 \cdot M_2 \cdot X_3 $. 

Thus we wish to deduce  
$P'_{n,n+1} ~=~ U_{{\rm hop},n} \otimes P_{n,n+1} $, where $P'_{n,n+1}$ is a new path product that has the same product form as $P_{n,n+1}$.  
However acting with the gate directly on the fully contracted path-product would require manipulating a tensor with all the intermediate bond indices exposed simultaneously, which is computationally impractical. Instead, we successively re-factorise the path by SVD so that the two physical tensors are brought into a position where they are adjacent to the LCA. The hopping gate may then be applied locally, after which the result is decomposed again and redistributed along the path to restore the original path-product ordering. In practice, all of this linear-algebraic manipulation is carried out on the extracted path-product without touching the TTN. Once the new path product has been calculated, the set of modified tensors can be sewn into place in the TTN.

With these local and hopping operators to hand we may now implement Hamiltonian time evolution by repeated application of the following Strang split operator:
\begin{equation}
\label{eq:Udt}
    {U} (\delta t) ~=~ \prod_n U_{{\rm loc},n}(\delta t/2) \, U_{{\rm hop},n}(\delta t) \, 
    U_{{\rm loc},n}(\delta t/2) ~.
\end{equation}
This completes all the machinery required for Schr\"odinger evolution of the TTN. 

\subsection{Vacuum preparation and properties}

\label{subsec:TTNvac_prep}

The crucial initial step in the set-up of scalar QFT on the TTN is the preparation of the vacuum. This is in fact the first real physics that we will do with the TTN, and the properties of the vacuum will as mentioned be the main focus of this study. In order to prepare the vacuum we will follow the method used in Ref.~\cite{Abel:2025pxa} which is to perform TEBD with imaginary time, that is with evolution operators of the form in Eq.~\eqref{eq:Udt} with $\delta t = - i \delta \tau$. The resulting evolution is of course non-unitary, and therefore after each step we rescale the state to normalise it. That is we evolve the state as
\begin{align}
    \ket{\Psi (\tau+\delta \tau) } ~=~ \frac{U(-i \delta\tau)\ket{\Psi(\tau)}  }{||U(-i \delta\tau ) \ket{\Psi(\tau)} || }~.
\end{align}
Since the imaginary time evolution acts as $U(-i \delta\tau)\ket{\Psi(\tau)}\,=\, e^{-H \delta \tau}\ket{\Psi(\tau)} $ under this evolution, $\ket{\Psi}$ will naturally converge to the state with smallest $H$ eigenvalue, namely the QFT ground state.

To test this method we adopted a lattice of $N=128$ points. A strongly gapped system is preferred for this study in order to avoid criticality and maintain a finite correlation length, and keep long-range correlations under control.
We therefore take as our benchmark model the free-field system with a bare mass of $m=1$, for which the dispersion relations and covariance matrices are known. 
In detail the covariance matrix of the system 
are 
$d \times d$ matrices characterising the correlations between sites which can simply be measured from the two-point expectation values:
\begin{align}
C_q &= \langle \hat{q} \hat{q}^T \rangle - \langle \hat{q} \rangle \langle \hat{q}^T \rangle~.
\end{align}
Its Fourier transform on the lattice is the structure factor,
\begin{align}
S(k) ~=~ 1/2\omega_k~,
\end{align} 
where $\omega_k$ is the free-field dispersion relation on the lattice, namely 
\begin{equation}\label{eqn:omegas}
\omega_k ~=~ a^{-1}\sqrt{ m^2 + {4}\sin^2 \left( k a/2 \right) }~.
\end{equation}
On the $N$-lattice the momentum values are discretised as 
\beq 
\label{eq:NinftyK}
k_n ~=~ 2\pi \frac{n}{aN}~~~; ~~n \in {\mathbbm Z} ~,
\eeq
where in the large $N$ limit we recover the usual relativistic relation, 
$\omega_k^2=m_{\rm phys}^2 +k^2$~.

\begin{figure}[t!]
\centering
\includegraphics[keepaspectratio, width=0.75\textwidth]{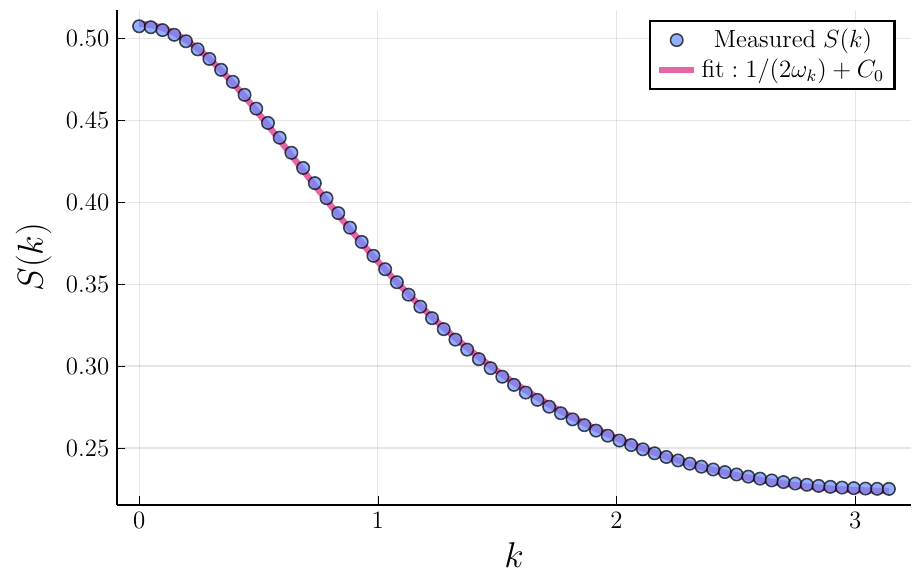}
\caption{Comparison of the structure factor of the QFT vacuum versus the theoretical one, for the free field with $N=128$ space points, and mass $m=1$ in the potential. Here we take local DVR Hilbert space of dimension 6, cut-off of $1\times 10^{-6}$ and maximum bond dimension of $20$. The best fit has $m_{\rm eff}=0.982$ and $C_0=-5\times 10^{-4}$. }
\label{fig:Sk_128}
\end{figure}

In \figref{fig:Sk_128} we show the resulting fit for $S(k)$ with imaginary time evolution.
In order to make the imaginary time TEBD efficient we use schedules for the evolution, with gradually decreasing  Trotter step in order to  improve  convergence. A suitable schedule with a cut-off of $10^{-6}$ and a maximum bond dimension of $20$ is as follows: 
\begin{center}
\begin{tabular}{c|c}
$\delta \tau$ & $n_{\rm sweeps}$  \\
\hline
\hline
0.05   & 30    \\
0.01   & 10  \\
0.005  & 10  \\
0.0025 & 10  \\
0.0001 & 10 
\end{tabular}
\end{center}

An important aspect of the schedule is to begin with a $\delta\tau $ step which is large enough that it allows the bond dimensions to inflate properly before any truncation occurs. Initial transient behaviour followed by slower convergence is preferable. Likewise a too hard truncation can lead to trivial bonds (\ie bonds of dimension one) throughout the network.

In the fit we allow two free parameters, following the approach of the MPS study in Ref.~\cite{Abel:2025pxa}: one is an {\it effective} mass in $\omega_k$ rather than the bare mass that appears in the potential. The other free parameter we allow is a constant shift $C_0$ which is also a function of the discrete lattice-spacing. Thus in total, setting $a=1$ we fit 
\begin{equation}
    S(k) ~=~ \frac{1}{2\sqrt{m_{\rm eff}^2 + {4} \sin^2 (k/2)  }} +C_0~.
    \label{eq:Skfit}
\end{equation}
In the $N\to\infty, \, \delta\tau\to 0$ limit the parameters $m_{\rm eff}$ and $C_0$ approach the continuum values of  $m=1$ and $C_0=0$. The fit in \figref{fig:Sk_128} is clearly a close match, and we can conclude that the TTN is able to learn the dispersion relation for the scalar vacuum of the free theory, and in fact for any arbitrary potential $V(\hat q)$. 

\begin{figure}[t!]
\centering
\includegraphics[keepaspectratio, width=0.75\textwidth]{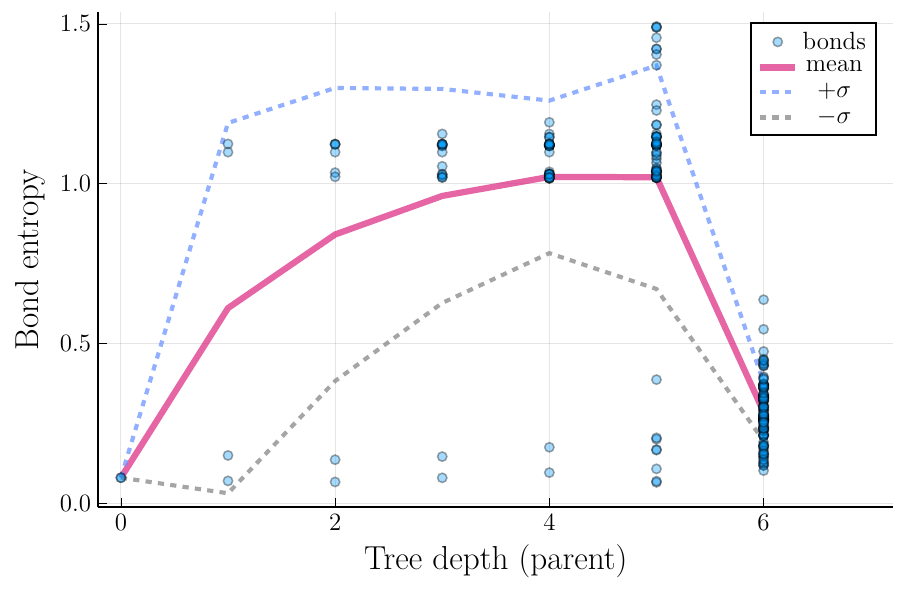}
\caption{Bond entropy with depth in the standard TTN, showing characteristic peaking in the centre, where the bonds are dividing the global system into roughly equally sized subsystems.}
\label{fig:ent_128}
\end{figure}

It is interesting at this point to examine  the bond entropies in the vacuum. This is easily done by canonicalising around each bond and computing the von Neumann entropy of the Schmidt eigenvalues going into the parent bond. In \figref{fig:ent_128} we show the resulting averaged entropies as a function of tree-depth. 
In this case the bond entropies peak in the middle of the tree. This is to be expected because this is where the canonicalised bond is dividing the tree into roughly two halves: hence the information contained in the ``missing half'' is maximised in the middle. This is a generic feature of TTNs, but as we shall later see not a universal one.

We also note that the bond entropies associated with the leaf nodes are small. This is partly because we have taken a free-field theory so that the fields $\phi_A $ and $\phi_B$ do not interact. However, if  an interaction such as $\lambda \phi_A^2 \phi^2_B$ were introduced, one would expect only a modest increase in the entropy at these nodes.
Significant enhancement  occurs only when excitations are injected into the system: for example localised wave-packets would lead to enhancement of the entanglement entropy near the interaction vertex in position space. 

It will be useful to understand  why this is the case. The field theoretic quartic coupling $\lambda \phi_A^2 \phi^2_B$ induces terms of the form $\lambda \hat q _{A,n}^2 \hat q _{B,n}^2 $ in the discretised Hamiltonian. 
Expanding around the vacuum state, this leads to terms that are schematically of the form  $H\supset \lambda \langle \hat q _{A,n}^2\rangle \delta \hat q _{B,n}^2 + \lambda \delta \hat q _{A,n}^2 \langle \hat q _{B,n}^2 \rangle + \lambda \delta \hat q _{A,n}^2 \delta \hat q _{B,n}^2 $. The first two of these terms constitute the na\"ive Hartree mean-field approximation, and generate a mutual squeezing of the $\hat q_A$ and $\hat q_B$ degrees of freedom. In the vacuum this leads primarily to fluctuations that are correlated between the $\phi_A$ and $\phi_B$ fields, rather than strong entanglement, and the state remains approximately factorisable (\ie into $A$ and $B$ factors) to leading order. Thus, the entropy associated with the parent bonds of the leaf-nodes would remain small in the vacuum in the absence of excitations.

\section{The BRTN}
\label{sec:BRTN}

\subsection{The BRTN framework}
\label{subsec:BRTNIntro}

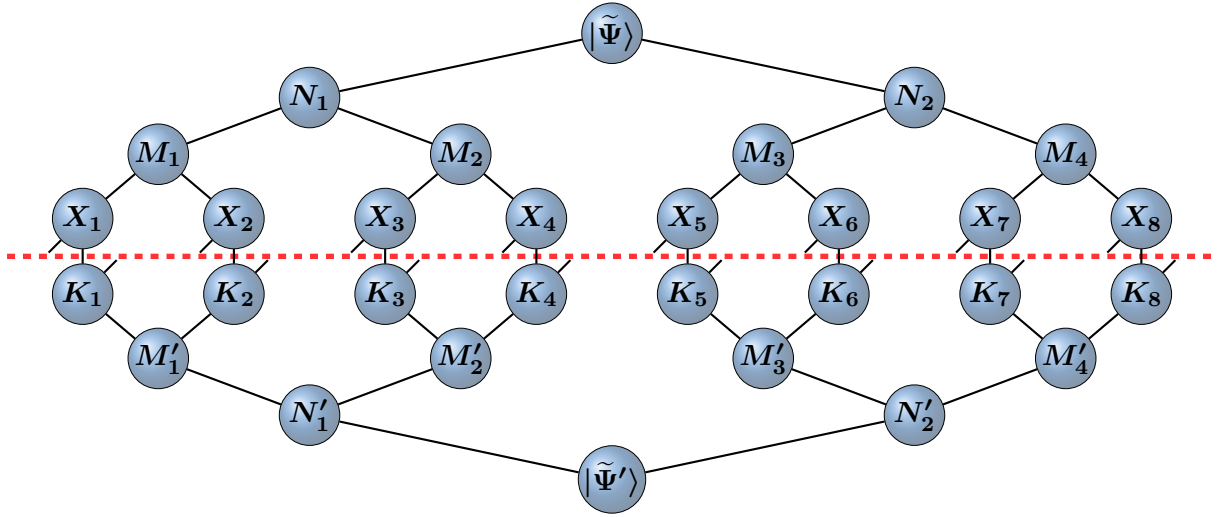
\begin{figure*}[t]
\centering
\begin{tikzpicture}[x=0.5cm, y=0.5cm,
font=\boldmath,
 tensor/.style={
    circle,
    draw=black,
    shade,
    ball color=tensorblue,
    fill opacity=0.5,
    text opacity=1,
    minimum size=8mm,
    inner sep=0pt,
    line width=0.2pt
  },
  leg/.style={line width=0.8pt},
  highlight/.style={line width=1.4pt, red!80},
  reflection/.style={line width=2.0pt, red!80, dashed},
  correl/.style={line width=1.4pt, green!60!black}
]

\draw[reflection] (-16,1.0) -- (16,1.0);

\foreach \x/\n in {-14/1,-10/2,-6/3,-2/4,2/5,6/6,10/7,14/8}
{
  \node[tensor] (X\n) at (\x,2.0) {$X_{\n}$};

  \node[tensor] (K\n) at (\x,0.0) {$K_{\n}$};

  \draw[leg] (X\n) -- (K\n);

  \draw[leg] (X\n) -- ++(-0.9,-0.9);

  \draw[leg] (K\n) -- ++(0.9,0.9);
}

\node[tensor] (M1) at (-12,3.7) {$M_1$};
\node[tensor] (M2) at (-4,3.7)  {$M_2$};
\node[tensor] (M3) at (4,3.7)  {$M_3$};
\node[tensor] (M4) at (12,3.7)  {$M_4$};

\draw[leg] (X1) -- (M1); \draw[leg] (X2) -- (M1);
\draw[leg] (X3) -- (M2); \draw[leg] (X4) -- (M2);
\draw[leg] (X5) -- (M3); \draw[leg] (X6) -- (M3);
\draw[leg] (X7) -- (M4); \draw[leg] (X8) -- (M4);

\node[tensor] (N1) at (-8,5.2) {$N_1$};
\node[tensor] (N2) at (8,5.2)  {$N_2$};

\draw[leg] (M1) -- (N1); \draw[leg] (M2) -- (N1);
\draw[leg] (M3) -- (N2); \draw[leg] (M4) -- (N2);

\node[tensor] (RootTop) at (0,6.9) {$\ket{\widetilde{\Psi}}$};
\draw[leg] (N1) -- (RootTop);
\draw[leg] (N2) -- (RootTop);

\node[tensor] (M1p) at (-12,-1.7) {$M'_1$};
\node[tensor] (M2p) at (-4,-1.7)  {$M'_2$};
\node[tensor] (M3p) at (4,-1.7)   {$M'_3$};
\node[tensor] (M4p) at (12,-1.7)  {$M'_4$};

\draw[leg] (K1) -- (M1p); \draw[leg] (K2) -- (M1p);
\draw[leg] (K3) -- (M2p); \draw[leg] (K4) -- (M2p);
\draw[leg] (K5) -- (M3p); \draw[leg] (K6) -- (M3p);
\draw[leg] (K7) -- (M4p); \draw[leg] (K8) -- (M4p);

\node[tensor] (N1p) at (-8,-3.2) {$N'_1$};
\node[tensor] (N2p) at (8,-3.2)  {$N'_2$};

\draw[leg] (M1p) -- (N1p); \draw[leg] (M2p) -- (N1p);
\draw[leg] (M3p) -- (N2p); \draw[leg] (M4p) -- (N2p);

\node[tensor] (RootBottom) at (0,-4.9) {$\ket{\widetilde{\Psi}'}$};
\draw[leg] (N1p) -- (RootBottom);
\draw[leg] (N2p) -- (RootBottom);

\end{tikzpicture}
\caption{Born-reciprocal renormalisation geometry: the BRTN. In a genuinely Born-reciprocal theory,  $\phi_B(K_i)$ is a dual field that is localised in momentum-space $K_{i\ldots 8}$, and is delocalised in position-space. The UV interface consists of pairwise links $X_i \leftrightarrow K_i$ between position- and momentum-space sites. These bonds form the entanglement bridge. Global Born-reciprocity consists of exchanging $X_{i}\leftrightarrow K_{i}$ {\it and} the fields $\phi_A$ and $\phi_B$, flipping the entire network about the red dashed line. This line delineates the furthest into the UV that the theory can go. The upper and lower trees flow to distinct infrared roots.}
\label{fig:brtn_8}
\end{figure*}

At this point we have gathered all the network technology that will be required for this discussion. 
I will now introduce the Born-reciprocal tensor network. Reciprocity is a symmetry which was proposed by Born~\cite{RevModPhys.21.463}, who noted that the Heisenberg uncertainty principle, and the canonical phase-space structure in Quantum Mechanics is invariant under a transformation which can be written  as $ x\to  k $ and $  k\to - x $. More generally the canonical commutation relation in the local Hilbert space, $[ x , k] = i $, allows a symplectic symmetry transforming $ z = (x,k)^T $, namely
\begin{equation}
z ~\to ~ S\cdot z 
\label{eq:symp}
\end{equation}
where 
$S^T \Omega  S=\Omega$ with 
$\Omega = i \sigma_2$. The simple mapping above has 
\begin{equation}
    S~=~ \begin{pmatrix}
        0 & 1\\
        -1 & 0
    \end{pmatrix} ~.
\end{equation}

Although Born-reciprocity was not widely developed in its original form, closely related ideas have appeared in the literature, partly inspired by the `minimal length' properties of string theory, in particular its $T$-duality which can be thought of as a compact-space version of Born-reciprocity. These include the related double-field theory, non-commutative geometry, metastring theory and modular space-time \cite{Majid:1988we,Low:2001rn,Freidel:2013zga, Freidel:2014qna,Freidel:2015pka,Freidel:2016pls}. These ideas touch on  Born-reciprocity when considered as an exact symmetry of the underlying spacetime phase space $(x,k)$. However in non-compact relativistic theories such approaches lead to locally modified phase-space or space-time structure (for a discussion see Ref.~\cite{Jarvis:2005yw}). 

By contrast, in the approach that we wish to pursue here, the reciprocal structure arises instead at the level of the Hilbert space organisation which is encoded by the tensor network, rather than as a result of an action on the phase space coordinates themselves. The exchange of the $x$ and $k$ sectors therefore acts on the global renormalisation geometry, avoiding the difficulties associated with imposing Born-reciprocity directly in spacetime. 
In a finite lattice approach, the   continuum statement of Born-reciprocity is not implemented by representing $x$ and $k$ as exact finite-dimensional canonical operators that obey $[x,k]=i$ exactly, but is implemented as an exchange symmetry between two finite, Fourier-dual lattice descriptions: that is accompanying the lattice coordinates $x_n$ we will introduce a set of reciprocal lattice coordinates $k_m$, with
\begin{equation}
e^{i k_m x_n}
~=~
e^{2\pi i mn/N}~.
\end{equation} 
The resulting approach can be used on any QFT whose states admit a regular TTN representation, to lift it to a Born-reciprocal form.

To begin to construct the BRTN, consider what such a global version of Born-reciprocity implies about the root node of the TTN. As we saw in the previous section, once truncation has been invoked in the TTN, the root node in some sense represents the deep IR: 
 on average, longer-range correlations probe higher parts of the tree, so that the upper layers of the ordinary TTN may be interpreted as increasingly infrared.
 In geometric terms the information stored in the upper layers of the tree relates to  correlations over increasingly large spatial unit cells of distance. If increasingly large distance cells lie at one extremity of the renormalisation geometry, then Born-reciprocity demands that there is a second root node that describes 
increasingly large momentum cells. To distinguish it in the network geometry we locate this root `beyond the UV' of the original network, 
with the two roots being mapped into each other by the reciprocity. 

Once the second root node has been fixed in place,  the rest of the tensor network follows, and we are led to the BRTN structure in \figref{fig:brtn_8}. The  entire network is forced to be Born-reciprocal, containing both the original coarse-graining space-side, and an image coarse-graining momentum-side. The dashed red line in \figref{fig:brtn_8} represents the UV interface of the theory, across which the Hilbert space is mapped into itself under Born-reciprocity. At this interface there are two sets of physical leaves, the $X_n$ leaves that latticise the space-side of the network, and $K_n$ leaves that latticise the momentum side. The 
$X_n$-$K_n$ type bonds  that cross the dashed red line will be referred to as {\it bridge bonds}.

There is very little freedom in this structure. One free choice is where and how many bridge bonds 
to put in place. This question is of less importance  because we will operate with this network just as we do the regular TTN and simply allow the bond entropy to grow in response to physical constraints. Therefore, to a certain extent whatever bond structure we decide to put between the two halves will adjust its entanglement  accordingly as long as sufficient bridge bonds are available. Our main simplifying assumption in this respect is that the entire network has at most three-index tensors.

The second degree of freedom is more crucial. To use orbifold language, it is the choice of whether to make the physical fields {\it twisted} or {\it untwisted} under the reciprocity. 
The choice taken here is to have independent $X_n$ leaves on the space-side of the network and $K_n$ leaves on the momentum-side representing physically distinct objects (\ie the untwisted choice). 
The alternative possibility, which is to consider the field to be self-reciprocal (\ie the twisted choice), has been rejected. To implement that, one would place single leaf nodes at the UV interface with reciprocity reflected in the physical indices and Hamiltonian.  It is obvious  why this kind of self-reciprocity should be rejected. If we were to adopt a self-reciprocal field at the UV interface then the leaf nodes would be describing a theory that was Born-reciprocal locally in phase-space. All the usual constraints that have to be met when we try to impose Born-reciprocity as a symmetry of the underlying dynamical theory would then apply, and the tensor network structure would not be adding anything new.
The novelty of the BRTN is the possibility to take the $X_n$ and $K_n$ leaf nodes to be describing physically distinct kinds of state which map into each other under Born-reciprocity. 

The interpretation of this configuration 
is that the root nodes delineate two possible equally valid IR's, with the opposing $K$-root node representing a reciprocal counterpart to the sector on the $X$-side of the network which we choose to represent the ordinary QFT. Of course from the viewpoint of the network, it simply resembles  two regular QFT descriptions that have been tied together at the UV. However, 
through the bridge bonds the $X$-side of the network learns about the reciprocal degrees of freedom. 
It is the physics across these bonds joining the two representations that will prove crucial: it can be chosen to imbue genuine physical reciprocal meaning to $K$-side of the BRTN. Or it can be chosen such that the BRTN indeed just describes two ordinary coupled QFTs (in which case the $K_n$ nodes can roughly speaking be thought of as a copy of the $X_n$ nodes, with some caveats -- to be  discussed later).

Having set up the general structure of the BRTN, let us consider how to introduce fields and couplings. 
We will proceed carefully in an attempt to avoid confusion.
If we continue to label the fields $\phi_A$ and $\phi_B$ respectively (with lattice quadratures $\hat q_{A,n}$ and  $\hat q_{B,n}$), then the mapping under Born-reciprocity includes the field operators themselves. Continuing to denote space-time momentum by $k$ (in order to avoid confusing it with the quadrature variables $\hat p_{A,n}$ and $\hat p_{B,n}$)  with the nodes being denoted by $K_n$, the map we wish to invoke is schematically of the form 
\begin{align}
      x \to    k ~;~~   k \to -  x ~ ; ~~ \phi_A ~\leftrightarrow ~\phi_B~. 
\end{align}
However, there is a complication in that  the space and momentum coordinates are related to the lattice points of our network by two different spacings. There is one physical spacing for the space-side and one for the wave-number on the reciprocal-side: 
\begin{align}
 x_n &~=~ \frac{2\pi n}{N a_K}=  n a_X\equiv  X_n a_X  ~,~\nonumber \\
 k_n &~=~ \frac{2\pi n}{N a_X} = n a_K \equiv K_n a_K  ~,
\label{eq:xnkn}
\end{align}
where $X_n, K_n$ imply the related dimensionless operators. Note that here and throughout our  discussion will be insensitive to a trivial  choice of
origin for the two reciprocal lattices. That is more generally, one may define
\begin{equation}
\bar x_n := x_n-x_0~, 
\qquad 
\bar k_m := k_m-k_0~,
\end{equation}
or equivalently
\begin{equation}
X_n := \frac{x_n-x_0}{a_X}~,
\qquad
K_m := \frac{k_m-k_0}{a_K}~,
\label{eq:origchange}
\end{equation}
but for ease of notation we will take $x_0=0$ and $k_0=0$. 

From the above we can read off 
\begin{equation} 
a_X a_K ~=~ \frac{2\pi}{N}~.
\label{eq:axak}
\end{equation} 
This is the required identification for Fourier conjugate lattices as for example it gives $\sum_n \exp(i  k_n ( x_m- x_\ell)  ) = \sum_n \exp(i 2\pi n (m-\ell)/N ) = N\delta_{m\ell} $, so that the finite $N$ BRTN is built with discrete phase-space cells of size 
\begin{equation}
    \Delta X\Delta K ~=~ \frac{2\pi}{N}~.
\end{equation}

In the BRTN as we built it above the symmetry involves the nodes  $X_n\to K_n$, and $K_n\to -X_n$. 
However on the finite lattice this becomes 
\begin{equation}
{\rm BR:~~~}    X_n\to K_n~;~~ K_n\to X_{N-n}~.
\label{eq:BRmap}
\end{equation}
Moreover this mapping is between dimensionless coordinates, while the mapping of the physical variables  involves the dimensionful lattice spacings. Indeed mapping $X_n\leftrightarrow K_n$ is equivalent to mapping $ x_n/a_X \leftrightarrow  k_n/a_K$. These lattice spacings can therefore be absorbed into the reciprocity matrix $S$ of  Eq.~\eqref{eq:symp}  as 
\begin{equation}
    S~=~ \begin{pmatrix}
        0 & a_X /a_K\\
        -a_K /a_X & 0
    \end{pmatrix} ~.
\end{equation}
This is the Born-reciprocity implied by the lattice: using it implies the straightforward symplectic map in Eq.~\eqref{eq:BRmap} between dimensionless variables. 

Now consider the evolution of this system. 
The central claim is that all this reorganisation of the  Hilbert space cannot alter the basic quantum-mechanical fact that the global state on the network must undergo Schr\"odinger evolution. The great advantage of the BRTN framework is that one is not obliged to construct a new quantum field theory on the reciprocal geometry to perform the  evolution of the system from a given Cauchy surface: one need only specify the  Hamiltonian on the network. 

Naturally the microscopic $\phi_A$  Hamiltonian encoded on the space-side of the BRTN is taken to be that of an ordinary scalar field theory lattice as in Eq.~\eqref{eq:disc_ham}. 
We wish to identify
\begin{equation}
\phi_A(x_n) ~\equiv ~  {\hat{q}_{A,n}}~~;~~~
\phi_B(k_n) ~\equiv ~ 
{\hat{q}_{B,n}} ~,
\end{equation}
but in a way that provides a map between dimensionless lattice quantities. On the $X$-side, since $\phi_A$ is dimensionless in 1+1D, we may therefore identify $\hat q_{A,n}$ with $\phi_A$ and $\hat p_{A,n}$ with $a_X \partial_t \phi_A$  directly.
Thus following Eq.~\eqref{eq:disc_ham} we have 
\begin{align}
H_X ~=~  a_X^{-1} \sum_{n=1}^N\left[   \frac{{\hat p}_{A,n}^2}{2} + 
\frac{\left({\hat q_{A,{n+1}} - \hat q_{A,n}}\right)^2}{2} + {V}\left(
 {\hat q_{A,n}}\right) \right]
\end{align}
where, $\hat p_{A,n}$ is also dimensionless, and where the second term can be regarded as the gradient-squared of $\phi_A$ written in $X_n$ derivatives.

In order to preserve the Born-reciprocity under $\phi_{A} \to \phi_{B}$ we must take the 
same functional form for the Hamiltonian of $\hat q_{B,n}$ on the momentum-side. 
The map to the $K$-side of the network acts on the dimensionless fields, and on the dimensionless coordinates $X_n$ and $K_n$, and therefore becomes trivially
\begin{align}
H_K ~=~ a_{X}^{-1} \sum_{n=1}^N\left[   \frac{{\hat p}_{B,n}^2}{2} + 
\frac{\left({\hat q_{B,{n+1}} - \hat q_{B,n}}\right)^2}{2} + {V}\left(
 {\hat q_{B,n}}\right) \right]~.
\end{align}
 Note that, in contrast with the $X$-side, if the $K_n$ nodes really represent momentum, then there is no continuum field theory for $\phi_B$ that we are trying to approximate at this point: the  $\hat q_{B,n}$ are regarded simply as the quadrature variables on the $K_n$ lattice sites, and the second term in $H_K$ is the gradient squared with respect to $K_n$ derivatives. Note also that the prefactor $a_X^{-1}$ by itself does not map, but rather it simply becomes the unit by which we measure time -- and we henceforth set $a_X=1$ where convenient. 

Finally there are the bridge couplings at the UV interface between $\hat q_{A,n}$ and $\hat q_{B,n}$ which for the moment we will write generically. 
Thus we have 
\begin{align}
H ~&=~\sum_{n=1}^N\left[   \frac{{\hat p}_{A,n}^2}{2 } + 
\frac{\left({\hat q_{A,{n+1}} - \hat q_{A,n}}\right)^2}{2} +{V}\left(
 {\hat q_{A,n}}\right) \right]~
 \nonumber \\
& ~~~~~~~~~~~~+\sum_{n=1}^N\left[   \frac{{\hat p}_{B,n}^2}{2}
 + 
\frac{\left({\hat q_{B,{n+1}} - \hat q_{B,n}}\right)^2}{2} +  {V}\left(
{\hat q_{B,n}}\right)  \right]
~+~ V_{\rm bridge} ( \{ \hat q_{A,n}\} ,\{ \hat q_{B,m}\} )~,
\label{eq:disc_ham2}
\end{align}
where 
\begin{equation}
    V_{\rm bridge} ( \{ \hat q_{A,n}\} ,\{ \hat q_{B,m}\} ) ~=~     V_{\rm bridge} ( \{ \hat q_{B,n}\} ,\{ \hat q_{A,m}\} )~.
    \label{eq:vbridgeconst}
\end{equation}
For a Born-reciprocal network the interactions in $V_{\rm bridge}$ must be chosen so that the $K_n$ variables on the $K$-side inherit the physical meaning of the QFT momentum on the $X$-side. We will return to this question in detail below, and for the moment continue the discussion under the assumption that such a choice of $V_{\rm bridge}$ will be possible.

Assuming that this is the case, one may ask if there is some other continuum theory that the second line of  Eq.~\eqref{eq:disc_ham2} could represent. This term defines an isomorphic scalar theory but on the discretised momentum lattice, in which we are interpreting the nearest-neighbour differences as momentum derivatives. 
A crucial aspect of this expression is that, while $H_X$ has a well understood 
continuum QFT limit 
(namely $\sum_n a_X^{-1} ( {\hat{q}_{A,n+1}} -\hat{q}_{A,n} )^2  ~\to  ~ \int dx (\partial_x \phi_A)^2   $) the $K$-side Hamiltonian does not admit the same ordinary field theory interpretation. Thus,  Eq.~\eqref{eq:disc_ham2} represents a map of the lattice Hamiltonian structure itself, and {\it not} a map to a second ordinary continuum QFT. We are of course at liberty to redesignate the $K$-side to represent displacements in space rather than momentum, in which case the converse is true and the $K$-side of the network {\it does} have a continuum QFT interpretation, but then the $X$-side does not. We may anticipate therefore that a $V_{\rm bridge}$ suitably chosen so as to enforce Born-reciprocity will  present an obstacle to simultaneously having a  QFT interpretation on both sides. 
This is the first sign that in the true Born-reciprocal network we may  choose only one side of the network to have asymptotic field-theory behaviour, while the other side behaves in an entirely different manner.

What kind of fundamental object then does $\phi_B$ represent? On the $X$-side of the network the $\phi_A$ leaf nodes can describe states that are narrow in space coordinates. The geometric interpretation of the coarse-graining on this side is that it corresponds to successively averaging over larger space cells, and information is lost about space location: hence $\phi_A$ describes particle-like objects which are narrow in $x$. 
Conversely $\phi_B$ coarse-graining corresponds to successively larger cells in {\it momentum}. Information about momentum resolution is progressively lost as we climb the tree on the $K$-side, and therefore the natural excitations described by $\phi_B$ are narrow in $k$-space and delocalised in $x$.
Thus, while the $K$-side of the BRTN
is highly non-local when expressed in position space, it describes delocalised objects that have a local Born-reciprocal description, and vice-versa. In other words, 
 from the perspective of one half of the tensor network, the other half appears to be delocalised.
This is a situation that is of course reminiscent of the $T$-duality mapping between winding modes and Kaluza-Klein (KK) modes which was partly responsible for the renewed interest in Born-reciprocity in the first place. It is important to realise however that this structural resemblance is superficial, because the Born-reciprocal relation has a different origin. The discreteness {\it does not} imply any kind of compactification, but simply represents discrete sampling of the Hilbert space. The lattice-spacing therefore represents the resolution scale of a presumed noncompact dimension, rather than any kind of compactification radius.
Note that  Born-reciprocity  {\it does not} imply that states on the $K$-side of the BRTN are the Fourier transform of those on the $X$-side either (if that were the case then the Hamiltonian would be the appropriate one to time evolve the Fourier modes). 
Rather, the BR map is a state-map / change of description. For example, we are allowed to excite a single node on just the $X$-side of the BRTN, which lifts the energy of the entire network by a single unit.

We can see all this in more detail if we examine the spectrum. Let us momentarily set $V_{\rm bridge}$ to zero, and take a harmonic oscillator potential, $V(\hat q_{A,n} ) = m^2 \hat q_{A,n}^2 /2$. On the $X$-side of the network we essentially copy the steps we took for the regular TTN. That is we expand in the usual momentum basis, 
\begin{equation}
    \hat q_{A,n} ~=~ \frac{1}{\sqrt{N}}\sum_m \hat q_{A,m } e^{i k_m x_n} ~;~~~
        \hat p_{A,n} ~=~ \frac{1}{\sqrt{N}}\sum_m \hat p_{A,m } e^{i k_m x_n} 
\end{equation}
where the letter $m$ (not to be confused with the dimensionless mass $m$) will be preferred for the momentum expansion, and where recall that $x_n = n a_X$ and $k_m = 2\pi m /Na_X$.  It is then straightforward to see that the Hamiltonian  $H_X$ is diagonal in this basis yielding the spectrum of field theoretic simple harmonic oscillators that we already presented for the regular TTN in Eq.~\eqref{eqn:omegas}. That is the space-side Hamiltonian becomes 
\begin{equation}
    H_X~=~ \sum_m  \omega_{K,m} \left(\hat N_m +\frac{1}{2} \right) 
\end{equation}
where 
\begin{equation}\label{eqn:omegas2}
\omega_{K,m} ~=~ a_X^{-1}\sqrt{ {m^2}  + {4}\sin^2 \left( { k}_m a_X/2 \right) }~.
\end{equation}
At small momenta, $m\ll N$, we recover 
\begin{equation}
    E_{X,m}^2 ~\approx ~{m_{\rm phys}^2 + { k}_m^2 } ~, 
\end{equation}
as expected, with explicit $a_X$ disappearing. 
This is the usual momentum mode spectrum: if $m_{\rm phys}\ll  k_m$ we have a KK-like spectrum of the form $H_{X,m} \approx (N+1/2) m /R$ where $R$ is the overall lattice size, given by 
\begin{equation}
    2\pi  R ~=~ N a_X ~.
    \label{eq:2piR}
\end{equation}
As one might expect, on the lattice this KK-like behaviour cuts off when $k_m a_X \sim 2\pi$ or $m \sim N$, when the energies are of order $1 /a_X$ and the structure of the lattice itself is being probed. 

By contrast, to diagonalise $H_K$ on the momentum-side of the BRTN
we must instead transform to the space basis using
\begin{align}
    \hat q_{B,m} ~&=~ \frac{1}{\sqrt{N}}\sum_n \hat q_{B,n } e^{-i k_m x_n} ;~~\nonumber \\
        \hat p_{B,m} ~&=~ \frac{1}{\sqrt{N}}\sum_n \hat p_{B,n } e^{-i k_m x_n} ~.
\end{align}
The rest of the treatment is identical to that on the space-side, yielding 
\begin{equation}
    H_K~=~ \sum_n \omega_{X,n}  \left(\hat N_n +\frac{1}{2} \right) 
\end{equation}
where now 
\begin{equation}\label{eqn:omegas3}
\omega_{X,n} ~=~ a_X^{-1}\sqrt{ {m^2}  + {4}\sin^2 \left({ x}_n a_K /2 \right) }~.
\end{equation}
At small lengths, $n\ll N$, we find 
\begin{equation}
    E^2_{K,n} ~\approx ~ {m_{\rm phys}^2 + a_{K}^{2} ({x}_na_X^{-1} )^2 } ~. 
    \label{eq:HK}
\end{equation}
Thus at small mass, $m_{\rm phys}\ll a_{ K}$, the Hamiltonian on the momentum side of the BRTN resembles a stretching energy, 
 \begin{equation} 
 H_{K,n} ~\approx ~ \frac{x_n}{a_{ X}} a_K ~.
 \label{eq:stretching}\end{equation}
 Note that this stretching energy behaviour also has a cut-off, when the size of the state approaches the size of the lattice itself, \ie at $x_n a_K \sim 2\pi$ or $n\sim N$. To compare the two towers of states, we can write the spacing on each side in terms of the single energy interval $a_{ K}$ using Eq.~\eqref{eq:xnkn}:
\begin{align} 
\delta E_X ~&\approx ~~~ a_{ K} ~=~ \frac{2\pi}{Na_X}~~\approx ~~ 
\delta E_K ~.
\end{align}
Not surprisingly the two spectra are equal, and at large $N$ the momentum-side of the spectrum is an equally dense band of stretched states, and not merely the discrete set of winding modes that one would expect in a compactified system. This must be the case because thus far we have not specified $V_{\rm bridge}$: without it the $K$-side of the BRTN could just as well be describing a second QFT sector. Therefore we now turn to the critical question of how to choose $V_{\rm bridge}$. 

\subsection{Choosing the bridge interaction, $V_{\rm bridge}$}
\label{subsec:FourierBridge}

Thus far we have seen how the doubled BRTN system admits a Born-reciprocal interpretation of the fields that live on it. However it can also admit an interpretation as two QFTs, and which situation prevails depends entirely on the interactions that we put into $V_{\rm bridge}$. 
For this reason the interactions appearing on the third line of the Hamiltonian in Eq.~\eqref{eq:disc_ham2} are perhaps its most interesting and crucial terms, because they completely determine what sort of theory we have.
These interactions also serve to generate entanglement across the interface between the reciprocal sectors, and as we shall see their form generates an interesting transmission of information through the bridge.  

The only requirement that we will impose on the bridge interactions is that,  rather than obeying ordinary spacetime locality, they should be local in the native coordinates on either side of the reciprocal lattice geometry, which constrains the interactions as follows:

\begin{definition}[Born-reciprocal locality] An interaction $V_{\rm BR}$ in the BRTN is said to be \emph{Born-reciprocal local} (BR-local) if it can be written as sums of field composites that are local in the native coordinates as
\begin{equation}
  V_{\rm BR}~=~  \sum_{m n} ~\sum_\alpha g^\alpha_{m n}\, {\mathcal O}^\alpha _A(x_n)\,{\mathcal O}^\alpha _B(k_m)~,
\end{equation}
where $\mathcal O^\alpha_A(x_n)$ are operators that are local in the $X$-sector coordinate $x_n$ and $\mathcal O^\alpha_B(k_m)$ are operators that are local in the reciprocal-sector coordinate $k_m$. 
\end{definition}

\noindent In the present context this implies that  the couplings  are of the form 
\begin{equation}
    V_{\rm bridge} ~=~ \sum_{nm} f_{m n}(\hat q_{A,n}, \hat q_{B,m} )~,  
\end{equation}
for some functions $f_{nm}$, with a sum over $n$ for the $X$-side and over $m$ for the $K$-side. 
This restriction is forced upon us by Eq.~\eqref{eq:vbridgeconst}  if the interactions of the QFT are initially defined to be local in the $x_n$ coordinates. In other words we are going to stipulate that the only possible source of any non-locality in the Born-reciprocal system will be the UV interface between the two halves of the Hilbert space. 

The functions $f_{m n}$ in the bridge couplings of the BR-local Hamiltonian are in general dependent on the indices, but they will have  a high degree of degeneracy governed by  the physical situation.
In this paper we will consider two possibilities for $V_{\rm bridge}$ in detail, one that will serve as a test case, where the BRTN indeed simply describes two coupled ordinary QFTs, and one that is a genuine Born-reciprocal system with coupled momentum modes and stretched modes. In addition I will introduce a third apparent possibility which is a theory that lies somewhere between these two extremes, parametrised by a {\it reciprocity angle} $\rho$. 

\subsubsection{Test case: the BRTN as a QFT of two coupled scalars}

First we present the `trivial' choice in which the BRTN produces the physics of two coupled scalars with an interaction of the form $V_{\rm bridge }\propto \phi_A(x)^2 \phi_B(x)^2$ (\ie the functions $f_{nm}$ are given by $f_{nm} ~=~ \delta_{nm} \hat q_{A,n}^2 q_{B,m}^2$). It  is the simplest nearest neighbour monomial on the network that does not disrupt the onsite DVR basis structure. We thus take the following for our total tensor network Hamiltonian: 
\begin{align}
H ~&=~\sum_{n=1}^N\left[   \frac{{\hat p}_{A,n}^2}{2 } + 
\frac{\left({\hat q_{A,{n+1}} - \hat q_{A,n}}\right)^2}{2} +{V}\left(
 {\hat q_{A,n}}\right) \right]
 ~\nonumber \\
 &
~~~~~~~~~~+~\sum_{n=1}^N\left[   \frac{{\hat p}_{B,n}^2}{2}
 + 
\frac{\left({\hat q_{B,{n+1}} - \hat q_{B,n}}\right)^2}{2} +  {V}\left(
{\hat q_{B,n}}\right)  \right]~
~+~ {\lambda} \sum_{n=1}^N \hat q_{A,n}^2\hat q_{B,n}^2~.
\label{eq:disc_ham2-lambda}
\end{align}
This theory is completely local, not just BR-local, and it will remain so. 
Indeed, although $\hat q_{A,n}$ and $\hat q_{B,n}$ talk to opposite halves of the lattice the zeroth order physical interpretation of this system is that it simply represents two coupled scalars in a single space-time with the identification $X_n \equiv K_n$. 
One then loses the relation for conjugate phase-space variables in Eq.~\eqref{eq:axak} and naturally we simply choose $a_K=a_X$.

A more accurate interpretation of this system is that it is two sectors confined to distinct subspaces that communicate through local bridge degrees of freedom. The reason for this distinction becomes clear if we recall the interpretation of the TTN depth as renormalisation scale. At low energy scales, or equivalently when measuring longer wavelength correlations, the path-product between nodes in the correlator goes  higher up in the tree on average, and so therefore does the corresponding entanglement. In a genuine two-scalar QFT this implies that generally there is entanglement between the two states throughout the tree that will affect correlations. By contrast in the local BRTN the two scalars are confined to opposite sides of the network. 
Of course modes of $\hat q_{B,n} $ propagate through the $K$-side of the BRTN 
and modes of $\hat q_{A,n} $ propagate through the $X$-side as usual. But they can only communicate with {\it each other} through the bridge, which is equivalently the UV. 
In other words the bridge remains the locus of intersection couplings at every energy scale, and consequently the effective $\lambda$ coupling is unlikely to run in the same way as it does in regular QFT. (The configuration is closer to scalars living on two displaced 1D hypersurfaces than ordinary 1+1D two-scalar QFT.)

\subsubsection{The Fourier bridge: Born-reciprocal systems}
\label{subsubsec:FourierBridge}
As we will see, for the BRTN to represent a Born-reciprocal system, the crux of the matter is to choose couplings such that the bridge correctly relates the native $k_m$ coordinate of the reciprocal half of the BRTN to the physical momentum of the space half. This further restricts the coupling to be of the following form:

\begin{definition}[Fourier bridge]
A BR-local interaction is said to define a \emph{Fourier bridge} if its coupling kernel is proportional to the canonical Fourier phase,
$    f_{m n} ~\propto~ \cos (  k_m x_n)~$,
such that 
  \begin{equation}
  V_{\rm Fourier}~=~  
  \frac{g}{N^{1/2}}
  \sum_{m n} 
  \sum_\alpha
 \cos  ( k_m x_n ) \, \,
 {\mathcal O}^\alpha _A(x_n)\,{\mathcal O}^\alpha _B(k_m) ~,
  \label{eq:fourierbridge}
\end{equation}
\end{definition}
\noindent where the operators $\cal O^\alpha_A ,~\cal O^\alpha_B  $ are assumed real.\\

\noindent  This definition, and recall all  our discussion, is understood to be up to a  change of $x_n$ and $k_m$ origin in Eq.~\eqref{eq:origchange} -- that is one could trivially shift coordinates such that 
$ f_{m n} ~\propto~ \cos (  (k_m-k_0)(x_n-x_0))~$ with the Born-reciprocity being shifted accordingly.

The function $f_{mn}$ is hermitian and it is chosen to be even in order to obey the Born-reciprocity condition, \ie invariance under 
 \begin{equation}
x_n\to a_X k_n/a_K~; ~~
k_m\to - a_K x_m/a_X~; ~~ A\leftrightarrow B~. 
\label{eq:hermitian_cond}
\end{equation}
The Fourier bridge is the ingredient that gives the reciprocal label $k_m$ its momentum interpretation, since it is this kernel that makes the $K_n$ nodes enter the interaction vertices in the same structural manner as conserved momenta. Note that the even structure causes a minor subtlety as we are obliged to identify $k_m$ and $-k_{m}$ momenta, and $x_n$ and $-x_{n}$ positions. We can think of this as giving a ${\mathbbm Z}_2$ orbifold projected meaning to the network: \ie we can interpret it as describing unconstrained physics in one half of the network. As this arises because we  
 are for simplicity assuming real fields, it is a subtlety that we can for the most part ignore: one could easily extend the BRTN to contain complex fields by employing  two sets of leaf nodes, with couplings which would then be proportional to $e^{i k_m x_n}$.

To summarise then, the Hamiltonian of the Born-reciprocal tensor network system can be written as follows: 
\begin{align}
H ~&=~\sum_{n=1}^N\left[   \frac{{\hat p}_{A,n}^2}{2 } + 
\frac{\left({\hat q_{A,{n+1}} - \hat q_{A,n}}\right)^2}{2} +{V}\left(
 {\hat q_{A,n}}\right) \right]
 ~+~\sum_{n=1}^N\left[   \frac{{\hat p}_{B,n}^2}{2}
 + 
\frac{\left({\hat q_{B,{n+1}} - \hat q_{B,n}}\right)^2}{2} +  {V}\left(
{\hat q_{B,n}}\right)  \right]
~\nonumber \\~&
\qquad\qquad\qquad\qquad\qquad\qquad\qquad\qquad ~+~ \frac{g}{N^{1/2}}
  \sum_{m n} 
  \sum_\alpha
 \cos  ( k_m x_n ) \, \,
 {\mathcal O}^\alpha _A(x_n)\,{\mathcal O}^\alpha _B(k_m) ~,
\label{eq:BR-fourierbridge}
\end{align}
 with the Fourier-bridge  interaction between momentum modes and stretched modes now forbidding  any  associated global continuum Hamiltonian. 

One of the most striking consequences of  non-trivial Born-reciprocal bridge couplings is that they generically induce some departure from ordinary translation invariance even  at leading order in $g$, albeit not in the bulk of the space lattice but at its boundary. As we shall see there is real physics in this phenomenon because it is the BRTN version of what we would simply call a counter-term in QFT: while in regular QFT such counter-terms are manifestly translation invariant, in the BRTN this is no longer possible. Ultimately this is all tied up with the question of how to correctly define the vacuum for the BRTN field theory, and once we resolve that question we will see that we also have to introduce  normal-ordering and localized counter-terms into the Hamiltonian. But before we get to this subtlety 
 let us first introduce some explicit examples of $V_{\rm bridge}$.\\

\listitem{Bi-quadratic bridge couplings} The most natural BR-local extension of the quartic two-scalar coupling of ordinary QFT is based on the following interaction 
\begin{equation}
  V^{(\rm biquad)}_{\rm bridge} ~=~ \frac{g}{N^{1/2}}
   \sum_{m  n} \cos (  k_m x_n ) ~\hat q _{B,m} ^2 \hat q _{A,n}^2~.
\label{eq:quarticfourierbad}
\end{equation}
 This coupling has a physical interpretation in terms of the physical field densities. If $\rho_B (x_n) = \frac{1}{N^{1/2}}\sum_n \cos (  k_m x_n )  \hat q _{B,n} ^2 $ is the even part of the $B$ field density as seen from $x$-space, then 
\begin{equation}
  V^{(\rm biquad)}_{\rm bridge} ~=~ {g}
   \sum_{n} ~\rho _{B}(x_n)  \rho_{A}(x_n)~.
\end{equation}
This of course does not mean that the interaction is a local one because the density is not simply the product of two fields in $x$-space, but half of it is the Fourier image of a quadratic density defined in the reciprocal coordinate.\\

\listitem{Trilinear bridge coupling} A coupling that will be an interesting test ground for later study is the following trilinear coupling:
\begin{equation}
   \frac{g}{N^{1/2}}\sum_{m  n} \cos(  k_m x_n) \phi_B(k_m )  \phi_A(x_n )^2~.
\end{equation}
By Fourier transforming (\ie using $\phi_B(x_n)=\frac{1}{N^{1/2}}\sum_{m} e^{i  k_m x_n} \phi_B(k_m ) $), this coupling is just a pointwise interaction in $x$-space between $\phi_A$ and the Fourier image of $\phi_B$ in $x$-space:  
\begin{equation}
g~\Re\left[\sum_{n}  \phi_B(x_n)   ~ \phi_A(x_n )^2\right] ~,
\end{equation}
but of course it is non-local  in the BRTN, because the rest of the reciprocal Hamiltonian on the $K$-side is local in the native $k_m$ coordinates. (Note the projection to the real part is required because, if the fields are real in their own half of the network, then they become hermitian-reflection symmetric in the reciprocal half, \ie $\phi_B(x)^*=\phi_B(-x)$). It is also straightforward to see that the coupling conserves momentum: in momentum-space it becomes 
\begin{align}
   \frac{g}{N^{3/2}} \Re \left[\sum_{m \mu\nu n} e^{i  x_n (k_m -k_\mu -k_\nu ) } \,\phi_B(k_m )   \phi_A(k_\mu  )  \phi_A(k_\nu  )\right] 
~=~\frac{g}{N^{1/2}} \Re \left[ \sum_{m \mu\nu} \delta _{m, \mu + \nu }\,  \phi_B(k_m )   \phi_A(k_\mu  )  \phi_A(k_\nu  )
   \right] ~,
\end{align}
by Eqs.~\eqref{eq:xnkn} and 
\eqref{eq:axak}.
Adding its Born-reciprocal partner, the trilinear BR-local coupling becomes 
\begin{equation}
V^{(\rm tri)}_{\rm bridge} ~=~    \frac{g}{N^{1/2}}\sum_{m n}  \cos(k_m x_n ) \left[  \hat q_{B,m} \hat q_{A,n}^2 + \hat q_{A,m} \hat q_{B,n}^2 \right] ~.
\label{eq:V_tri}
\end{equation}
Note that the Born-reciprocal term in the Hamiltonian is local in momentum \ie in the coordinates native to the $K$-side of the BRTN.

We may give the following physical interpretations to these two pieces in the interaction. The first is quite standard: it says that two $\phi_A$ quanta of the QFT may combine into a single excitation of $\phi_B$ in a momentum-conserving way.
This is the precise point where the designated $K_n$ nodes assume their meaning as
$X$-side momentum. To elucidate what happens on the BRTN, this means that if two $\phi_A$ momentum modes on the $X$-side with momenta $k_\mu$ and $k_\nu$ combine, they will produce an excitation on the single $K_{m = \mu+\nu}$ node on the $K$-side. Moreover if one imagines colliding wavepackets, then at tree-level there is a local interaction vertex.

However the second interaction is less familiar and makes more sense in the language of the reciprocal sector. It describes an annihilation that preserves  length, that is two stretched modes of length $x_n$ and $x_m$ can annihilate into a single excitation on the $x_{n+m}$ node on the $X$-side. This behaviour is reminiscent of string-theoretic joining interactions. 

Note however that in  both of these interactions two near energy eigenstates are being converted into a product state that is far from being an eigenstate of energy in its half of the BRTN: the product state is therefore expected to rapidly evolve. In other words the bridge vertex should be viewed as a vertex  into a decaying intermediate state rather than into a stable asymptotic mode.
(We also know that, throughout, the total energy must  be conserved because the Hamiltonian is time independent.) \\

\listitem{The odd bridge coupling}
There is precisely one bridge Hamiltonian that has factors of only odd powers of fields up to quartic order, namely  
\begin{equation}
V^{(\rm odd)}_{\rm bridge} ~=~    \frac{g}{N^{1/2}}\sum_{m n}  \cos( k_m x_n )\left[ 
\hat q_{B,m} \hat q_{A,n}^3 +   \hat q_{A,m} \hat q_{B,n}^3 \right]~.
\label{eq:V_fourier_13}
\end{equation}
As will will see this gives its vacuum slightly different properties.
This bridge potential has a very similar interpretation to that of the trilinear coupling. For example it corresponds to interactions that are local expressed in the native coordinates of either side of the BRTN. That is, the first term can be understood as a local interaction 
\begin{equation}
\frac{g}{N^{1/2}}\sum_{m n}  \cos (k_m x_n ) \hat q_{B,m} \hat q_{A,n}^3~\equiv ~ g\Re\left[ \sum_{n}  \phi_B(x_n)  \, \phi_A(x_n )^3\right] ~,
\label{eq:quadrilinear_contin}
\end{equation}
with the non-locality in the system again being attributed to the fact that the propagators of the $\phi_B(x)$ field are natural in the native $k_m$ coordinates. 

\subsubsection{A continuum of theories: Born-deformation with reciprocity angle $\rho$ }
There is a generalisation of the two kinds of Born-reciprocal theories that we have considered thus far, namely a theory that lies  between these two extremes. Indeed, given the two kernels $f_{nm}\propto \delta_{nm}$ and $f_{nm}\propto e^{i k_mx_n}$, the natural extension is BR-local theories with a kernel that is a linear combination of the two:
\begin{definition}[Born-deformed theory with reciprocity angle $\rho$]
A theory with \emph{reciprocity angle $\rho$} has a coupling  kernel $    f_{m n} ~\propto~ \cos \rho\delta_{mn} + \sin\rho \, e^{i k_m x_n}$,
such that 
\begin{equation}
  V_\rho~=~  
  \frac{g}{N^{1/2}}\Re
  \sum_{m n} 
 (\cos \rho\, \delta_{nm}+ \sin\rho e^{i k_m x_n} )
 ~ \mathcal O_A(x_n)\,\mathcal O_B(k_m)~.
\end{equation}
\end{definition}

Our previous examples had reciprocity angle $\rho=0$ and $\pi/2$ respectively. In the TTN language these  appear to merely be special cases of a family of theories parametrised by $\rho$. The behaviour at general reciprocity angle is expected to interpolate between these two extremes.
However the physical interpretation is not clear. At $\rho=0$ the $K_n$ nodes  manifestly correspond to a second position lattice, whereas at $\rho=\pi/2$ they manifestly correspond to momentum. That is they can be identified with the operator $\partial_x \phi$, and we may choose  couplings that conserve this momentum. However, at generic $\rho$ they are simply degrees of freedom  that exist prior to any choice of phase space polarisation. We will leave further discussion of the general case to future work.

\subsection{Localised counter-terms and normal ordering}

\begin{figure}
    \centering
\begin{tikzpicture}[thick, line cap=round, line join=round]

  \def\Rout{1.0}   
  \def\Rin{0.6}    

  \coordinate (C) at (0,\Rout-0.2);
  \coordinate (C1) at (0,\Rout+0.075);
  \coordinate (C2) at (0,\Rout+0.125);
  \coordinate (C3) at (0,\Rout+0.175);
  \coordinate (C4) at (0,\Rout+0.20);
  \coordinate (V) at (0,0.14);   
   \coordinate (g) at V + (0.2,-0.0);
  \coordinate (L) at (0,-1.0);
  \coordinate (R) at ( 1.8,-0.5);

  \fill[gray!20, even odd rule] (C1) circle (\Rout) (C) circle (\Rin);

  \draw[dashed] (L) -- (V);
  \fill[gray] (V) circle (3.0pt);

  \node[left]  at (L) {$\phi_A$};
  
  \node[below=0pt] at (g) {$g$};

  \node at (-0.0,1.75) {$\phi_B$};

\end{tikzpicture}
\hspace{0.5cm}
\begin{tikzpicture}[thick, line cap=round, line join=round]

  \def\Rout{1.0}   
  \def\Rin{0.6}    

  \coordinate (C) at (0,\Rout-0.2);
  \coordinate (C1) at (0,\Rout+0.075);
  \coordinate (C2) at (0,\Rout+0.125);
  \coordinate (C3) at (0,\Rout+0.175);
  \coordinate (C4) at (0,\Rout+0.20);
  \coordinate (V) at (0,0.14);         
  \coordinate (L) at (-1.8,-0.5);
  \coordinate (R) at ( 1.8,-0.5);

  \fill[gray!20, even odd rule] (C1) circle (\Rout) (C) circle (\Rin);

  \draw[dashed] (L) -- (V) -- (R);
  \fill[gray] (V) circle (3.0pt);

  \node[left]  at (L) {$\phi_A$};
  \node[right] at (R) {$\phi_B$};

  \node[below=4pt] at (V) {$g$};

  \node at (-0.0,1.75) {$\phi_B$};

\end{tikzpicture}

    \caption{
    One-loop contribution to a $\phi_A(x)$ tadpole  from the reciprocal degrees of freedom in the trilinear model (left), and to $\phi_A-\phi_B$ mixing (right). Both of these effects are cancelled by normal-ordering which corresponds to a localised counter-term.}
    \label{fig:smeared_bubble}
\end{figure}
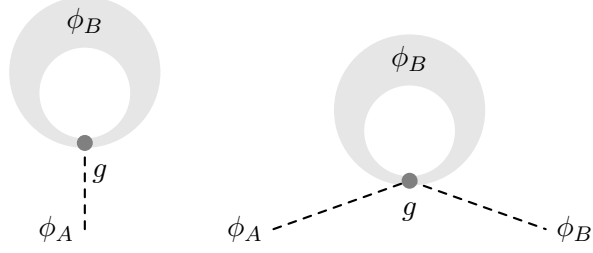

We now turn to the question of counter-terms and the correct definition of the vacuum. This issue becomes relevant once we realise that leading order corrections will in general break translation invariance. 

To see why, consider for example the trilinear bridge coupling. In its present form it can be shown to produce an inhomogeneous vacuum as follows. Suppose that the vacuum were homogeneous with the VEVs on all the nodes being degenerate. The Hartree approximation indicates a  tadpole for 
$ \hat q_A $ of the form 
$\frac{g}{\sqrt{N}}\sum_m \cos( k_m x_n)\, \hat q_{A,n} \langle q^2_{B,m} \rangle $. Perturbatively this corresponds to the first diagram in \figref{fig:smeared_bubble}.
However degeneracy of VEVs implies that this  contribution is proportional to $\sum_m \cos (  k_m x_n )  $, and we know that 
\begin{equation}
    \sum_m e^{i k_m x_n  } ~=~ \sum_m e^{i m n a_X a_K} ~=~ \sum_m e^{i mn  \frac{2\pi }{N}} ~=~ N\delta_{n,0}~.
\end{equation}
Therefore there is a tadpole of order $V_{\rm eff} ~\sim~ \sqrt{g^2 N} \phi_A(0)  $. This term induces a VEV for just the $n=0$ node of the $\phi_A$ field, and we conclude that translational invariance is broken at linear order in $g$. 
Indeed the term corresponds to a  divergent tadpole in the corresponding regular field theory that has the same coupling. Unlike regular field theory however this divergence is not degenerate in space but is located at the origin, and this is no surprise given the  BRTN structure. We recall that a change of $x_n$ and $k_m$ origin is allowed as per Eq.~\eqref{eq:origchange}, and therefore more generally we conclude that the diagram generates a translation breaking ``defect'' somewhere in space. 

This loss of homogeneity can in fact be seen to be a generic feature. Wherever one would find a UV divergence in the equivalent regular field theory with that coupling, in the BRTN we find a term that grows with $N$ which is located on the $n=0$ boundary node. And by reciprocity this is mirrored on the $m=0$ node on the reciprocal side.

In the bi-quadratic model the same effect generates a space-dependent contribution to the mass-squared of the $\phi_A$-field  proportional to $\sum_{m} \cos (  k_m x_n )  \langle \hat q_{B,m}^2\rangle $. This turns into an $x$-dependent mass-profile for the scalar peaked on the $n=0$ boundary. Meanwhile the breaking of homogeneity in the odd bridge coupling model is more subtle but still present.
In the two previous examples, leading order inhomogeneity was caused by the non-zero factors $\langle \hat q_{B,m}^2\rangle $ and $\langle \hat q_{A,n}^2\rangle $ which appeared in the couplings. In the QFT vacuum, parity arguments imply that the  expectation values of such operators are non-zero because they are even. By contrast, parity arguments imply that the VEVs of odd powers of fields are zero.  As the Hamiltonian is invariant under the ${\mathbbm Z}_2 $
 symmetry $(\hat q_{A,n},\hat q_{B,m}) \to -(\hat q_{A,n},\hat q_{B,m})$ a tadpole cannot be produced at any order because it would be accompanied by the contraction of an odd number of $q_{B,m}$ operators, and there can be no diagram with only $\hat q_{A,n}$ external legs with a purely $K$-local contraction either. There is one diagram however that breaks translation invariance maximally. This is the second diagram in \figref{fig:smeared_bubble}.
 The diagram generates a mixing between the $\hat q_{A,0}$ and $\hat q_{B,0}$ fields, just on that particular node. 

This breaking of translation invariance can be removed by cancelling counter-terms that live on the $n=m=0$ boundary node. The effect of such counter-terms is reproduced in the potential by normal ordering. Thus our three Fourier bridge potentials become 
\begin{align}
V^{(\rm biquad)}_{\rm bridge} ~&=~    \frac{g}{N^{1/2}}\sum_{m n}  \cos( k_m x_n )~ 
:\hat q^2_{B,m}:~ : \hat q_{A,n}^2 : ~\nonumber \\
V^{(\rm tri)}_{\rm bridge} ~&=~    \frac{g}{N^{1/2}}\sum_{m n}  \cos( k_m x_n )\left[ 
\hat q_{B,m} : \hat q_{A,n}^2: +  ~ \hat q_{A,m} : \hat q_{B,n}^2 : \right]~\nonumber \\
V^{(\rm odd)}_{\rm bridge} ~&=~    \frac{g}{N^{1/2}}\sum_{m n}  \cos( k_m x_n )\left[ 
\hat q_{B,m} : \hat q_{A,n}^3: +  ~ \hat q_{A,m} : \hat q_{B,n}^3 :\right]~,
\label{eq:V_fourier_summary}
\end{align}
 where the normal ordering is equivalent to $:\hat q^2: ~\equiv~ \hat q^2 - \langle \hat q^2 \rangle $ and $:\hat q^3:~ \equiv ~\hat q^3 - 3\hat q \,\langle \hat q^2 \rangle $.
In general higher order counter terms may be required, always on the $n=m=0$ boundary. 

Localised space-dependent counter-terms are perhaps most familiar  in bulk/brane set-ups, and it is completely natural for them to appear in a Born-reciprocal framework. Indeed the discussion above establishes a general geometric principle for what happens to the UV divergences of the erstwhile QFT once we embed it in the BRTN. Unfortunately the BRTN does not make such divergences magically disappear, nor does it remove the need for their counter-terms. Instead the reciprocal structure shunts them to the boundary where they must be handled by local counter terms. 

This is the BRTN analogue of  renormalisation in conventional field theory. There the counter-terms are uniform, reflecting the translation invariance of the UV regulator (which is a function only of momentum), while here they are localised at the origin, reflecting the geometry of the Fourier bridge.
By contrast effects such as  the running of couplings, which reflects scaling in the  effective theory, remain as bulk contributions and do not grow with $N$.  
This geometric separation of the UV sensitive aspects of the theory from the IR physics is a distinctive feature of renormalisation in the Born-reciprocal framework. The unphysical nature of these UV divergences is made manifest because they sit on the boundary of the network where they are unable to play a role in IR physics. One interesting aspect of this separation is that what would normally correspond to
renormalisation ``scheme'' (\eg $\overline{\rm MS}$ versus on-shell) becomes the freedom to leave finite boundary terms. Such terms must propagate into the bulk through the equations of motion, adjusting the parameterisation of the effective theory (\ie the extracted values of running couplings), without altering physical observables. However the cancellation of boundary terms that normal ordering gives us has a natural physical implication: namely we are  simply requiring that the coupling of the reciprocal sector vacuum to the QFT vacuum should not manifestly break translational invariance. 

\section{Implementing the BRTN}
\label{sec:BRTNImp}

The above structure can be  implemented in {\tt ITensor} with a relatively straightforward modification of the ordinary TTN. 
The steps for doing this are as follows (in increasing order of difficulty). 

\begin{enumerate}
\item 
The topology of the regular TTN is modified to carry the mirror tree-structure with special bridge bonds introduced. As two sets of physical fields were already present in the regular TTN, one need simply relocate the $\hat q_B$ local Hilbert space indices to the opposing $K_n$ leaf nodes.

\item
One crucial aspect of the BRTN is that we must now define and keep track of two  orthogonality centres, one on each side of the BRTN. These centres do not clash however, because it is never necessary to move the orthogonality centre across the bridge. 

\item
The onsite evolution gates for the TEBD are also a relatively straightforward extension of those of the regular TTN. They do not require special treatment as they do not involve  moves of the orthogonality centres. Hence they do not require any SVD or truncation, and are simply applied locally at the leaf nodes. 

\item
The hopping gates of the BRTN are performed on either the space-side or momentum-side of the network as appropriate, with the LCA and path-product being determined on that side as described above for the regular TTN. 

\item  The Fourier bridge gates of the genuinely Born-reciprocity symmetric system must be applied as part of each Trotter step. For this we will keep the same Strang split as above, however each bridge gate is now of the form 
\begin{equation}
    U_{{\rm bridge},n } ~=~ \prod_m e^{-i\delta t \frac{g\cos (k_m x_n) }{\sqrt{N}}
    \sum_\alpha {\mathcal O}^\alpha _A(x_n)\,{\mathcal O}^\alpha _B(k_m)
    }
    ~,
\end{equation}
where the index $m$ is a separate product over all $K_m$ indices for each $n$ index. As the ${\cal O}^\alpha_A(x_n)$ and ${\cal O}^\alpha_B (k_m)$  are actual operators that need to act on their respective physical leaves, in order to implement a single gate at position $X_n$, we are therefore obliged to sweep over all the $K_m$ nodes, and vice-versa. 

\end{enumerate} 

For the test case modifications 1 -- 4 above are all that is required. The same Strang splitting as in Eq.~\eqref{eq:Udt} may be used, with the straightforward inclusion of the extra bridge factor,
\begin{equation}
    U_{{\rm bridge},n }(\delta t) ~=~ e^{-i\delta t \, \lambda  \hat q _{A,n}^2\,  \hat q _{B,n}^2 }~,
\end{equation}
where the measured VEVs are subtracted to implement the normal ordering. 
An important resource saving is to order the application of these gates so as to reduce the number of SVD operations that are made. While no SVD is involved in the onsite gates which are simply applied locally, the bridge and hopping gates should be applied as follows in order to maximise efficiency:
\begin{equation}
    U_{{\rm bridge+hop} }(\delta t) ~=~ \prod_{n=1}^N [U_{{\rm bridge},n }(\delta t) ~U_{{\rm hop},n }(\delta t)]~.
\end{equation}
Note that the operators in this product all commute, and therefore  the bridge gates can be factored in with the hopping gates in the Strang split, and their ordering does not matter. However, importantly in each time step we are performing a single sweep over the lattice sites (as opposed to for example performing the bridge gates and the hopping gates in separate sweeps), so that the orthogonality centre moves are minimised. One additional modification is to reverse the direction of the sweep over lattice sites every time step, in order to remove biasing. 

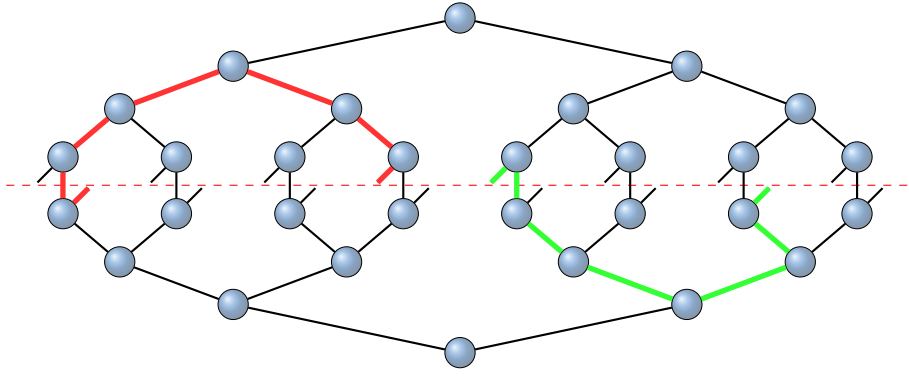
\begin{figure}[t]
\centering
\begin{tikzpicture}[scale=1.5,x=0.25cm, y=0.25cm,
font=\boldmath,
 tensor/.style={
    circle,
    draw=black,
    shade,
    ball color=tensorblue,
    fill opacity=0.5,
    text opacity=1,
    minimum size=4mm,
    inner sep=0pt,
    line width=0.2pt
  },
  leg/.style={line width=0.8pt},
  highlight/.style={line width=2.0pt, red!80},
  highlightK/.style={line width=2.0pt, green!80},
  reflection/.style={line width=0.5pt, red!80, dashed},
  correl/.style={line width=1.4pt, green!60!black}
]

\draw[reflection] (-16,1.0) -- (16,1.0);

\foreach \x/\n in {-14/1,-10/2,-6/3,-2/4,2/5,6/6,10/7,14/8}
{
  \node[tensor] (X\n) at (\x,2.0) {};

  \node[tensor] (K\n) at (\x,0.0) {};

  \draw[leg] (X\n) -- (K\n);

  \draw[leg] (X\n) -- ++(-0.9,-0.9);

  \draw[leg] (K\n) -- ++(0.9,0.9);
}

\node[tensor] (M1) at (-12,3.7) {};
\node[tensor] (M2) at (-4,3.7)  {};
\node[tensor] (M3) at (4,3.7)  {};
\node[tensor] (M4) at (12,3.7)  {};

\draw[leg] (X1) -- (M1); \draw[leg] (X2) -- (M1);
\draw[leg] (X3) -- (M2); \draw[leg] (X4) -- (M2);
\draw[leg] (X5) -- (M3); \draw[leg] (X6) -- (M3);
\draw[leg] (X7) -- (M4); \draw[leg] (X8) -- (M4);

\node[tensor] (N1) at (-8,5.2) {};
\node[tensor] (N2) at (8,5.2)  {};

\draw[leg] (M1) -- (N1); \draw[leg] (M2) -- (N1);
\draw[leg] (M3) -- (N2); \draw[leg] (M4) -- (N2);

\node[tensor] (RootTop) at (0,6.9) {};
\draw[leg] (N1) -- (RootTop);
\draw[leg] (N2) -- (RootTop);

\node[tensor] (M1p) at (-12,-1.7) {};
\node[tensor] (M2p) at (-4,-1.7)  {};
\node[tensor] (M3p) at (4,-1.7)   {};
\node[tensor] (M4p) at (12,-1.7)  {};

\draw[leg] (K1) -- (M1p); \draw[leg] (K2) -- (M1p);
\draw[leg] (K3) -- (M2p); \draw[leg] (K4) -- (M2p);
\draw[leg] (K5) -- (M3p); \draw[leg] (K6) -- (M3p);
\draw[leg] (K7) -- (M4p); \draw[leg] (K8) -- (M4p);

\draw[highlight] (K1) -- (X1);
\draw[highlight] (X1) -- (M1);
\draw[highlight] (M1) -- (N1);
\draw[highlight] (N1) -- (M2);
\draw[highlight] (M2) -- (X4);
\draw[highlight] (X4) -- ++(-0.9,-0.9);
\draw[highlight] (K1) -- ++(0.9,0.9);

\node[tensor] (N1p) at (-8,-3.2) {};
\node[tensor] (N2p) at (8,-3.2)  {};

\draw[leg] (M1p) -- (N1p); \draw[leg] (M2p) -- (N1p);
\draw[leg] (M3p) -- (N2p); \draw[leg] (M4p) -- (N2p);

\draw[highlightK] (X5) -- (K5);
\draw[highlightK] (K5) -- (M3p);
\draw[highlightK] (M3p) -- (N2p);
\draw[highlightK] (N2p) -- (M4p);
\draw[highlightK] (M4p) -- (K7);
\draw[highlightK] (X5) -- ++(-0.9,-0.9);
\draw[highlightK] (K7) -- ++(0.9,0.9);

\node[tensor] (RootBottom) at (0,-4.9) {};
\draw[leg] (N1p) -- (RootBottom);
\draw[leg] (N2p) -- (RootBottom);

\end{tikzpicture}
\caption{Applying full Fourier bridge gates. The upper red path is the path that has to be traversed to implement the gate $\exp(-i\delta t \,  
{\mathcal O}^\alpha _A(x_4)\,{\mathcal O}^\alpha _B(k_1)\, )$ biased on the $X$-side. 
The lower green path has to be traversed to implement the gate $\exp(-i\delta t \, ({\mathcal O}^\alpha _A(x_5)\,{\mathcal O}^\alpha _B(k_7))$ biased on the $K$-side. Biasing is alternated with sweeps. For certain applications we can perform the full path for only the `near-field' gates, and use the (trivial) Hartree approximation for the `far-field' gates.
}
\label{fig:brtn_bridge_gates}
\end{figure}

The most difficult aspect of the BRTN implementation is step 5, namely the application of the Fourier gates.
 This is computationally intensive. To see why, consider for example just the $\hat q_{A,4}^2 \hat q_{B,1}^2$ contribution to the $U_{\rm bridge,4}$ gate. To implement it we must find a path-product that crosses over the UV interface in such a way as to minimise the number of tensors it contains. It is straightforward to show that these minimal path-products cross over the UV interface at either of the two physical nodes that appear in the coupling itself, \ie in this case at either the $X_1-K_1$ bridge or the $X_4-K_4$ bridge. This leads to two orientations for the minimal path: either it has the LCA on the $X$-side of the BRTN, or on the $K$-side. 
There are thus two kinds of  associated path-products that we must implement which are both shown in \figref{fig:brtn_bridge_gates}, with the red path having $X$-side biasing, and the green path having $K$ side biasing (where in order to maintain Born-reciprocity we alternate between $X$-side and $K$-side biasing with Trotter step). We can now see the difficulty with these gates: the path has two sets of $X$ (or $K$) indices {\it and} a set of $K$ (or $X$) indices. Therefore there are in principle $d^2$ dimensions of physical indices that need to be carried up the path to the LCA (in other words one is performing SVD with at least $d^2\times d^2$ matrices), and truncation during this process can damage the entanglement already in the network.

For certain applications there is no short-cut to the full implementation of Fourier gates. This is clear from the interaction in Eq.~\eqref{eq:quadrilinear_contin} which is entirely local in $x$-coordinates. Thus one can imagine an interaction vertex involving two incoming wavepackets on the $X$-side of the BRTN that via the bridge interaction produce a wavepacket on the $X$-side and a wavepacket on the $K$-side at a relatively localised vertex.  
 In such an interaction the mode on the $K$-side of the BRTN would also be localised in $X$ coordinates, and its creation therefore entails a sum across the entire dimension of $K_n$ nodes. If any kind of truncation were attempted the physics would be artificially altered. 

 However there are situations in which it is possible to find an alternative approximation for the Fourier gates. 
 For example to verify the appearance of boundary terms discussed in the previous section the leading effect may be described by the Hartree approximation (which suffices for one-loop leading corrections), in which we may use the mean field approximation for part of the potential. That is in the $U_{\rm bridge}$ gate we may for example approximate the (un-normal-ordered) bi-quadratic bridge potential as 
\begin{equation}
  V_{\rm bridge} ~\approx ~ \frac{g}{N^{1/2}}
   \sum_{m  n} \cos (  k_m x_n ) \left[ \, \langle \hat q _{B,m} ^2\rangle \, \hat q _{A,n}^2~+~ \hat q _{B,m} ^2\,  \langle \hat q _{A,n}^2\rangle ~-~
   \langle \hat q _{B,m} ^2 \rangle \langle \hat q _{A,n}^2\rangle\, \right]~. 
\label{eq:quarticfourierHartree_approx}
\end{equation} 
This approximation greatly simplifies the application of the Fourier gates because we may first collect all the expectation values at each of the nodes, and then apply the gate locally at the leaves without ever having to implement the path-products in \figref{fig:brtn_bridge_gates}. 

One modification that improves the performance of the network with Fourier gates is to use a schedule that includes the coupling $g$ as well. That is, the system is first allowed to evolve to the regular QFT vacuum with $g=0$. This allows one to choose a $\delta \tau $ that is large enough to allow the bond dimensions to inflate. (This is indeed a lesson that we learned with the ordinary TTN.) A non-zero $g$ coupling  at this stage would stall the evolution.  Once the system is in the standard QFT vacuum 
 the $g$ coupling may then be gradually turned on to deform away from it. 
 
A final remark on normalisation: the BRTN provides a double cover of the Born-reciprocal Hilbert space and thus involves the two separate fields $\phi_A$ and $\phi_B$, however we may still continue to normalise the BRTN to 
\begin{equation} 
\langle \Psi|\Psi\rangle_{\rm BRTN} ~=~1~.
\end{equation}
In this sense, the fields $\phi_A$ and $\phi_B$ should be viewed as describing distinct sectors of the enlarged BRTN Hilbert space. They do not correspond to states being projected out or being identified with one another, but rather to different regimes of the global state.

\begin{figure}[t!]
\centering
\includegraphics[keepaspectratio, width=0.75\textwidth]{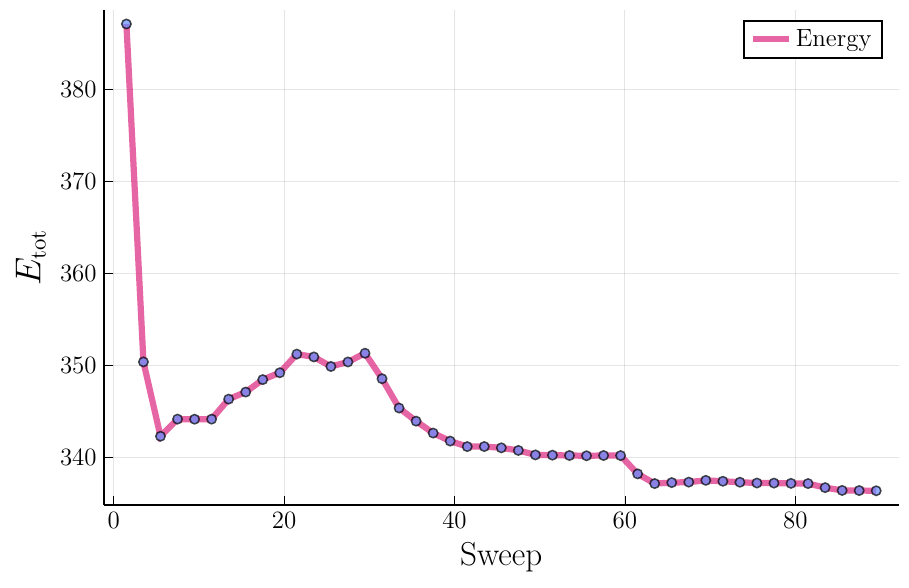}
\caption{Typical evolution of the averaged local energy estimator, used as a proxy to track convergence to the QFT vacuum: this example is for the $\lambda=1.5$ vacuum in \figref{fig:BRTN_Sk_128}. As the time step $\delta \tau$ is dialed down initial transients give way to smooth convergence.}
\label{fig:BRTN_conv_128}
\end{figure}

\begin{figure}[t!]
\centering
\includegraphics[keepaspectratio, width=0.75\textwidth]{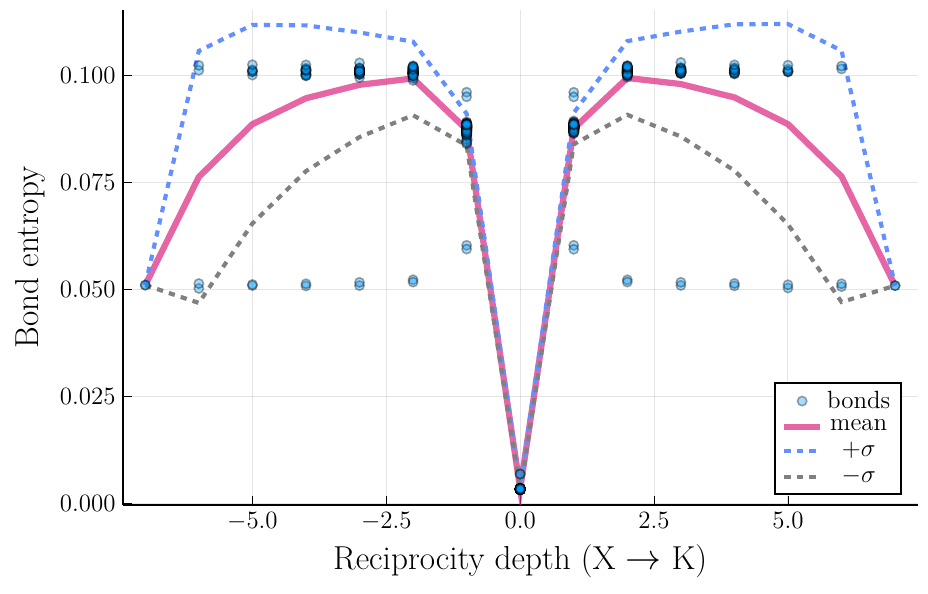}
\caption{A typical example of average bond entropy versus depth in the BRTN, with bridge coupling $\lambda=1.5$.}
\label{fig:BRTN_ent_128}
\end{figure}

\begin{figure}[t]
    \centering
\includegraphics[width=0.33\textwidth,angle=-90]{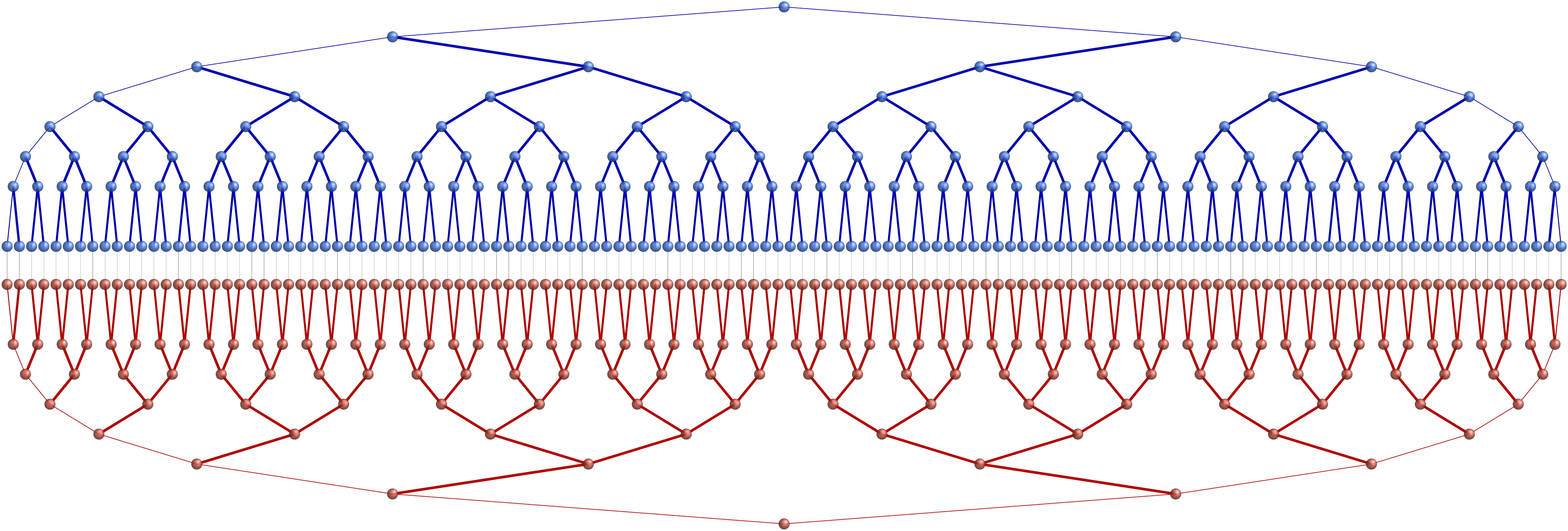}
    \caption{The 128 space-node BRTN in its ground state for the test case, showing the bond-entanglement. (The physical legs emanating from the leaf nodes are not shown.) Line thickness indicates bond entropy. Bond entropy is largest for mid-level bonds and small at the bridge bonds and the exterior bonds.}
    \label{fig:brtn_entropy}
\end{figure}

\section{Test case: Vacuum preparation and radiative mass corrections in a QFT of two scalars}
\label{sec:testimp}
\subsection{Vacuum preparation}
We now turn to the preparation of the  vacuum with the test potential, which is in principle equivalent to a regular QFT of two quartically coupled scalars. This will again be done by imaginary time TEBD, as for the regular TTN. We will see that this procedure is an important probe of the radiative corrections that come from the states on the reciprocal side of the network. In this case the radiative corrections depend on the $\lambda$ bridge coupling which generates contributions to the mass-squared that can be measured from the dispersion relation $\omega_{k}$. 

As $\omega_k$ is one of the quantities  of the BRTN that we will be probing, we cannot use it as a measure of the convergence of the imaginary time TEBD as we did for the ordinary TTN: therefore we first need to devise an independent method of checking that the system has properly converged to its vacuum state.  
The most natural diagnostic for this would be the 
total energy  $\langle{H}\rangle$ of the BRTN, which we expect to approach a global energy minimum under imaginary time evolution. Unfortunately, it is prohibitively expensive to calculate $\langle{H}\rangle$ exactly for the entire system. Therefore a local estimator can be used for this purpose instead, which is defined as
\begin{equation} 
E_{{\rm loc},n } ~=~  \frac{\langle \psi_n |H_n  |\psi_n\rangle }{\langle \psi_n | \psi_n \rangle }
\end{equation}  
where $\ket{\psi_n}$ is the system canonicalised about the space point $X_n$, and $H_n$ is the  Hamiltonian associated with that site (including the adjacent link and bridge terms). Because this construction neglects the full tree, it should be thought of as a convergence diagnostic that merely reflects the total energy. In particular, both the truncation pattern and the canonical form depend on the Trotter step,  $\delta \tau$, and therefore we expect the total energy estimator, \begin{equation}
E_{\rm tot} ~=~ \sum_n E_{{\rm loc},n},
\end{equation}
to acquire a $\delta \tau$-dependent offset, and also to depend on the `sweep direction'. The latter dependence is reduced by the aforementioned procedure of reversing sweep direction at each time step. 

In this study we will again take $N=128$ space points, and apply imaginary time TEBD. For the test case we will again use a schedule 
at fixed values of the bridge coupling $\lambda$, and with a cut-off of $10^{-6}$ and a maximum bond dimension of $\chi=20$.
A typical evolution of the energy estimator is shown in \figref{fig:BRTN_conv_128} for $\lambda =1.5$.
As might have been anticipated  the energy estimator exhibits finite jumps as $\delta \tau$ is stepped down. Nevertheless within each stage of the schedule it displays the required convergence when $\delta \tau$ is sufficiently small.

An interesting initial observation is the distribution of average bond entropies by network depth in the converged system. These are shown in \figref{fig:BRTN_ent_128}. As can be seen the average bond entropies are a Born-reciprocity symmetric  version of those we found in the regular TTN, but are smaller overall. This is chiefly because there is only one field on either side of the BRTN, and hence only a $d=6$ dimensional Hilbert space feeding into each leaf node rather than a $d^2 =36$ dimensional one. Note however that the relatively large maximum bond dimension of $\chi=20$ was required in order to give good convergence. 

The distribution of entropy in the tree that leads to this distribution of bond entropies is shown in \figref{fig:brtn_entropy}, where the bond entropy is represented by the width of the bonds. As can be seen bond entropy is maximised in the mid-level bonds, and falls away at the edges of the tree.  
The bridge bonds themselves do not carry significant entropy in the test case. This is the same effect that we discussed in the context of the normal TTN in Subsection~\ref{subsec:TTNvac_prep}. 
As there, the leading effect of a large bridge coupling is to squeeze the ground states rather than entangle them. It is for this reason that one should think of the set of $X_n-K_n$  couplings in the BRTN as a bridge rather than a source of entropy. It mediates entanglement between the fields and their reciprocals but in the vacuum does not need to store a great deal of entanglement itself. Note that this statement concerns the entropy per bond: each bridge bond carries only a parametrically small entropy of order $\lambda^2$, which is independent of $N$. The total entanglement between the two halves of the network, being the sum over all $N$
bridge bonds, is therefore extensive even though every individual bond remains close to trivial. This is a point we will return to in the context of the BRTN in Section~
\ref{sec:breaking}.\\

\begin{figure}
    \centering
\begin{tikzpicture}[scale=1.3]
\begin{feynman}
    \vertex (L) at (-1.4,-0.5) {$\phi_A$};
    \vertex [dot] (V) at (0,0) {};
    \vertex (R) at ( 1.4,-0.5) {$\phi_A$};

    \vertex (T) at (0,1.8);

    \diagram*{
        (L) -- [scalar] (V) -- [scalar] (R),

        (V) -- [half left, looseness=1.7, scalar, edge label'=$\phi_B$] (T),
        (T) -- [half left, looseness=1.7, scalar] (V),
    };
\end{feynman}

\node[below=2pt] at (V) {$\lambda_{\rm eff}$};
\end{tikzpicture}

    \caption{The radiative corrections to the mass-squared from the `reciprocal' degrees of freedom in the test case corresponds to just a standard bubble diagram for a theory of two scalars with $\lambda_{\rm eff}=\lambda/4$.}
    \label{fig:bubble}
\end{figure}
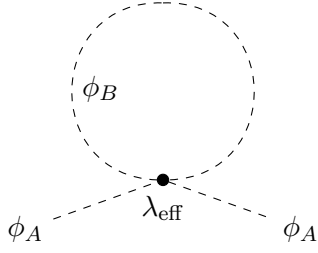

\begin{figure}[t!]
\centering
\includegraphics[keepaspectratio, width=0.75\textwidth]{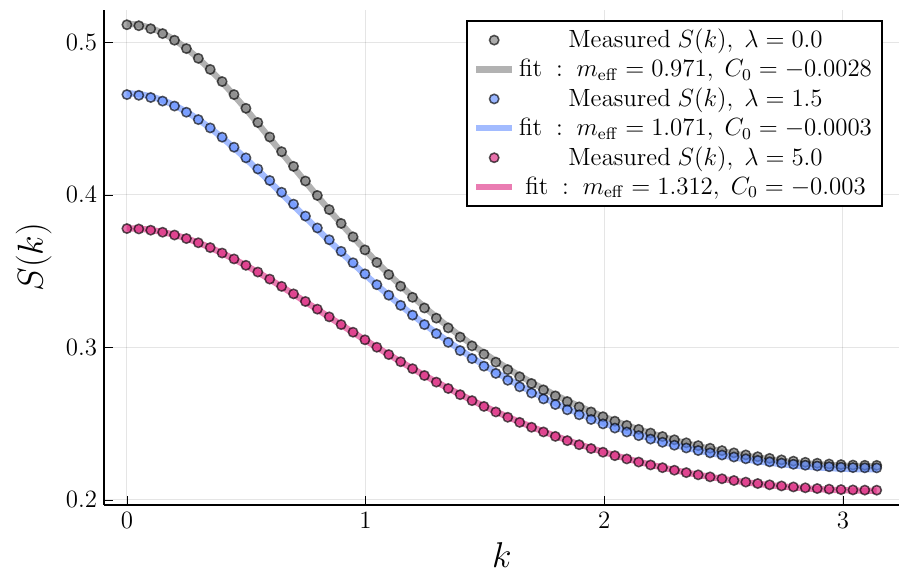}
\caption{Comparison of the structure factor of the QFT vacuum in the BRTN, for the field with bare mass $m=1$, and various values of bridge coupling, $\lambda$. Here we take local DVR Hilbert space of dimension 6, cut-off of $1\times 10^{-6}$, and a maximum bond dimension of $20$. The best fit values of the effective mass and shift $C_0$ are shown for $\lambda =0,~1.5,~5.0$. The near perfect fit indicates dominant mass renormalisation.}
\label{fig:BRTN_Sk_128}
\end{figure}

\begin{figure}[t!]
\centering
\includegraphics[keepaspectratio, width=0.75\textwidth]{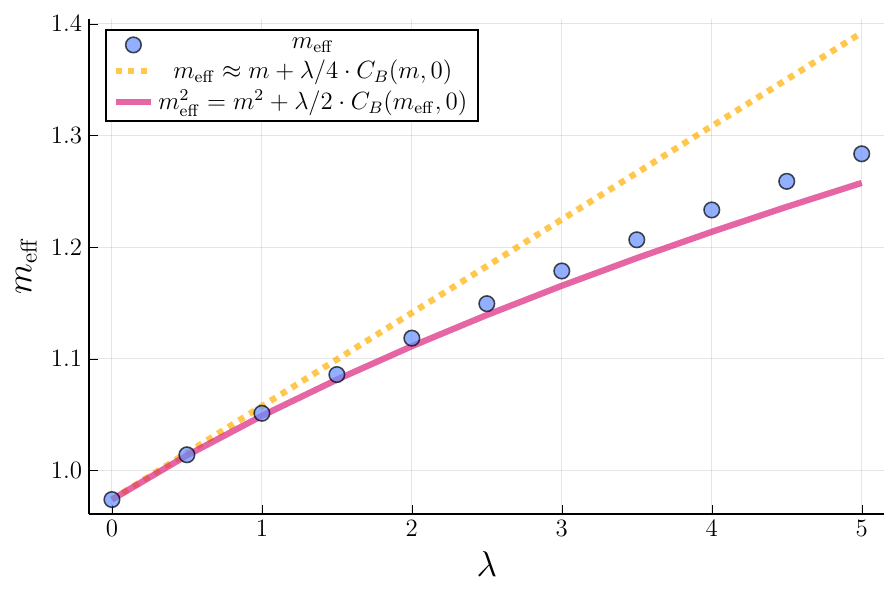}
\caption{The mass $m_{\rm eff}$ in the two scalar field test case, deduced by measuring the structure factor, and compared with  the mass obtained using the bubble diagram of \figref{fig:bubble}. The yellow dashed line is the linear approximation, the pink solid line is the solution to the gap equation in Eq.~\eqref{eq:gap}.}
\label{fig:meff_vs_lambda}
\end{figure}

\subsection{Radiative corrections in the test case}  
In the test case ({\it a.k.a.}~Quartic bridge coupling with reciprocity angle of $\rho=0$), the bridge coupling $\lambda \sum_n \hat q _{A,n} ^2  \hat q _{B,n}^2 $
simply acts as a direct quartic interaction between two scalar modes on the $X$-side and two scalar modes on the $K$-side, with the $K_n$ nodes being interpreted as
providing a copy of the $X_n$ coordinate nodes.

Thus we expect to find radiative corrections to the effective mass-squared whose leading order contribution would come from the same one-loop bubble diagram that one would find in ordinary two-scalar QFT, which is shown in \figref{fig:bubble}.
One could at this point choose to normal order the potential, \ie take $  V_{\rm bridge}
    ~=~
    \lambda_{\rm eff}\sum_{n=1}^N : \hat q_{A,n}^2 : ~: \hat q_{B,n}^2:~,
$
which would remove this leading order contribution. However we will instead take advantage of the fact that if we retain the entire term, the resulting radiative corrections are  significant, and measuring them will constitute a useful diagnostic test.

To translate the BRTN bridge into ordinary two-scalar language, we need to compensate for the double cover of the space coordinate. A 
fully symmetric realisation would carry both fields on both sides, 
$\phi_A = \tfrac{1}{\sqrt2}(q_{A,X}+q_{A,K})$ and similarly for $\phi_B$, so 
that the physical coupling $\lambda\,\phi_A^2\phi_B^2$ expands into four 
sectoral terms, $\tfrac{\lambda}{4}(q_{A,X}+q_{A,K})^2(q_{B,X}+q_{B,K})^2$. Only one of these, the single cross-sector term 
$\tfrac{\lambda}{4}\,q_{A,X}^2 q_{B,K}^2$, is retained, while  the others are images that 
normalisation renders redundant. Thus it is convenient to introduce the effective bridge coupling
\begin{equation}
    \lambda_{\rm eff}=\frac{\lambda}{4}~,
\end{equation}
which accounts for this doubled covering of the two sectors in the BRTN parametrisation. 
The leading bubble correction to the $A$-sector mass is equivalent to replacing one pair of $B$-fields by its vacuum expectation value:
\begin{equation}
    \frac{\delta m_A^2}{2} \;\approx\; \lambda_{\rm eff}\,\langle \phi_B(x)^2\rangle
    \;=\;
    \lambda_{\rm eff}\, C_B(0),
    \label{eq:dma2}
\end{equation}
where
\begin{equation}
    C_B(0)
    ~\equiv~
    \langle \phi_B(x)\phi_B(x)\rangle_{t=t'}
\end{equation}
is the coincident equal-time two-point correlator (with the  equal-time correlator being $   C_B(x-x')
    =
    \langle \phi_B(x)\phi_B(x')\rangle_{t=t'}$ generally). When we do not implement any kind of normal ordering or counter-term, we expect the contributions on the finite lattice to be dominated by this  Hartree approximation. 
Assuming translational invariance of the vacuum, the local variance is independent of $n$, and we may use 
\begin{equation}
   C_B(0)~=~\langle \hat q_{B,n}^2\rangle ~, 
\end{equation}
for some $n$, 
where the equal-time two-point correlator is simply the Fourier transform of the structure factor $S_B(k)=1/2\omega_k$ (\ie the momentum propagator after performing the $k_0$ integral): 
\begin{equation}
    C_B(0)
    ~=~
    \frac{1}{N}\sum_{\ell=0}^{N-1}\frac{1}{2\omega_\ell}~.
    \label{eq:CB0_disc}
\end{equation}
Noting that for the trivial local BRTN, the $B$-sector has the same free lattice dispersion given in Eq.~\eqref{eqn:omegas2} as the $A$-sector,  Eq.~\eqref{eq:dma2} gives the discrete prediction,
\begin{equation}
    m_{A,\rm eff}^2
    ~\approx~
    m^2
    +
 \frac{ \lambda}{2}
    C_B(m,0)~,
        \label{eq:mA2_disc}
\end{equation}
where 
\begin{equation}
   C_B(m,0)~=~  \frac{1}{N}\sum_{\ell=0}^{N-1}
    \frac{1}{2\sqrt{m^2+4\sin^2(\pi \ell/N)}}~.
    \label{eq:CBm_disc}
\end{equation}
For comparison, for the parameters used in the numerics, $m=1$ and $N=128$, Eq.~\eqref{eq:mA2_disc} gives
$    \delta m_A^2 \approx 0.1606\,\lambda. 
$

There is a further improvement that can be made to the calculation of $\delta m_A^2 $ which is to correct for the fact that the above computation of the mass of $\phi_A$ did not also consider what happens to the mass of $\phi_B$, and this  is an implicit breaking of the symmetry of the network. To correct for this, both masses should be treated simultaneously, yielding coupled gap equations. That is, the mass appearing in the correlator $C_B$ should be the already corrected mass $m_A$ (recalling that the reciprocity also swaps $A$ and $B$ field labels), and vice versa, which gives 
\begin{align}
m_{A,{\rm eff}}^2 &~=~ m^2 +  \frac{ \lambda}{2}C_B(m_{B,{\rm eff}},0) \nonumber \\
m_{B,{\rm eff}}^2 &~=~ m^2 +  \frac{ \lambda}{2}C_A(m_{A,{\rm eff}},0)~.
\end{align}
Imposing Born-reciprocity implies an effective mass of  $m_A=m_B \equiv m_{\rm eff}$, with the degenerate mass $m_{\rm eff}$ solving 
\begin{align}
m_{\rm eff}^2 &~=~ m^2 + \frac{ \lambda}{2}C(m_{\rm eff} , 0) ~.
\label{eq:gap}
\end{align}\\

Let us now  probe these radiative corrections to $m^2$ by measuring the structure factor  $S_A(k)$ in the vacuum. \figref{fig:BRTN_Sk_128} shows  $S_A(k)$ for the three values of $\lambda = 0,\,1.5,\,5.0$. The same figure shows clearly excellent fits to the function Eq.~\eqref{eq:Skfit} as solid lines which allows us to extract the effective mass-squared $m_{\rm eff}^2$. Importantly there is no extra momentum dependence in these fits, in agreement with there being no leading order field renormalisation in the theory. As in regular $\lambda\phi^4$ theory one might expect it to appear at higher order. 

In \figref{fig:meff_vs_lambda} we compare the theoretical prediction of the radiatively corrected mass, with the value of $m_{\rm eff}$ deduced as above by making fits to the measured covariance matrix. We show two theoretical curves. The first dashed yellow line is the linear approximation to Eq.~\eqref{eq:mA2_disc}. The second solid pink line is the value of $m_{\rm eff}$ deduced from the gap equation \eqref{eq:gap}. The remaining discrepancy in the result could be due to several factors. At $N=128$, the discreteness of the lattice plays almost no discernible role, however the truncation to $d=6$ of the local Hilbert space is likely to be more important. In addition two-loop corrections (which we can estimate by for example considering the one-loop corrected estimate for $\lambda$ in $\delta m^2$) are of roughly the same order as the remaining discrepancy. We conclude that to this order the network reproduces the theoretically expected results. 

\section{Vacuum preparation and radiative corrections in the Born-reciprocal case}
\label{sec:BRTNObs}
We have now gathered all the tools that we will need for the genuinely Born-reciprocal system with Fourier bridge. Again our focus is on the properties of the vacuum, and the radiative contributions to the self-energy of the field $\phi_A\equiv \hat q _A$. As for the test case we will begin by discussing and calculating the expected contributions, in order to decide precisely what it is that we wish to measure. This now becomes a rather subtle issue, because we now encounter  the breaking of  translation invariance that was alluded to earlier. As we shall see this will give us a useful way to probe the radiative corrections.  

\subsection{The perturbative approach} 

To calculate the expected radiative corrections we note that Born-reciprocity allows us to operate  perturbatively, even in the absence of a conventional local continuum description of the full theory, under the assumption that the bridge coupling  behaves as contact interactions between QFT $\phi_A$ modes and   $\phi_B$ modes. 

Under this assumption we construct the corrections as Feynman diagrams in the usual manner, however with the propagators for the $\phi_B$ fields having a form that is appropriate when $K_n$ is the native coordinate. In other words, 
the $N\to\infty$ limit of the structure factor and correlator of the $\phi_A$ free theory are the same as for ordinary scalar QFT,
\begin{align}
    S_A(k) ~&=~ \frac{1}{2\omega_k}~; ~~
    \omega_k = \sqrt{m_{\rm phys}^2 + k^2} \nonumber \\
    C_A(x) ~&=~ 
    \int \frac{dk}{2\pi}\,
    \frac{\cos{k x}}{2\omega_k}
   ~ =
~    \frac{1}{2\pi}K_0(m_{\rm phys}|x|)~,
\end{align}
where $K_0$ is the modified Bessel function of the second kind, but for the $\phi_B$ field the functions are swapped around;
\begin{align}
    C_B(x) ~&=~ \frac{1}{2\omega_x}~; ~~
    \omega_x = \sqrt{ m_{\rm phys}^2 + x^2 a^2_K/a^2_X } \nonumber \\
    S_B(k) ~&=~
~    \frac{1}{2\pi}K_0(m_{\rm phys}|k|a_X/a_K)~.
\label{eq:SBk}
\end{align}
A crucial aspect of these equations which is very different from regular perturbation theory is that the propagators are defined with respect to  a bi-local system that is translationally invariant and BR-local only at tree-level. In this unperturbed theory one can consistently introduce diagonal two-point functions such as the structure factors $S_A(k)$ and reciprocal correlator $C_B(x)$. In general these quantities will not remain diagonal in the full interacting theory however, reflecting the breakdown of translational invariance. Perturbation theory therefore allows us to calculate the {\it leading} departure from translational invariance, but it does not provide a systematically improvable iterative expansion which can replace normal perturbation theory.   

\begin{figure}[h]
\centering
\includegraphics[keepaspectratio, width=0.75\textwidth]{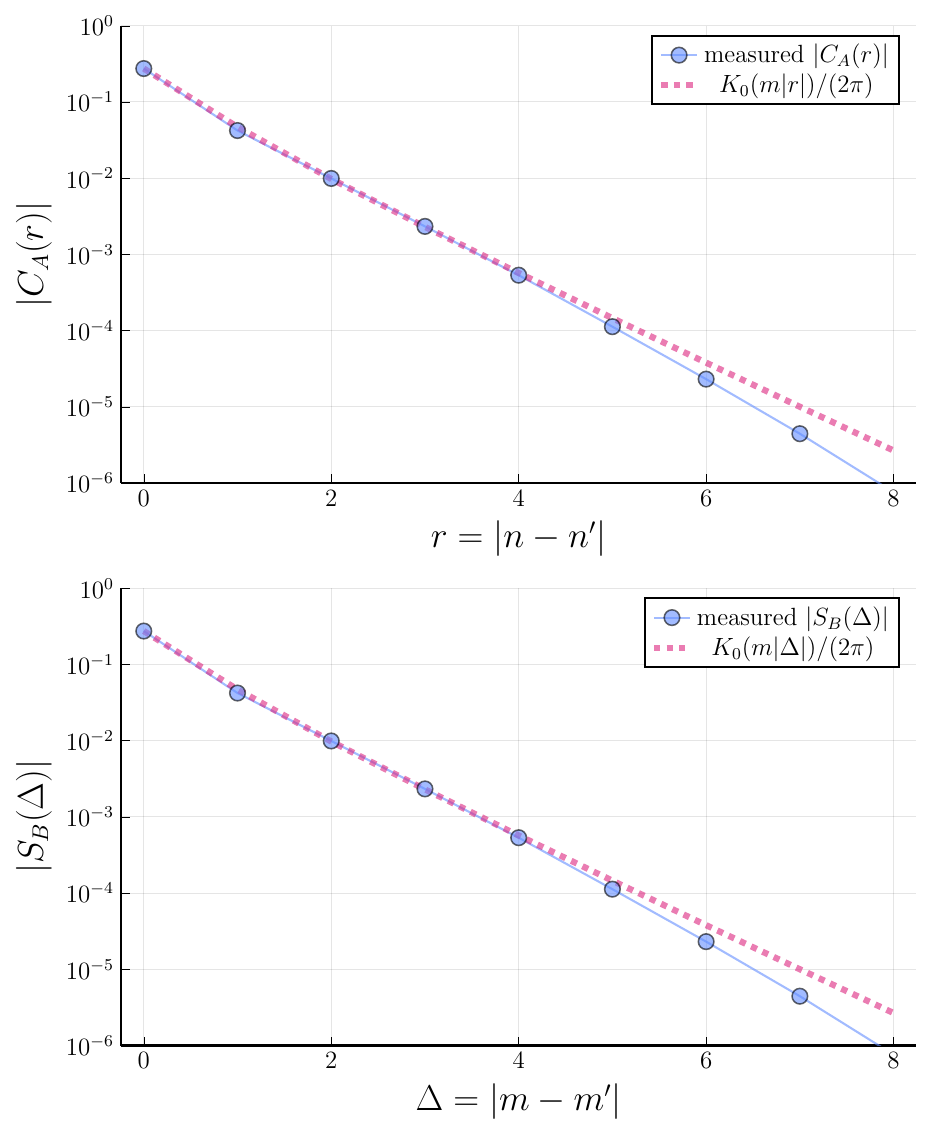}
\caption{ Correlators measured in the native coordinates on the $g=0$ BRTN lattice compared to the ideal. Deviation is due to truncation on the lattice. Further small deviations will arise due to radiative corrections caused by non-zero $g$. This can be used to motivate a near-field approach to measuring the self-energy, $\Sigma$. }
\label{fig:CACB_diag}
\end{figure}

At finite $N$  these functions take the obvious generalisation of the discrete form that we saw already in Eqs.~\eqref{eq:HK}, with discrete $\omega$ dispersion relations. For convenience let us collect them and write them in the dimensionless coordinates $X_n\equiv n$ and $K_m\equiv m$: 
\begin{align}
S_A(K_m) ~=~ \frac{1}{2\omega_{K,m}} ~;~~~C_A(X_n)
~&=~
\frac{1}{N}\sum_r
\,
\frac{e^{ik_r x_n}}{2\omega_{K,r}}~,
\nonumber \\
 C_B(X_n) ~=~ \frac{1}{2\omega_{X,n}} ~;~~~S_B(K_m)
~&=~
\frac{1}{N}\sum_{n'}
\,
\frac{e^{i k_m x_{n'}}}{2\omega_{X,n'}}~,
\label{eq:CA_SB_defs}
\end{align}
where 
\begin{align}\label{eqn:omegas4}
\omega_{X,n} ~&=~ \sqrt{ {m^2}  + {4}\sin^2 \left(\pi X_n /N) \right) }~\nonumber \\
\omega_{K,m} ~&=~ \sqrt{ {m^2}  + {4}\sin^2 \left(\pi K_m /N \right) }~,
\end{align}
and where recall that $k_m x_n = nma_X a_K = 2\pi nm/N$. 
As a first step we may measure these correlation functions on the BRTN, which we show in \figref{fig:CACB_diag} for an $N=32$ lattice. An important aspect of these functions is that they decay exponentially fast in the native coordinate. 

\subsection{Radiative corrections in the trilinear model: boundary tadpole}

In this work we will study radiative corrections in the trilinear theory. A useful test is to study the  boundary terms. According to our previous arguments, in the non-normal ordered theory, such terms appear at linear order in $g$: to recap, a Hamiltonian of the form given in Eq.~\eqref{eq:fourierbridge}
makes a leading contribution to the effective theory of $\phi_A$ proportional to  
$$ ~\propto~ g\sum_\alpha \sum_{nm} \cos(k_m x_n) ~{\cal O}_A^\alpha (x_n) \langle {\cal O}_B^ \alpha (k_m)  \rangle~,$$
and then tree-level translational invariance of the expectation values allows us to sum over $m$, giving a contribution proportional to $g \delta _{0,n} {\cal O}_A^\alpha (x_n) $. Thus we expect the leading contribution to be boundary terms 
encapsulating the ``divergences'' (\ie the terms that grow with $N$) which the bridge couplings {\it would} give rise to in the corresponding regular QFT. 

The trilinear theory without normal ordering has a bridge potential of the form $V_{\rm bridge}^{\rm (tri)}$ given in Eq.~\eqref{eq:V_tri}. The tadpole comes from the previously introduced diagram of \figref{fig:smeared_bubble}, which is straightforward to calculate:
 \begin{align}
     V_{\rm eff} ~\supset~ \hat q_{A,0}~ g \sqrt{N} S_B(0)~, 
 \end{align}
where $S_B(0)$ is determined from Eq.~\eqref{eq:CA_SB_defs} using the effective mass. If we also wish to include a $\lambda_4 \hat q_A ^4 $ term in the potential,  the effective mass depends on 
the connected variance 
\begin{equation}
    v_A(n)
    ~=~
    \langle \hat q_{A,n}^2\rangle -    \langle \hat q_{A,n}\rangle^2~, 
\end{equation}
as 
\begin{align} 
m_{\rm eff}^2(n)~=~ m^2 + 12\lambda_4 ~v_A(n) ~,
\end{align}
which is in general position dependent. 

\begin{figure}[t!]
\centering
\includegraphics[keepaspectratio, width=0.75\textwidth]{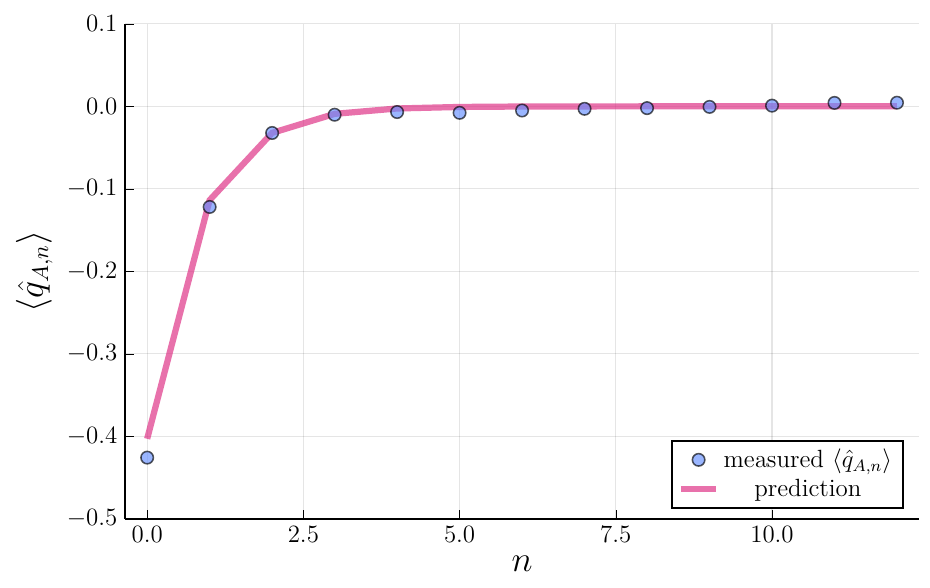}
\caption{The field profile induced at the boundary due to the one-loop tadpole in \figref{fig:smeared_bubble} with the non-normal-ordered trilinear bridge coupling. Here the coupling is $g=0.5$ with a quartic coupling $\lambda_4 \phi_A^4 $ with $\lambda_4 = 0.25$ on an $N=64$ BRTN lattice. The solid line is the theoretical profile deduced from Eq.~\eqref{eq:solving}. The scatter plot is of the measured expectation values.  }
\label{fig:defect}
\end{figure}

The localised tadpole causes a defect located at the boundary, and we will  measure its profile. This defect can be modelled using a local Hartree approximation. Writing $
    q_n \equiv \langle \hat q_{A,n}\rangle ~,$
we define open-chain Hartree operators which act on the defect profile as 
\begin{align}
    (\mathcal L_A q)_n ~&=~
    m_{\rm eff}^2(n)\,q_n +
   {a_X^{-2}} 
    \left(
        2q_n-q_{n-1}-q_{n+1}
    \right)~,
    \qquad
    1\le n\le N-2 ~,\nonumber \\
    (\mathcal L_A q)_0
    ~&=~
   m_{\rm eff}^2(0)\,q_0 +  {a_X^{-2}}(q_0-q_1),
    \nonumber \\
    (\mathcal L_A q)_{N-1}
    ~&=~m_{\rm eff}^2(N-1) \, q_{N-1} +    {a_X^{-2}}(q_{N-1}-q_{N-2})~.
\end{align}
The profile of the defect is then determined by solving
\begin{equation}
    (\mathcal L_A q)_n+    4\lambda_4 q_n^3 +g \sqrt{N} S_B(0)\delta_{n,0}
    ~=~
    0 ~.\label{eq:solving}
\end{equation}
An example of the resulting profile is shown in \figref{fig:defect} as the solid line, with $\lambda_4 = 0.25$ and a coupling $g=0.5$. In the same figure we show the expectation values $\langle \hat q_{A,n}\rangle$ measured on the same network, which are in agreement modulo a small discrepancy in the peak at the $n=0$ node of order a few percent. 

\begin{figure}[t!]
\centering
\includegraphics[keepaspectratio, width=0.75\textwidth]{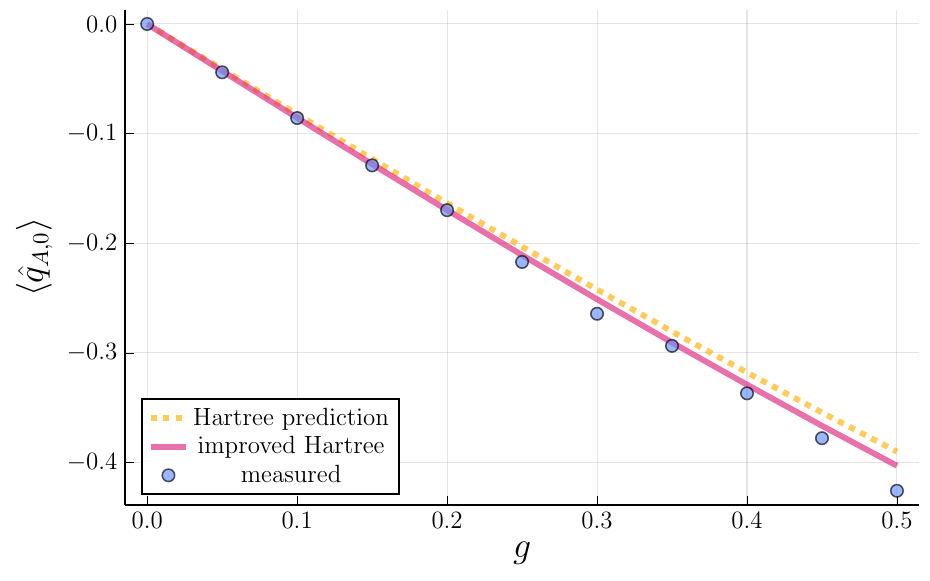}
\caption{Boundary field value $\langle q_{A,0}\rangle$ versus coupling  $g$ with the non-normal-ordered trilinear bridge coupling, and with quartic coupling $\lambda_4 \phi_A^4 $ with $\lambda_4 = 0.25$ on an $N=64$ BRTN lattice. The yellow dashed line shows the predicted field value using the Hartree approximation, \ie solving Eq.~\eqref{eq:solving}, with the decoupled effective mass $m_{\rm eff}$ with $g=0$, \ie  with $m^2 _{\rm eff} = m^2 + 12\lambda _4  v_A^{(g=0)}$. The solid red line shows the improved Hartree approximation, made using the empirically measured variance,  $v_A^{(g)}=\langle {\hat q_{A,n}^2\rangle}-\langle {\hat q_{A,n}\rangle}^2$,  in $m_{\rm eff}$. The scatter plot is of the measured  values. }
\label{fig:tadpole_versus_g}
\end{figure}

This discrepancy grows with $g$, and is likely due to back reaction of the reciprocal  sector on $S_B(0)$. To illustrate, \figref{fig:tadpole_versus_g} which plots the field value versus $g$, shows the effect of back-reaction on the expected field value if we improve how we determine the effective mass. The yellow dashed line shows the expected result found solving Eq.~\eqref{eq:solving} with constant effective mass determined from $m_{\rm eff}^2 = m^2 + 12\lambda_4 v_A^{(g=0)}$, \ie $m_{\rm eff}$ when there is no bridge term. The solid red line shows the improved Hartree approximation made using the 
measured local connected variances
$m_{\rm eff}^2(n) = m^2 + 12\lambda_4 v^{(g)}_A(n))$.

\begin{figure}
\begin{minipage}{0.5\textwidth}
\centering
\begin{tikzpicture}[thick, line cap=round, line join=round]
  \coordinate (V1) at (-1.3,0);
  \coordinate (V2) at ( 1.3,0);
  \coordinate (L) at (-2.3,0);
  \coordinate (R) at ( 2.3,0);

  \fill[gray!20] 
    (V1) .. controls (-1.0,1.5) and (1.0,1.5) .. (V2)
    -- (V2) .. controls (0.7,0.9) and (-0.7,0.9) .. (V1) -- cycle;

  \fill[gray!20] 
    (V1) .. controls (-1.0,-1.5) and (1.0,-1.5) .. (V2)
    -- (V2) .. controls (0.7,-0.9) and (-0.7,-0.9) .. (V1) -- cycle;

  \draw[dashed] (L) -- (V1);
  \draw[dashed] (V2) -- (R);

  \fill[gray] (V1) circle (2.5pt);
  \fill[gray] (V2) circle (2.5pt);

  \node[left] at (L) {$\phi_A$};
  \node[right] at (R) {$\phi_A$};
  \node at (0,0.90) {$\phi_B$};
  \node at (0,-0.88) {$\phi_B$};

\end{tikzpicture}
\end{minipage}
\hfill
\begin{minipage}{0.5\textwidth}
\centering
\begin{tikzpicture}[thick, line cap=round, line join=round]
  \coordinate (V1) at (-1.3,0);
  \coordinate (V2) at ( 1.3,0);
  \coordinate (L) at (-2.3,0);
  \coordinate (R) at ( 2.3,0);

  \fill[gray!20] 
    (V1) .. controls (-1.0,1.5) and (1.0,1.5) .. (V2)
    -- (V2) .. controls (0.7,0.9) and (-0.7,0.9) .. (V1) -- cycle;

  \draw[dashed] (L) -- (V1);
  \draw[dashed] (V2) -- (R);

  \fill[gray] (V1) circle (2.5pt);
  \fill[gray] (V2) circle (2.5pt);

  \draw[dashed] (V1) .. controls (-0.6,-1.3) and (0.6,-1.3) .. (V2);

  \node[left] at (L) {$\phi_A$};
  \node[right] at (R) {$\phi_A$};
  \node at (0,0.90) {$\phi_B$};
  \node at (0,-0.7) {$\phi_A$};

\end{tikzpicture}
\end{minipage}
\hfill

\begin{minipage}{1.0\textwidth}
\centering
\begin{tikzpicture}[thick, line cap=round, line join=round]
  \coordinate (V1) at (-1.3,0);
  \coordinate (V2) at ( 1.3,0);
  \coordinate (L) at (-2.3,0);
  \coordinate (R) at ( 2.3,0);

  \fill[gray!20] 
    (V1) .. controls (-1.0,1.5) and (1.0,1.5) .. (V2)
    -- (V2) .. controls (0.7,0.9) and (-0.7,0.9) .. (V1) -- cycle;

  \draw[dashed] (L) -- (V1);
  \draw[dashed] (V2) -- (R);

  \fill[gray] (V1) circle (2.5pt);
  \fill[gray] (V2) circle (2.5pt);

  \draw[dashed] (V1) .. controls (-0.6,-1.3) and (0.6,-1.3) .. (V2);

  \node[left] at (L) {$\phi_A$};
  \node[right] at (R) {$\phi_B$};
  \node at (0,0.90) {$\phi_B$};
  \node at (0,-0.7) {$\phi_A$};

\end{tikzpicture}
\end{minipage}

    \caption{One-loop contributions to the self-energy $\Sigma_A(K,K')$ and to the mixing term $\Sigma_{AB}(K,K')$ from the reciprocal degrees of freedom in the trilinear model. }
    \label{fig:fat_two_tailed_fish}
\end{figure}
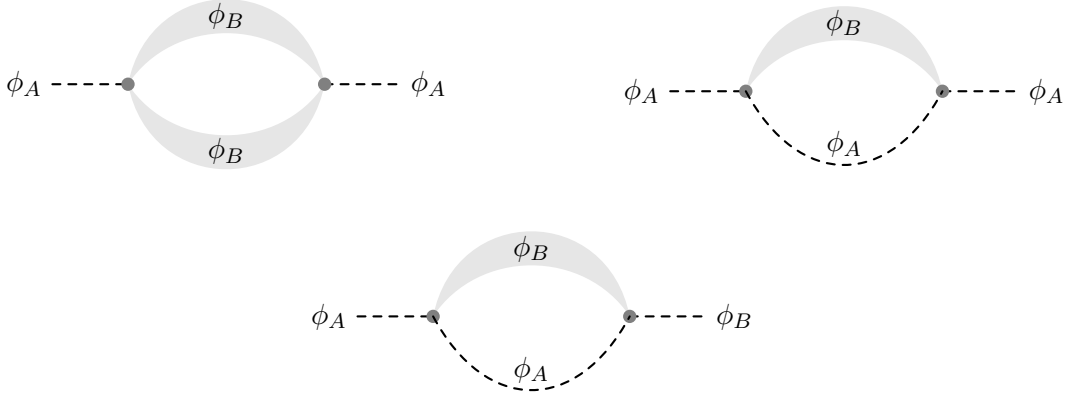

\subsection{Radiative corrections in the trilinear model: bulk self-energy}

As outlined, the defect of the previous subsection can be removed by normal ordering the theory. However there are higher order bulk contributions in the trilinear theory that remain, which we now consider.  For these we may consider the bridge potential to now be normal ordered to subtract the boundary term.

Let us focus on the self-energy $\Sigma(K,K')$. This operator receives
contributions through the diagrams in \figref{fig:fat_two_tailed_fish}, which are the leading terms in the bulk, occurring  at order $g^2$.
These contributions will of course be far more interesting than the tadpole, because they involve non-coincident reciprocal propagators. In general they will give contributions to the self-energy that are off-diagonal, $\Sigma(K_m,K_{m'})$, and they will also mix the $\phi_A$ and $\phi_B$ fields. 

It is this latter effect that will produce the cross-bridge entanglement in the trilinear system. It is worth noting that such $\phi_A-\phi_B$ mixing is not mandatory. For example in the bi-quadratic theory with the bridge potential $V_{\rm bridge}^{\rm (biquad)}$ given in Eq.~ \eqref{eq:V_fourier_summary}, it trivially cannot happen because of symmetry arguments (for example the Hamiltonian is invariant under ${\mathbbm Z}_2$ transformations with  $\hat q_A$ and $\hat q _B$  having charges $-1$ and $+1$ respectively). (In the theory with potential $V_{\rm bridge}^{\rm (odd)}$ there is in principle a three loop diagram that gives $\phi_A-\phi_B$ mixing.) 

Let us estimate these terms using the same perturbative approximations that we used for the tadpole. The first diagram gives the following contribution in native $X_n$ coordinates: 
\begin{equation}
    \Sigma^{\rm (ABB)}_{n,n'} ~=~ \frac{2g^2 }{N}\sum_{m,m'} \cos(k_m x_n)\cos(k_{m'}x_{n'})\; 
    \langle \hat q_{B,m}\, \hat q_{B,m'}\rangle^2 ~,
\end{equation}
where the factor of two is a symmetry factor. Setting $m' = m + \Delta$ and performing the sum over $m$ gives a single-site operator and its image:
\begin{equation}
\label{eq:sigma_onsite}
    \Sigma^{\rm (ABB)}_{n,n'} ~=~ g^2 \delta_{n,n'} \sum_{\Delta\in {\mathbb Z}} S_B(\Delta)^2\; e^{-i\frac{2\pi \Delta}{N} n}
    + g^2 \delta_{n,-n'} \sum_{\Delta\in {\mathbb Z}} S_B(\Delta)^2\; e^{-i\frac{2\pi \Delta}{N} n}~.
\end{equation}
Transforming to $K_m$-space we find 
\begin{equation}
\Sigma^{\rm (ABB)}(K_m,K_{m'})
~=~
g^2
\left[
S_B(K_m-K_{m'})^2
+
S_B(K_m+K_{m'})^2
\right]~.
\label{eq:Sigma_full_cos}
\end{equation}
Meanwhile the second diagram of 
\figref{fig:fat_two_tailed_fish}
is given by 
\begin{equation}
    \Sigma^{\rm (AAB)}_{n,n'} ~=~ \frac{2g^2 }{N}\sum_{m,m'} \cos(k_m x_n)\cos(k_{m'}x_{n'})\; 
    \langle \hat q_{B,m}\,\hat  q_{B,m'}\rangle \langle \hat q_{A,n}\, \hat q_{A,n'}\rangle  ~.
\end{equation}
A similar treatment
gives 
\begin{equation}
\Sigma^{\rm (AAB)}(K_m,K_{m'})
~=~
g^2
\left[
C_A(0) S_B(K_m-K_{m'})
+
\frac{1}{N} \sum_{r}S_A(K_r) S_B(K_m+K_{m'}- 2K_r)
\right]~.
\label{eq:Sigma_full_cos2}
\end{equation}
In total then we now find a ``structure kernel'' matrix in momentum space,  which we can write as 
\begin{equation}
S_{AA}(K_m,K_{m'})
~=~
\frac{\delta_{m,m'}}{2\omega_{K,m}}
~-~
\frac{
\Sigma_{AA}(K_m,K_{m'})
}{
\omega_{K,m}\omega_{K,m'}
\left(\omega_{K,m}+\omega_{K,m'}\right)
}\left( 1+{\cal O}(g^2/N)\right) 
~,
\label{eq:SA_full_cos3}
\end{equation}
where $\Sigma_{AA}(K_m,K_{m'}) ~=~ 
\Sigma^{\rm (ABB)}(K_m,K_{m'}) + \Sigma^{\rm (AAB)}(K_m,K_{m'})
$. 
There is of course an equivalent contribution to $S_B$ that may get by simply reciprocating the result in 
Eq.~\eqref{eq:SA_full_cos3}.  

Finally the third mixing diagram in \figref{fig:fat_two_tailed_fish} involves one of each kind of propagator. It gives
\begin{equation}
    \Sigma_{AB}(X_n,K_m) ~=~ \frac{2g^2 }{N}\sum_{n',m'} \cos(k_{m'} x_n)\cos(k_{m}x_{n'})\; 
    \langle \hat q_{B,m'}\,\hat  q_{B,m}\rangle \langle \hat q_{A,n}\, \hat q_{A,n'}\rangle  ~,
\end{equation}
where the external indices are now one from the $X$-sector ($n$) and one from the $K$-sector ($m$). Performing the sum over $m'$ yields $C_B(X_n)\cos(k_m x_n)$, and the sum over $n'$ gives $S_A(K_m)\cos(k_m x_n)$, which yields
\begin{equation}
    \Sigma_{AB}(X_n,K_m) ~=~ \frac{g^2 }{N}\; C_B(X_n)\; S_A(K_m)\; 
    \left[1 + \cos(2k_{m}x_{n})\right]  ~.
    \label{eq:Sigma_AB}
\end{equation}
Transforming to momentum space this becomes 
\begin{equation}
\Sigma_{AB}(K_m,K_{m'})
~=~
\frac{g^2}{\sqrt{N}}\,
S_A(K_{m'})
\left[
S_B(K_m)
+
\frac{1}{2}S_B(K_m-2K_{m'})
+
\frac{1}{2}S_B(K_m+2K_{m'})
\right]~.
\label{eq:Sigma_AB_momentum}
\end{equation}
The full structure kernel of the BRTN is then a $2N\times 2N$ matrix of the form 
\begin{equation}
S_{\rm BRTN}  ~=~ \begin{pmatrix} S_{AA}(K_m,K_{m'}) & S_{AB}(K_m, K_{m'})  \\ S_{BA}(K_m, K_{m'}) & S_{BB}(K_m,K_{m'}) \end{pmatrix} ~,
\end{equation}
where the within-sector parts are given by \Eq{eq:SA_full_cos3} and its Born-reciprocal image, and where the cross-sector part is 
\begin{equation}
S_{AB}(K_r,K_s) ~=~ - \frac{
\Sigma_{AB}(K_r,K_s)
}{
\omega_{K,r}\omega_{X,s}
\left(\omega_{K,r}+\omega_{X,s}\right)
} \left( 1 + {\cal O}(g^2/N) \right) ~,
\end{equation}
where $\Sigma_{AB}(K_r,K_s)$ is given by \Eq{eq:Sigma_AB_momentum}. 

The within-sector self-energies $\Sigma_{AA}$ and $\Sigma_{BB}$ are 
${\cal O}(g^2)$ and encode the physical effects of integrating out the 
reciprocal degrees of freedom. The associated structure kernel 
$S_{AA}(K,K')$ acquires non-diagonal pieces in $K$-space, marking a departure 
from ordinary QFT. These pieces break translational invariance (as we discuss 
in detail below), but crucially they remain within a single sector: they are 
absorbable, in the sense that they can be reproduced by a local, 
$X$-dependent modification of the $\phi_A$ parameters, so they do not by 
themselves prevent the $K$-sector from being integrated out. By contrast the 
cross-sector kernel $S_{AB}$ represents genuine $A$--$B$ entanglement, and it 
is {\it this} term that appears to obstruct any Wilsonian procedure of 
integrating out the $K$-sector.  As we shall now see, the non-standard effects 
generated by both kinds of term are diluted in the large-volume limit, such 
that the bulk Wilsonian picture is restored.

\section{When renormalisation forgets: restoration of translational invariance and locality at large volume}

\label{sec:breaking}

The non-diagonal form of the structure kernel is in stark contrast to normal QFT, in which the structure kernel is $S(k,k') = S(k) \delta (k-k')$. This has important physical implications that we will now address. 

Essentially such off-diagonality means that once there is propagation via modes of $\phi_B$, neither momentum conservation nor locality is preserved in the system. Crucially however this off-diagonal structure is controlled by the reciprocal propagator, $S_B(\Delta)$, which decays exponentially rapidly in the lattice index. This behaviour is very general: if $C_B(X_n) = 1/(2\omega_{X,n})$ is analytic and periodic, its Fourier coefficients 
$S_B(\Delta)$ decay as $e^{-\eta|\Delta|}$, where $\eta$ is determined by the nearest 
complex singularity of $1/\omega_X$. Indeed, we verified this exponential decay in \figref{fig:CACB_diag} which  agrees with the  large-$N$ Bessel function approximation in Eq.~\eqref{eq:SBk}. Thus although the self-energy has unfamiliar off-diagonal contributions, it is {\it nearly} diagonal. The Born-reciprocity has smeared the usual delta function by an amount that is of order $\eta \sim 1/m$, which in the example we have been considering is barely wider than the lattice spacing itself.

What is the meaning of such smearing? Recall that the normal diagonal structure of $
\Sigma_A(K_m,K_{m'}) 
$ 
is synonymous with translational invariance: suppose for simplicity that we had \begin{equation}
\Sigma_A(K,K')~=~ \rho\!\left({|K-K'|m} \right)  ~,
\end{equation}
where the function $\rho$ encapsulates our smearing of the $\delta$ function with its width in the dimensionless momentum-space coordinate $K$  being order $1/m$. The quadratic correction to the Hamiltonian is then  
\begin{align}
\Delta H_{\rm eff}
&~=~
\sum_{K,K'} \phi_A(K)\,\rho\!\left({|K-K'|m}\right) \,\phi_A(K')~.
\end{align}
By straightforward manipulation we may write this in terms of the  field in $X$-space $\phi_A(X)$ as 
\begin{align}
\Delta H_{\rm eff}
&~=~
\sum_{X} 
  \frac{\delta m_{\rm eff}^2(X)}{2} \phi_A(X)^2
\end{align}
where
\begin{align}
   \frac{\delta m_{\rm eff}^2(X)}{2}~=~ \sum_K   
  e^{-i 2\pi K X/N} \,\rho\!
  \left({|K|m}\right) ~.
\end{align}
In other words the field has picked up an $X$ dependent mass which is the Fourier transform of $\rho$. 
It is this breaking of translational invariance (even after the subtraction of the boundary defect) that leads to momentum non-conservation. As the $\phi_A$ particle propagates it experiences a force due to its interaction with the reciprocal background, which is able to absorb momentum without producing an asymptotic state. This will be seen as momentum non-conservation in amplitudes. For example, the amplitude for an $s$-channel exchange $\phi_A\phi_A\to \phi_A\phi_A$ with an intermediate $\phi_B$, which ordinarily would take the form
\begin{equation}
    {\cal M}(k_1,k_2\to k_3,k_4) ~\sim ~ g^2 \delta (k_1+k_2 - k_3 - k_4)\frac{1}{2\omega_{k_1+k_2}} ~,
\end{equation}
is instead 
\begin{equation}
    {\cal M}(k_1,k_2\to k_3,k_4) ~\sim ~ g^2 S_B(k_1+k_2 - k_3 - k_4) + \mbox{image term}~.
\end{equation}
Thus whenever $\phi_A$ encounters an intermediate $\phi_B$ state, the momentum distribution in the amplitude becomes smeared. 
So far we have for simplicity been considering $m=1$ which is as large a mass as can be handled by the network -- its Compton wavelength is of order $1/a_X$ and its structure kernel decays very rapidly. The usual Fourier duality tells us that the width of the mass-squared profile $m_{\rm eff} ^2(X)$ in $X$-space is then maximally large, of order $mN/2\pi$. 
Thus for such a heavy state the breaking of translational invariance happens on length scales that are comparable to the entire  network.

This breaking of translational invariance is much less alarming than one might suppose, as becomes clear once one considers how the physics scales with volume. For this purpose let us consider arbitrary finite masses $m$. For typical phenomenological applications  for example we might wish to consider a mass much smaller than the scale associated with the lattice spacing $a_X$. Adopting the lattice spacing as our fundamental unit, $M_*=1/a_X$, we will keep the physical mass fixed such that $m_{\rm phys}/M_*= {\rm constant}$. The other scales in the set-up are the physical size of the `box' $2\pi R$ which was defined earlier in Eq.~\eqref{eq:2piR}, and the minimum momentum exchange $a_K$:  in summary we have 
\begin{align} 
a_X ~=~ \frac{1}{M_*} ~,~~m ~=~ \frac{m_{\rm phys}}{M_*}~,~~2\pi R ~=~ N a_X =\frac{N}{M_*}~,~~ a_K ~=~ \frac{1}{R}~. 
\label{eq:lims}
\end{align} 
As we have seen the structure kernel has width $\Delta K ~\sim ~1/m = \frac{M_*}{m_{\rm phys}}$. This could now be very large in dimensionless momentum units if the mass happens to be much smaller than $M_*$, however crucially the physical momentum violation scales as  
\begin{align} 
\Delta k ~=~ a_K \Delta K ~= ~
\frac{2\pi M_* }{N}  \times 
\frac{M_*}{m_{\rm phys}}~.
\label{eq:Deltak}
\end{align} 
Thus the QFT of the $\phi_A$ particle will appear to be approximately momentum conserving provided that $\Delta k \ll m_{\rm phys}$, which according to  Eq.~\eqref{eq:Deltak} provides the following  criterion: 
\begin{equation}
    2\pi R ~\gg ~ a_X \, \left( \frac{M_*}{m_{\rm phys}}\right)^2~.
    \label{eq:Nconst}
\end{equation}
In other words, the momentum violation depends  parametrically on the size of the network, and in the large $N$ limit normal translationally invariant QFT will be restored in the effective theory. (Note that this equation holds for space lattices with arbitrary dimension.)
In this limit the smallest unit of momentum exchange, namely $1/R$, goes to zero, while the largest possible momentum, $Na_K=N/R=2\pi M_*$, remains fixed and of order the fundamental scale. Therefore this is a large-volume / thermodynamic IR limit of the network and not a UV limit, as the lattice spacing  $a_X=1/M_*$ remains finite.  

In summary then the breaking of translational invariance that one finds in the BRTN should in fact be no more disconcerting than that found when performing QFT in a large box with Dirichlet boundary conditions. In both cases one finds non-diagonal structure kernels, and in both cases translational invariance is restored in the large volume limit. 

What about non-locality? This is  also induced by our structure kernel, at two orders higher in $g$,   but it turns out to be in a sense even less problematic. We may approximate this effect by utilising the induced operator $\Sigma_{AB}$ of Eq.~\eqref{eq:Sigma_AB} as a vertex (\ie using the `mass-insertion' approximation) in order to calculate an effective $\phi(X_n)\phi(X_{n'})$ coupling in the Hamiltonian. The result is schematically of the form 
\begin{equation} 
\Delta \Sigma ^{\rm (A)}(X_n,X_{n'})~\sim~-M_*^2~
\frac{g^4}{N} C_B(X_n) C_B(X_n') \sum_\ell {\cal V}_{n \ell }\Delta_B(\ell)  {\cal V}_{\ell n' } 
\end{equation} 
where ${\cal V}_{n \ell } = C_A(X_\ell) + \frac{1}{2}C_A(X_\ell - 2 X_n) +\frac{1}{2} C_A(X_\ell + 2 X_n) $, and where $\Delta_B(\ell)  = 1/\omega_{X,\ell}^2$ is the intermediate propagator. The physical interpretation of this operator is that a $\phi_A$ state can produce a  $\phi_B$ state  of length $a_X \ell$ which stretches between $X_n$ and $X_{n'}$. Indeed if the criterion for the theory to appear local is that this operator is much smaller than the physical mass-squared, $m_{\rm phys}^2$, then we find $m_{\rm phys}^2 \gg \frac{M_*^2g^2}{N}$ which is essentially the constraint in Eq.~\eqref{eq:Nconst} with more powers of $g$.

The ${\cal V}_{n\ell}$ vertices appearing in this operator play an important role. Due to the aforementioned exponential decay of the correlators,  $C_A\sim e^{-m_{\rm phys} x} $, these vertices serve to localise the contributions in the regions where they overlap. These are the boundary, and the locations where $X_\ell \pm 2 X_n = X_\ell \pm 2 X_{n'} $. The latter implies a smeared quasi-local contribution at $X_n\sim X_{n'}$ and its image at $X_n\sim  - X_{n'}$. Thus each particle, its image, and the boundary acquire a non-perturbative reciprocal dressing cloud. Unlike the tadpole,
these localised
contributions are not coherently enhanced at large $N$. Instead using Eq.~\eqref{eq:lims} we see that they are weighted
by the fraction of the volume occupied by the dressed region,
\begin{equation}
\Delta \Sigma^{(A)}
~\sim~
M_*^2 g^4\,\frac{r_{\rm phys}}{R} ~ ,
\end{equation}
where $r_{\rm phys}$ is the effective  radius of the dressed particle. 
Therefore this source of non-locality also disappears from the effective Hamiltonian in the large volume / thermodynamic limit.

Finally we should remark on the explicit manifestations of UV/IR mixing. 
The primary example of this is the  boundary tadpole that we have subtracted. The coefficient of this would-be UV contribution grows coherently with the size of the reciprocal lattice: in the large-volume limit of 
Eq.~\eqref{eq:lims}, with 
$a_X=M_*^{-1}$ fixed and 
\(R=N a_X\), the tadpole is of order
\begin{equation}
g\sqrt{N}
~=~
g\sqrt{RM_*}~ .
\end{equation}
In other words the ultraviolet boundary subtraction is controlled by the infrared 
volume parameter. Moreover this UV divergence appears on the boundary, which is the part of the network  corresponding to low-momentum on the reciprocal side. This is the sharpest manifestation of UV/IR mixing in the 
construction. 

Another manifestation of UV/IR mixing is the entropy contained in the bridge between the two halves of the system. We may estimate this from the von Neumann entropy associated with the tracing out of the reciprocal sector, as follows. Our perturbative approximations amount to an off-diagonal coupling between three excited modes on either of the network, of the form 
\begin{equation}
  {V}  ~=~ 
  \frac{g}{N}
  \sum_{nmm'}
  \cos \Big(\frac{2\pi(m+m')n}{N}\Big)\,
  \hat q_{B,n}\, \hat q_{A,m} \hat q_{A,m'}\; .
  \label{eq:Vmode}
\end{equation}
Here the modes $\hat q_{A,m}$ are the low lying modes of the $\phi_A$ field in the momentum basis, while $\hat q_{B,n}$ is the reciprocal equivalent for $\phi_B$ in the position basis. 
This gives a vacuum that can schematically be written 
\begin{equation}
  \ket{\Psi_0} ~=~ \ket{0}_{g=0} \;+\;
  \sum_{n}\sum_{m\le m'} c_{mm',n}\,
  \ket{1_{-m}1_{-m'}}_A \ket{1_{-n}}_B \;+\; \mathcal O(g^2),
\end{equation}
where the coefficient of each creation channel is
\begin{align}
  c_{mm'n} &~=~
  -\,\frac{g}{N}\,
  \cos\!\Big(\frac{2\pi(m+m')n}{N}\Big)\,
  F(m,m',n)~,\nonumber \\
  F(m,m',n) &~=~
  \frac{\kappa_{mm'}}
{\sqrt{8\,\omega_{K,m}\omega_{K,m'}\omega_{X,n}}\;
\big(\omega_{K,m}+\omega_{K,m'}+\omega_{X,n}\big)} ,
  \label{eq:cmunua}
\end{align}
with $\kappa_{mm'}$ an $\mathcal O(1)$ combinatorial factor. We may now perform exact same tracing out procedure that began our discussion in Section~\ref{sec:TTNintro}, by 
grouping the excited terms by their $B$-content:
\begin{equation}
  \ket{\Psi_0} ~=~ \sqrt{1-\varepsilon}\;\ket{0}_A\ket{0}_B
  \;+\; \sum_n \sqrt{\gamma_n}\;\ket{\chi_n}_A\,\ket{1_{-n}}_B ,
\end{equation}
where $\ket{\chi_n}_A \propto \sum_{m\le m'} c_{mm',n}
\ket{1_{-m}1_{-m'}}_A$ is normalised, and where the reduced density matrix  has Schmidt eigenvalues
\begin{equation}
  \lambda_A ~=~
  \{\,1-\varepsilon,\; \gamma_1,\,\dots,\,\gamma _N\,\} + \mathcal O(g^4),
  \qquad
  \gamma_n ~=~ \sum_{m\le m'} |c_{m m',n}|^2 ,
  \qquad
  \varepsilon ~=~ \sum_n \gamma_n~ .
\end{equation}
Performing the sum over $m\le m'$ yields
$  \gamma_n \,\sim\, g^{2}$ for every $n$. Hence $
  \varepsilon ~=~ \sum_{n} p_n \,\sim\, N g^{2}$, and we find a total
$X|K$ entropy that is {\it extensive},
namely \begin{equation}
  S_{X|K} ~\simeq~ \bar c\, N\, g^{2}\,\log(1/g^{2}) ~,
\end{equation}
with $\bar c$ an $N$-independent lattice constant, and with a constant   entropy per
bridge bond which is intensive and parametrically small, $S_{\rm bond} = S_{X|K}/N \sim
g^{2}\log(1/g^{2})$. We can conclude that at small bridge coupling no individual bridge bond ever carries significant Schmidt weight, and this is indeed what is observed in the actual BRTN.

 This near-triviality of the bridge bonds is expected to be a general structural consequence of the Fourier bridge itself. 
That is the total cross-bridge entanglement is extensive as each node couples to $N$ partners, but it is democratically distributed, and no individual bond ever carries significant Schmidt weight. Thus the same Fourier kernel that endows the reciprocal label with its momentum interpretation protects the bridge from local entanglement concentration. This contrasts with the test case, where the site-diagonal coupling acts at full strength on each bond, and the smallness of the bridge entropy is instead due to the mean-field (squeezing) nature of the leading response.

Finally we should remark on the nature of this entropy. One might worry that the extensive scaling of the bridge entanglement, 
$S_{X|K}\approx N\, S_{\rm bond}$, might come into tension with holographic or 
black-hole entropy bounds, particularly if $M_*$ is identified with a fundamental scale which is close to the 
Planck scale. However, the bridge cut is not a spatial bipartition of the 
QFT: detaching a region of size $\ell$ on the $X$-tree severs only 
$\sim\log_2\ell$ bonds, whereas the $X|K$ split severs all $N$ bridge bonds 
at once. Thus, despite being extensive, the bridge entropy carries no dependence on a spatial sub-region size, $\ell$. 
The integrated-out 
$K$-sector cannot act as independent matter in the $X$-frame and cannot gravitate, and it is not therefore a quantity that any entropy bound would 
constrain. In other words, bounds of Bekenstein type, do not constrain the bare entanglement entropy but rather 
the relative entropy of a state with respect to the vacuum.

\section{Discussion}
\label{sec:physicalmeaning}

This, then, is the BRTN picture of UV/IR mixing. The main benefit of this formulation is its ability to incorporate UV/IR mixing at the level of the Hilbert space structure, without requiring a global continuum field-theory description. This is not merely a technical convenience. In a genuinely UV/IR mixed system, Wilsonian behaviour may well appear in an appropriate limit as we have seen here, but a global description may be difficult to obtain, or indeed one may not exist at all.
Examples in which such global descriptions {\it do} exist typically achieve this in conjunction with very constraining symmetries, for example world-sheet modular invariance in string theory, which place strong restrictions on the properties of the theory as a whole. Or they imply a drastic modification of space-time.  

By contrast, in the BRTN approach we jettison the idea of implementing additional symmetry at the level of the Hamiltonian, and instead realise it at the level of the Hilbert space itself. In the Born-reciprocal case the symmetry in question is realised as an automorphism within a doubled Hilbert space, which gives it a renormalisation geometry. 
Tensor networks are a natural framework for this idea because they are a direct expression of the Hilbert space structure, with very little extra baggage. It may well be that a continuum global description of the theory exists, but one is not necessary. All that is needed is a network to support the Hilbert space, which is then able to encode  long range entanglement.
What the network formalism adds is a way to treat such dynamics. It tells us how correlations propagate, how coarse-graining acts, and how renormalisation acts in such systems. In other words, the network continues to determine how  information flows through Hilbert space even in the presence of UV/IR mixing.

This provides the criterion anticipated in the Introduction. Namely, an effective 
field theory exists precisely when the cross-sector entanglement is diagonal 
in the reciprocal momentum: such contributions are local and 
translation-invariant, and can be absorbed into localised counter-terms and 
the parameters of the effective theory. The genuine obstruction to Wilsonian physics is 
off-diagonal entanglement $S_{AB}(K,K')$ with $K\neq K'$, which is non-local 
and breaks translation invariance, and which no local counter-term can reach. 
A well defined effective description therefore exists if and only if this 
off-diagonal entanglement scales away in the large-volume limit.
As we saw the BRTN is a theory of this kind.  

Thus, while the BRTN is conceptually related to earlier Born-reciprocal and Planck-scale phase-space frameworks (\eg \cite{Majid:1988we,Jarvis:2005yw}) its organising principle is very different. This gives a novel overall picture, with a number of distinctive features, and several open questions. One may well ask for example what physical meaning should be ascribed to it, and how does it relate to other descriptions of UV/IR mixing in the literature. We will now discuss these in turn.     \\ 

\listitem{The physical meaning of the reciprocal sector} 
In the BRTN picture the reciprocal  $K$-sector degrees of freedom are not ordinary particle modes of the low-energy QFT. They belong to the reciprocal description that sits across the UV interface of the BRTN. When one constructs the Wilsonian effective theory for the $X$-sector, these reciprocal modes correspond to  degrees of freedom that are integrated out.

Their physical imprint therefore appears only indirectly, for example through the renormalisation of couplings.
Because the reciprocal sector is integrated out, one should not expect propagator poles corresponding to light particles arising from its modes: they are not expected to survive as asymptotic excitations in the $X$-description. Instead, as we saw above, these modes contribute to the boundary values of the parameters such as the masses of the effective QFT theory, and play no further role in the low energy dynamics.

Which side of the BRTN is to be integrated out is, in the end, a matter of choice: it hinges entirely on which half one designates to be the $X$-side (\ie the QFT side). 
This choice of course amounts to a breaking of the Born-reciprocal symmetry which we are obliged to make ourselves. This is analogous to the situation described in Ref.~\cite{Abel:2021tyt} for modular invariant theories. Such a breaking  is inevitable for systems with a renormalisation geometry -- we will return to this analogy shortly.

The role of the bridge couplings, and the physics happening there is of particular interest. We may think of the UV of the regular TTN as being the place where the modes are maximally localised in the $x$-direction. Their reciprocals on the momentum-side of the BRTN are, as we have already noted, localised in the $k$-direction and delocalised in $x$. In a phase-space (Wigner function) sense, localisation corresponds to squeezing along one direction. The bridge then relates states that are maximally squeezed in the 
$x$-direction to states that are maximally squeezed in the reciprocal 
$k$-direction, with this dual structure leading to correlations between the two sectors. \\  

\listitem{BRTN as a UV completion, and the importance of the large-volume limit}
One of the most interesting aspects of Born reciprocity is that it introduces a granularity to space-time~\cite{Jarvis:2005yw} which represents a physical duality scale in place of a conventional continuum limit. Indeed, as we saw in Eq.~\eqref{eq:lims} 
the physical interpretation and scaling of the lattice spacings is 
\begin{align} 
a_X ~=~ \frac{1}{M_*} ~,~~~ a_K ~=~ \frac{1}{R}~,~~~ N~=~2\pi R M_*~.
\label{eq:lims2}
\end{align} 
(To be sure there is no confusion, one should resist tying $a_X$ and $a_K$ through a canonical relation 
$a_X a_K \sim \hbar$: $X_n$ and $K_m$ are the native coordinates of two 
different fields, not conjugate variables of one, and their spacings are fixed 
purely by the Fourier-lattice reciprocity $a_X a_K = 2\pi/N$.)
The fundamental energy scale
\begin{equation}
M_* ~\equiv~ \frac{Na_K}{2\pi} ~=~ \frac{1}{a_X}
\end{equation}
then marks a crossover between the ordinary QFT description and the reciprocal description. In this sense the conventional limit $a_X\to0 $, or equivalently $M_*\to\infty$, is not  a natural scaling limit in the BRTN. In that limit, since $a_K\to\infty$, the reciprocal sector is simply pushed to infinite energy and its finite-distance response disappears. A non-trivial Born-reciprocal UV completion instead requires maintaining a finite duality scale $M_*$, with ordinary QFT being recovered in the large volume limit as an effective description which is valid below this scale, as we saw in Section~\ref{sec:breaking}.

As we also saw this finiteness does not remove the usual UV sensitivity in the BRTN, but rather recasts it as  IR sensitivity. Would-be UV divergences in the ordinary QFT are instead reorganised into boundary defects whose size increases with the size of the network. Such terms are the BRTN avatar of the usual QFT counterterms: they may be removed by cancelling boundary terms (which in practice may be achieved by normal-ordering the potential). Once these counterterms have been added one is  free to take the large volume $a_K\to 0$ limit while keeping $M_*$ fixed, whereupon regular Wilsonian QFT with continuous momenta emerges. \\

\listitem{UV/IR mixing and renormalisation in string theory}
String theory is a prime example of a theory that is UV/IR mixed, and yet resembles a Wilsonian theory at low scales.
Recent work has explored the interplay between UV/IR mixing and renormalisation in  modular invariant theories~
\cite{Abel:2021tyt,Abel:2023hkk,Abel:2024twz}. 
The apparent paradox of finding Wilsonian behaviour in a UV/IR-mixed theory was resolved in that work by observing that a consistent and unique definition of energy scale $\mu$ requires a renormalisation geometry: namely the effective theory must be  invariant under the transformation $\mu\to M_s^2/\mu$, where $M_s$ is in this case the string scale. Without such a symmetry there would effectively be two energy scales in the system, and so it is this symmetry that makes a sensible effective low-energy description possible. It can be thought of as a cousin of the space time T-dualities which relate  KK  modes  to dual winding modes. 
Thus a global picture emerges where the IR of the theory has an image IR'. 

The BRTN clearly has an intriguing overlap with that picture although it is not an exact one. Certainly if we identify the root nodes of a tree tensor network with the IR of the theory, then the presence in the BRTN of delocalised degrees of freedom that have their own image root node is very suggestive.
Moreover in both pictures the crucial feature that appears is renormalisation geometry: namely the idea that the definition of renormalisation scale must commute with the symmetry that mixes UV and IR  (which as mentioned is worldsheet modular invariance in the case of closed string theory). 

However the analogy is imperfect. In particular the renormalisation scale in the TTN is not as well defined as  it is for example in the MERA. It can be loosely identified with tree depth but this identification is warranted only because correlations over increasing distances $\xi$ in space involve excursions along paths that on average go higher up the tree. As we saw, there is a roughly logarithmic relation between the two. In the BRTN scheme it is not immediately obvious how reciprocity translates to a symmetry on energy scale of the form $\mu\to M_s^2/\mu$, as there  is no direct concept of a renormalisation scale that is higher than the lattice scale $1/a_X$. 

One possible route to obtaining one is to leverage the fact that correlations of length $\xi_X$ on the space-side of the BRTN, must be exactly the same as correlations over Born-reciprocal momentum differences on the momentum side. This is guaranteed by the Born-reciprocity of the network.

In detail, to introduce a renormalisation scale into the BRTN one could imagine measuring correlations over dimensionless length scales on the lattice of $\xi_X$. We may associate a corresponding energy scale in terms of inverse lattice units, 
$\xi_X a_X \equiv  1/\mu_X$. By Born-reciprocity this corresponds to the same dimensionless scale on the momentum side of $\xi_K = {\xi_X}$. (Indeed if we forget all of the physical interpretation for a moment we are ultimately just measuring correlations on a symmetric network.) It is natural to {\it define} this dual resolution scale to correspond to the energy scale in 
the reciprocal sector given by Eq.~\eqref{eq:stretching}: that is 
\begin{equation}
\mu_K ~=~  \xi_K a_K  ~=~ \frac{a_K}{a_X\mu_X} ~=~ \frac{M_*}{R\mu_X}~,
\end{equation}
using Eq.~\eqref{eq:lims2}.
Thus we recover a stringy $\mu_K = M_s^2 /\mu_X$ relation if we identify the ``string scale'' with the geometric mean of the lattice-spacing energy scales on either side, $M_s = \sqrt{M_* /R}$.

What does this mean in practice? Suppose that we decide to introduce a renormalisation scale $\mu$ into the lattice by quenching wavelengths larger than the corresponding length scale on the space-side (assuming that this were possible): \ie we quench modes with wavelengths $\xi_Xa_X > 1/\mu$, and call $\mu$ the energy scale of the theory. In order to maintain the reflection symmetry of the BRTN we must then also quench modes on the momentum-side of the network that have $\xi_Ka_K  > M_*/R\mu$, \ie modes that live in the deep UV or equivalently the IR'. If we were to choose a different scale on the $K$-side of the network,  then, just as in modular invariant theories, we would effectively have two different energy scales at play in the system. 

This argumentation closely parallels the line of thinking in modular invariant theories which led to $\mu \to M_s^2/\mu$ symmetry.
However the discussion here adds a new twist to that story, because KK modes and winding modes are acting in string theory like the QFT and its reciprocal sector in the BRTN. Our findings suggest that an important factor in Wilsonian behaviour emerging in the context of string theory is that KK modes and winding modes do not become strongly entangled. \\

\section{Conclusions}\label{sec:conclusions}

Few ideas in physics are more suggestive and less manageable than UV/IR mixing. It may well  be of relevance to several important problems, and yet it is difficult to treat in a generic way.   
It is crucial therefore to  develop better and more generally applicable language for describing UV/IR mixed theories. This paper has presented an approach based on tensor networks, introducing Born-reciprocity at the level of the Hilbert space. This sidesteps some of the difficulties associated with finding a global continuum theory that can accommodate such a symmetry, by instead building it directly into the renormalisation geometry. 

The overall picture that emerges is one in which we abandon the ambition of a global continuum UV/IR mixed theory, and instead work with the Hamiltonian of a Born-reciprocal tensor network (BRTN): this is enough to capture the time evolution of a system that has continuum QFT-like behaviour in certain limits, but no global field theoretical description. 

This provides a realisation of UV/IR mixing that is compatible with modern network approaches to renormalisation in QFT, and which  is practical and calculable. For example it was possible to measure radiative corrections coming from what is essentially the UV-completion of the theory. These agreed with the theoretical one-loop computations that were made by invoking Born-reciprocity. This shows the BRTN to be a genuine calculational framework, rather than merely a conceptual construct, which does not require a continuum field theory Hamiltonian.

The UV/IR mixing in the system is manifest in several ways. For example the UV divergences of regular QFT become boundary defects, which must be subtracted by boundary counter terms. Following these subtractions we find that in the large volume / thermodynamic  limit all further evidence of Born-reciprocity scales away: the reciprocal sector can be integrated out and normal Wilsonian QFT behaviour is restored. In this scaling limit the fundamental scale $M_*$ (given by the lattice-spacing as $M_*=1/a_X$) remains finite, while  momentum transfer becomes continuous.  The criterion for the theory to be in this Wilsonian regime is that the lattice size $R$ obeys
\begin{equation}
    2\pi R   ~\gg ~  \,  \frac{M_*}{m^2_{\rm phys}}~.
    \label{eq:Nconst2}
\end{equation}
Without wishing to overstate its significance, a curious dimensional observation is that if we assume the fundamental scale to be the Planck mass, and take the physical mass to be the lightest oscillation-derived neutrino mass scale, ${m^2_{\rm phys}}=\Delta m_{12}^2 $, then the  length scale
$R \sim  {M_{\rm Pl}}/{m^2_{\rm phys}}$ is cosmological, of order a Gpc. This coincidence is structurally the same as that appearing (albeit with the dark energy scale in the denominator) in the context of UV/IR mixing in Ref.~\cite{Cohen:1998zx}.

The present work focussed on the set-up of the Born-reciprocal network and on the preparation and properties of its vacuum. 
Future work \cite{tocome} will consider dynamics using real time Hamiltonian evolution employing the machinery that has been constructed here. 
There are several aspects of physics that may prove interesting, for example scattering at energies approaching and exceeding the UV scale, phase transitions driven by the bridge coupling $g$, and bridge couplings as Higgs portal couplings. All of these phenomena cannot easily be treated using normal perturbative methods. 

\bigskip

\subsection*{Acknowledgements} This work was supported by the STFC under the IPPP grant ST/T001011/1.  I would like to thank the CERN Theory group for its support and hospitality during the preparation of this work, and Keith Dienes, Luca Nutricati, Arttu Rajantie, Michael Spannowsky and Simon Williams for discussions and for collaborations connected with the broader themes in this paper. AI assistance was used in the coding of the numerical ITensor routines employed in this work. The author takes full responsibility for the implementation and verification of all results.

\bibliographystyle{inspire}
\bibliography{refs}{}

@article{Seiberg:1999vs,
    author = "Seiberg, Nathan and Witten, Edward",
    title = "{String theory and noncommutative geometry}",
    eprint = "hep-th/9908142",
    journal = "JHEP",
    volume = "09",
    pages = "032",
    year = "1999"
}

@article{Minwalla:1999px,
    author = "Minwalla, Shiraz and Van Raamsdonk, Mark and Seiberg, Nathan",
    title = "{Noncommutative perturbative dynamics}",
    eprint = "hep-th/9912072",
    journal = "JHEP",
    volume = "02",
    pages = "020",
    year = "2000"
}

@article{Matusis:2000jf,
    author = "Matusis, A. and Susskind, Leonard and Toumbas, N.",
    title = "{The IR/UV connection in noncommutative gauge theories}",
    eprint = "hep-th/0002075",
    journal = "JHEP",
    volume = "12",
    pages = "002",
    year = "2000"
}

@article{Arkani-Hamed:2021xlp,
    author = "Arkani-Hamed, Nima and Harigaya, Keisuke",
    title = "{Naturalness and the muon magnetic moment}",
    eprint = "2106.01373",
    archivePrefix = "arXiv",
    primaryClass = "hep-ph",
    doi = "10.1007/JHEP09(2021)025",
    journal = "JHEP",
    volume = "09",
    pages = "025",
    year = "2021"
}

@article{Kephart:2022vfr,
    author = {Kephart, Thomas W. and P{\"a}s, Heinrich},
    title = "{UV/IR mixing, causal diamonds and the electroweak hierarchy problem}",
    eprint = "2209.03305",
    archivePrefix = "arXiv",
    primaryClass = "hep-ph",
    doi = "10.1142/S0217732325500762",
    journal = "Mod. Phys. Lett. A",
    volume = "40",
    number = "21n22",
    pages = "2550076",
    year = "2025"
}

@article{Swingle:2009bg,
    author = "Swingle, Brian",
    title = "{Entanglement Renormalization and Holography}",
    eprint = "0905.1317",
    archivePrefix = "arXiv",
    primaryClass = "cond-mat.str-el",
    doi = "10.1103/PhysRevD.86.065007",
    journal = "Phys. Rev. D",
    volume = "86",
    pages = "065007",
    year = "2012"
}

@article{Harlow:2018fse,
    author = "Harlow, Daniel",
    title = "{TASI Lectures on the Emergence of Bulk Physics in AdS/CFT}",
    eprint = "1802.01040",
    archivePrefix = "arXiv",
    primaryClass = "hep-th",
    doi = "10.22323/1.305.0002",
    journal = "PoS",
    volume = "TASI2017",
    pages = "002",
    year = "2018"
}

@article{Pastawski:2015qua,
    author = "Pastawski, Fernando and Yoshida, Beni and Harlow, Daniel and Preskill, John",
    title = "{Holographic quantum error-correcting codes: Toy models for the bulk/boundary correspondence}",
    eprint = "1503.06237",
    archivePrefix = "arXiv",
    primaryClass = "hep-th",
    doi = "10.1007/JHEP06(2015)149",
    journal = "JHEP",
    volume = "06",
    pages = "149",
    year = "2015"
}

@article{tocome,
    author = "Abel, Steven",
    title = "{From Entanglement to Dynamics: Scattering and Decay in UV/IR-Mixed Theories}"
}

@article{Cohen:1998zx,
    author = "Cohen, Andrew G. and Kaplan, David B. and Nelson, Ann E.",
    title = "{Effective field theory, black holes, and the cosmological constant}",
    eprint = "hep-th/9803132",
    archivePrefix = "arXiv",
    reportNumber = "BUHEP-98-7, DOE-ER-40561-358, INT-98-00-6, UW-PT-97-24",
    doi = "10.1103/PhysRevLett.82.4971",
    journal = "Phys. Rev. Lett.",
    volume = "82",
    pages = "4971--4974",
    year = "1999"
}

@article{Dienes:2001se,
    author = "Dienes, Keith R.",
    title = "{Solving the hierarchy problem without supersymmetry or extra dimensions: An Alternative approach}",
    eprint = "hep-ph/0104274",
    archivePrefix = "arXiv",
    doi = "10.1016/S0550-3213(01)00344-3",
    journal = "Nucl. Phys. B",
    volume = "611",
    pages = "146--178",
    year = "2001"
}

@article{Lust:2017wrl,
    author = "Lust, Dieter and Palti, Eran",
    title = "{Scalar Fields, Hierarchical UV/IR Mixing and The Weak Gravity Conjecture}",
    eprint = "1709.01790",
    archivePrefix = "arXiv",
    primaryClass = "hep-th",
    doi = "10.1007/JHEP02(2018)040",
    journal = "JHEP",
    volume = "02",
    pages = "040",
    year = "2018"
}

@article{Craig:2019zbn,
    author = "Craig, Nathaniel and Koren, Seth",
    title = "{IR Dynamics from UV Divergences: UV/IR Mixing, NCFT, and the Hierarchy Problem}",
    eprint = "1909.01365",
    archivePrefix = "arXiv",
    primaryClass = "hep-ph",
    doi = "10.1007/JHEP03(2020)037",
    journal = "JHEP",
    volume = "03",
    pages = "037",
    year = "2020"
}

@article{Abel:2021tyt,
    author = "Abel, Steven and Dienes, Keith R.",
    title = "{Calculating the Higgs mass in string theory}",
    eprint = "2106.04622",
    archivePrefix = "arXiv",
    primaryClass = "hep-th",
    reportNumber = "IPPP/20/113",
    doi = "10.1103/PhysRevD.104.126032",
    journal = "Phys. Rev. D",
    volume = "104",
    number = "12",
    pages = "126032",
    year = "2021"
}

@article{Castellano:2021mmx,
    author = "Castellano, Alberto and Herr{\'a}ez, Alvaro and Ib{\'a}{\~n}ez, Luis E.",
    title = "{IR/UV mixing, towers of species and swampland conjectures}",
    eprint = "2112.10796",
    archivePrefix = "arXiv",
    primaryClass = "hep-th",
    doi = "10.1007/JHEP08(2022)217",
    journal = "JHEP",
    volume = "08",
    pages = "217",
    year = "2022"
}

@article{Craig:2022eqo,
    author = "Craig, Nathaniel",
    title = "{Naturalness: past, present, and future}",
    eprint = "2205.05708",
    archivePrefix = "arXiv",
    primaryClass = "hep-ph",
    doi = "10.1140/epjc/s10052-023-11928-7",
    journal = "Eur. Phys. J. C",
    volume = "83",
    number = "9",
    pages = "825",
    year = "2023"
}

@article{Abel:2023hkk,
    author = "Abel, Steven and Dienes, Keith R. and Nutricati, Luca A.",
    title = "{Running of gauge couplings in string theory}",
    eprint = "2303.08534",
    archivePrefix = "arXiv",
    primaryClass = "hep-th",
    reportNumber = "CERN-TH-2023-044, IPPP/23/16",
    doi = "10.1103/PhysRevD.107.126019",
    journal = "Phys. Rev. D",
    volume = "107",
    number = "12",
    pages = "126019",
    year = "2023"
}

@article{Abel:2024twz,
    author = "Abel, Steven and Dienes, Keith R. and Nutricati, Luca A.",
    title = "{New nonrenormalization theorem from UV/IR mixing}",
    eprint = "2407.11160",
    archivePrefix = "arXiv",
    primaryClass = "hep-th",
    reportNumber = "IPPP/24/46",
    doi = "10.1103/PhysRevD.110.126021",
    journal = "Phys. Rev. D",
    volume = "110",
    number = "12",
    pages = "126021",
    year = "2024"
}

@article{Nortier:2025gmc,
    author = "Nortier, Florian",
    title = "{Light Scalars in Light of UV/IR Mixing: Classicalization via Synergy between Vainshtein {\&} Chameleon Screenings}",
    eprint = "2511.01739",
    archivePrefix = "arXiv",
    primaryClass = "hep-ph",
    month = "11",
    year = "2025"
}

@article{Cribiori:2025oek,
    author = "Cribiori, Niccol{\`o} and Tonioni, Flavio",
    title = "{Cosmological constraints from UV/IR mixing}",
    eprint = "2507.02738",
    archivePrefix = "arXiv",
    primaryClass = "hep-th",
    doi = "10.1007/JHEP02(2026)035",
    journal = "JHEP",
    volume = "02",
    pages = "035",
    year = "2026"
}

@article{Aoufia:2026bau,
    author = "Aoufia, Christian and Basile, Ivano and Leone, Giorgio and Lotito, Matteo",
    title = "{UV/IR relations from the worldsheet}",
    eprint = "2603.11157",
    archivePrefix = "arXiv",
    primaryClass = "hep-th",
    month = "3",
    year = "2026"
}

@article{RevModPhys.21.463,
  title = {Reciprocity Theory of Elementary Particles},
  author = {Born, Max},
  journal = {Rev. Mod. Phys.},
  volume = {21},
  issue = {3},
  pages = {463--473},
  numpages = {0},
  year = {1949},
  month = {Jul},
  publisher = {American Physical Society},
  doi = {10.1103/RevModPhys.21.463},
  url = {https://link.aps.org/doi/10.1103/RevModPhys.21.463}
}

@article{Majid:1988we,
    author = "Majid, S.",
    title = "{Hopf Algebras for Physics at the Planck Scale}",
    doi = "10.1088/0264-9381/5/12/010",
    journal = "Class. Quant. Grav.",
    volume = "5",
    pages = "1587--1606",
    year = "1988"
}

@article{Low:2001rn,
    author = "Low, Stephen G.",
    title = "{Representations of the canonical group: (The Semidirect product of the unitary and Weyl-Heisenberg groups), acting as a dynamical group on noncommuting extended phase space}",
    eprint = "math-ph/0101024",
    archivePrefix = "arXiv",
    doi = "10.1088/0305-4470/35/27/312",
    journal = "J. Phys. A",
    volume = "35",
    pages = "5711--5730",
    year = "2002"
}

@article{Jarvis:2005yw,
    author = "Jarvis, Peter D. and Morgan, S. O.",
    title = "{Born reciprocity and the granularity of space-time}",
    eprint = "math-ph/0508041",
    archivePrefix = "arXiv",
    doi = "10.1007/s10702-006-1006-5",
    journal = "Found. Phys. Lett.",
    volume = "19",
    pages = "501--517",
    year = "2006"
}

@article{Freidel:2013zga,
    author = "Freidel, Laurent and Leigh, Robert G. and Minic, Djordje",
    title = "{Born Reciprocity in String Theory and the Nature of Spacetime}",
    eprint = "1307.7080",
    archivePrefix = "arXiv",
    primaryClass = "hep-th",
    doi = "10.1016/j.physletb.2014.01.067",
    journal = "Phys. Lett. B",
    volume = "730",
    pages = "302--306",
    year = "2014"
}

@article{Freidel:2014qna,
    author = "Freidel, Laurent and Leigh, Robert G. and Minic, Djordje",
    title = "{Quantum Gravity, Dynamical Phase Space and String Theory}",
    eprint = "1405.3949",
    archivePrefix = "arXiv",
    primaryClass = "hep-th",
    doi = "10.1142/S0218271814420061",
    journal = "Int. J. Mod. Phys. D",
    volume = "23",
    number = "12",
    pages = "1442006",
    year = "2014"
}

@article{Freidel:2015pka,
    author = "Freidel, Laurent and Leigh, Robert G. and Minic, Djordje",
    title = "{Metastring Theory and Modular Space-time}",
    eprint = "1502.08005",
    archivePrefix = "arXiv",
    primaryClass = "hep-th",
    doi = "10.1007/JHEP06(2015)006",
    journal = "JHEP",
    volume = "06",
    pages = "006",
    year = "2015"
}

@article{Freidel:2016pls,
    author = "Freidel, Laurent and Leigh, Robert G. and Minic, Djordje",
    title = "{Quantum Spaces are Modular}",
    eprint = "1606.01829",
    archivePrefix = "arXiv",
    primaryClass = "hep-th",
    doi = "10.1103/PhysRevD.94.104052",
    journal = "Phys. Rev. D",
    volume = "94",
    number = "10",
    pages = "104052",
    year = "2016"
}

@article{Haegeman:2011,
  author = {Haegeman, Jutho and Cirac, J. Ignacio and Osborne, Tobias J. and Pizorn, Iztok and Verschelde, Henri and Verstraete, Frank},
  title = {Time-Dependent Variational Principle for Quantum Lattices},
  journal = {Phys. Rev. Lett.},
  volume = {107},
  pages = {070601},
  year = {2011},
  doi = {10.1103/PhysRevLett.107.070601},
  eprint = {1103.0936},
  archivePrefix = {arXiv},
  primaryClass = {cond-mat.str-el}
}

@article{Haegeman:2016,
  author = {Haegeman, Jutho and Lubich, Christian and Oseledets, Ivan and Vandereycken, Bart and Verstraete, Frank},
  title = {Unifying time evolution and optimization with matrix product states},
  journal = {Phys. Rev. B},
  volume = {94},
  pages = {165116},
  year = {2016},
  doi = {10.1103/PhysRevB.94.165116},
  eprint = {1408.5056},
  archivePrefix = {arXiv},
  primaryClass = {quant-ph}
}

@article{Lubich:2015,
  author = {Lubich, Christian and Oseledets, Ivan V. and Vandereycken, Bart},
  title = {Time integration of tensor trains},
  journal = {SIAM Journal on Numerical Analysis},
  volume = {53},
  number = {2},
  pages = {917--941},
  year = {2015},
  doi = {10.1137/140976546},
  eprint = {1407.2042},
  archivePrefix = {arXiv},
  primaryClass = {math.NA}
}

@article{Paeckel:2019yjf,
    author = {Paeckel, Sebastian and K{\"o}hler, Thomas and Swoboda, Andreas and Manmana, Salvatore R. and Schollw{\"o}ck, Ulrich and Hubig, Claudius},
    title = "{Time-evolution methods for matrix-product states}",
    eprint = "1901.05824",
    archivePrefix = "arXiv",
    primaryClass = "cond-mat.str-el",
    doi = "10.1016/j.aop.2019.167998",
    journal = "Annals Phys.",
    volume = "411",
    pages = "167998",
    year = "2019"
}

@article{Lubich:2013,
  author = {Lubich, Christian and Rohwedder, Thorsten and Schneider, Reinhold and Vandereycken, Bart},
  title = {Dynamical approximation by hierarchical Tucker and tensor-train formats},
  journal = {SIAM Journal on Matrix Analysis and Applications},
  volume = {34},
  number = {2},
  pages = {470--494},
  year = {2013},
  doi = {10.1137/120885713},
  eprint = {1206.4203},
  archivePrefix = {arXiv},
  primaryClass = {math.NA}
}

@article{Holtz:2012,
  author = {Holtz, Sebastian and Rohwedder, Thorsten and Schneider, Reinhold},
  title = {The Alternating Linear Scheme for Tensor Optimization in the Hierarchical Tucker Format},
  journal = {SIAM Journal on Scientific Computing},
  volume = {34},
  number = {2},
  pages = {A683--A713},
  year = {2012},
  doi = {10.1137/100818893},
  eprint = {1104.2260},
  archivePrefix = {arXiv},
  primaryClass = {math.NA}
}

@article{10.1063/1.462100,
    author = {Colbert, Daniel T. and Miller, William H.},
    title = {A novel discrete variable representation for quantum mechanical reactive scattering via the S-matrix Kohn method},
    journal = {The Journal of Chemical Physics},
    volume = {96},
    number = {3},
    pages = {1982-1991},
    year = {1992},
    month = {02},
    issn = {0021-9606},
    doi = {10.1063/1.462100},
    url = {https://doi.org/10.1063/1.462100}
}

@article{Calabrese:2004eu,
    author = "Calabrese, Pasquale and Cardy, John L.",
    title = "{Entanglement entropy and quantum field theory}",
    eprint = "hep-th/0405152",
    archivePrefix = "arXiv",
    doi = "10.1088/1742-5468/2004/06/P06002",
    journal = "J. Stat. Mech.",
    volume = "0406",
    pages = "P06002",
    year = "2004"
}

@article{Calabrese:2005zw,
    author = "Calabrese, Pasquale and Cardy, John L.",
    title = "{Entanglement Entropy and Quantum Field Theory: a Non-technical Introduction}",
    eprint = "quant-ph/0505193",
    archivePrefix = "arXiv",
    doi = "10.1142/s021974990600192x",
    journal = "Int. J. Quant. Inf.",
    volume = "04",
    number = "03",
    pages = "429--438",
    year = "2006"
}

@article{Casini:2011kv,
    author = "Casini, Horacio and Huerta, Marina and Myers, Robert C.",
    title = "{Towards a derivation of holographic entanglement entropy}",
    eprint = "1102.0440",
    archivePrefix = "arXiv",
    primaryClass = "hep-th",
    doi = "10.1007/JHEP05(2011)036",
    journal = "JHEP",
    volume = "05",
    pages = "036",
    year = "2011"
}

@article{Casini:2012ei,
    author = "Casini, H. and Huerta, Marina",
    title = "{On the RG running of the entanglement entropy of a circle}",
    eprint = "1202.5650",
    archivePrefix = "arXiv",
    primaryClass = "hep-th",
    doi = "10.1103/PhysRevD.85.125016",
    journal = "Phys. Rev. D",
    volume = "85",
    pages = "125016",
    year = "2012"
}

@article{itensor,
	title={{The ITensor Software Library for Tensor Network Calculations}},
	author={Matthew Fishman and Steven R. White and E. Miles Stoudenmire},
	journal={SciPost Phys. Codebases},
	pages={4},
	year={2022},
	publisher={SciPost},
	doi={10.21468/SciPostPhysCodeb.4},
	url={https://scipost.org/10.21468/SciPostPhysCodeb.4}
}

@article{itensor-r0.3,
	title={{Codebase release 0.3 for ITensor}},
	author={Matthew Fishman and Steven R. White and E. Miles Stoudenmire},
	journal={SciPost Phys. Codebases},
	pages={4-r0.3},
	year={2022},
	publisher={SciPost},
	doi={10.21468/SciPostPhysCodeb.4-r0.3},
	url={https://scipost.org/10.21468/SciPostPhysCodeb.4-r0.3}
}

@article{White:1992zz,
  author = {White, Steven R.},
  title = {Density matrix formulation for quantum renormalization groups},
  journal = {Phys. Rev. Lett.},
  volume = {69},
  pages = {2863-2866},
  year = {1992}
}

@article{White:1993zz,
  author = {White, Steven R.},
  title = {Density-matrix algorithms for quantum renormalization groups},
  journal = {Phys. Rev. B},
  volume = {48},
  pages = {10345-10356},
  year = {1993}
}

@article{Ostlund:1995zz,
  author = {Ostlund, Stellan and Rommer, Stefan},
  title = {Thermodynamic Limit of Density Matrix Renormalization},
  journal = {Phys. Rev. Lett.},
  volume = {75},
  pages = {3537-3540},
  year = {1995}
}

@article{Fannes:1992zz,
  author = {Fannes, Mark and Nachtergaele, Bruno and Werner, Reinhard F.},
  title = {Finitely correlated states on quantum spin chains},
  journal = {Commun. Math. Phys.},
  volume = {144},
  pages = {443-490},
  year = {1992}
}

@article{Verstraete:2004cf,
  author = {Verstraete, Frank and Porras, David and Cirac, J. Ignacio},
  title = {Density Matrix Renormalization Group and Periodic Boundary Conditions: A Quantum Information Perspective},
  journal = {Phys. Rev. Lett.},
  volume = {93},
  pages = {227205},
  year = {2004},
  eprint = {cond-mat/0404706}
}

@article{Verstraete:2004uw,
  author = {Verstraete, Frank and Cirac, J. Ignacio},
  title = {Renormalization algorithms for Quantum-Many Body Systems in two and higher dimensions},
  journal = {arXiv preprint},
  eprint = {cond-mat/0407066},
  year = {2004}
}

@article{PerezGarcia:2006rz,
  author = {Perez-Garcia, David and Verstraete, Frank and Wolf, Michael M. and Cirac, J. Ignacio},
  title = {Matrix Product State Representations},
  journal = {Quantum Inf. Comput.},
  volume = {7},
  pages = {401-430},
  year = {2007},
  eprint = {quant-ph/0608197}
}

@article{Shi:2006zz,
  author = {Shi, Y.-Y. and Duan, L.-M. and Vidal, G.},
  title = {Classical simulation of quantum many-body systems with a tree tensor network},
  journal = {Phys. Rev. A},
  volume = {74},
  pages = {022320},
  year = {2006}
}

@article{Vidal:2007hda,
  author = {Vidal, G.},
  title = {Entanglement Renormalization},
  journal = {Phys. Rev. Lett.},
  volume = {99},
  pages = {220405},
  year = {2007},
  eprint = {cond-mat/0512165}
}

@article{Vidal:2008zz,
  author = {Vidal, G.},
  title = {Class of Quantum Many-Body States That Can Be Efficiently Simulated},
  journal = {Phys. Rev. Lett.},
  volume = {101},
  pages = {110501},
  year = {2008}
}

@article{Levin:2006jai,
  author = {Levin, Michael and Nave, Cody P.},
  title = {Tensor Renormalization Group Approach to Two-Dimensional Classical Lattice Models},
  journal = {Phys. Rev. Lett.},
  volume = {99},
  pages = {120601},
  year = {2007},
  eprint = {cond-mat/0611687}
}

@article{Evenbly:2015uca,
  author = {Evenbly, Glen and Vidal, Guifre},
  title = {Tensor Network Renormalization},
  journal = {Phys. Rev. Lett.},
  volume = {115},
  pages = {180405},
  year = {2015},
  eprint = {1412.0732}
}

@Article{10.21468/SciPostPhysLectNotes.8,
	title={{The Tensor Networks Anthology: Simulation techniques for many-body quantum lattice systems}},
	author={Pietro Silvi and Ferdinand Tschirsich and Matthias Gerster and Johannes Jünemann and Daniel Jaschke and Matteo Rizzi and Simone Montangero},
	journal={SciPost Phys. Lect. Notes},
	pages={8},
	year={2019},
	publisher={SciPost},
	doi={10.21468/SciPostPhysLectNotes.8},
	url={https://scipost.org/10.21468/SciPostPhysLectNotes.8},
}

@article{Appelquist:1974tg,
    author = "Appelquist, Thomas and Carazzone, J.",
    title = "{Infrared Singularities and Massive Fields}",
    reportNumber = "Print-74-1486 (HARVARD)",
    doi = "10.1103/PhysRevD.11.2856",
    journal = "Phys. Rev. D",
    volume = "11",
    pages = "2856",
    year = "1975"
}

@article{Wilson:1983xri,
    author = "Wilson, K. G.",
    title = "{The renormalization group and critical phenomena}",
    doi = "10.1103/RevModPhys.55.583",
    journal = "Rev. Mod. Phys.",
    volume = "55",
    pages = "583--600",
    year = "1983"
}

@article{Donoghue:2009mn,
    author = "Donoghue, John F.",
    editor = "Pich, A. and Portoles, J. and Rodrigo, G.",
    title = "{When Effective Field Theories Fail}",
    eprint = "0909.0021",
    archivePrefix = "arXiv",
    primaryClass = "hep-ph",
    doi = "10.22323/1.069.0001",
    journal = "PoS",
    volume = "EFT09",
    pages = "001",
    year = "2009"
}

@article{Abel:2025pxa,
    author = "Abel, Steven and Spannowsky, Michael and Williams, Simon",
    title = "{Qumode Tensor Networks for False Vacuum Decay in Quantum Field Theory}",
    eprint = "2506.17388",
    archivePrefix = "arXiv",
    primaryClass = "quant-ph",
    reportNumber = "IPPP/25/37",
    month = "6",
    year = "2025"
}

@article{Abel:2025zxb,
    author = "Abel, Steven and Spannowsky, Michael and Williams, Simon",
    title = "{Real-Time Scattering Processes with Continuous-Variable Quantum Computers}",
    eprint = "2502.01767",
    archivePrefix = "arXiv",
    primaryClass = "quant-ph",
    reportNumber = "IPPP/24/82",
    month = "2",
    year = "2025"
}

@article{Abel:2024kuv,
    author = "Abel, Steven and Spannowsky, Michael and Williams, Simon",
    title = "{Simulating quantum field theories on continuous-variable quantum computers}",
    eprint = "2403.10619",
    archivePrefix = "arXiv",
    primaryClass = "quant-ph",
    reportNumber = "IPPP/24/10",
    doi = "10.1103/PhysRevA.110.012607",
    journal = "Phys. Rev. A",
    volume = "110",
    number = "1",
    pages = "012607",
    year = "2024"
}

@article{Rigobello:2021fxw,
    author = "Rigobello, Marco and Notarnicola, Simone and Magnifico, Giuseppe and Montangero, Simone",
    title = "{Entanglement generation in (1+1)D QED scattering processes}",
    eprint = "2105.03445",
    archivePrefix = "arXiv",
    primaryClass = "hep-lat",
    doi = "10.1103/PhysRevD.104.114501",
    journal = "Phys. Rev. D",
    volume = "104",
    number = "11",
    pages = "114501",
    year = "2021"
}

@article{pjm/1103039709,
author = {H. F. Trotter},
title = {{Approximation of semi-groups of operators.}},
volume = {8},
journal = {Pacific Journal of Mathematics},
number = {4},
publisher = {Pacific Journal of Mathematics, A Non-profit Corporation},
pages = {887 -- 919},
year = {1958},
}

@article{10.1063/1.526596,
    author = {Suzuki, Masuo},
    title = {Decomposition formulas of exponential operators and Lie exponentials with some applications to quantum mechanics and statistical physics},
    journal = {Journal of Mathematical Physics},
    volume = {26},
    number = {4},
    pages = {601-612},
    year = {1985},
    month = {04},
    issn = {0022-2488},
    doi = {10.1063/1.526596},
    url = {https://doi.org/10.1063/1.526596},
    eprint = {https://pubs.aip.org/aip/jmp/article-pdf/26/4/601/19120226/601\_1\_online.pdf},
}

@article{PhysRevResearch.6.033057,
  title = {Simulating $(2+1)\mathrm{D}$ SU(2) Yang-Mills lattice gauge theory at finite density with tensor networks},
  author = {Cataldi, Giovanni and Magnifico, Giuseppe and Silvi, Pietro and Montangero, Simone},
  journal = {Phys. Rev. Res.},
  volume = {6},
  issue = {3},
  pages = {033057},
  numpages = {23},
  year = {2024},
  month = {Jul},
  publisher = {American Physical Society},
  doi = {10.1103/PhysRevResearch.6.033057},
  url = {https://link.aps.org/doi/10.1103/PhysRevResearch.6.033057}
}

@article{PRXQuantum.5.037001,
  title = {Quantum Computing for High-Energy Physics: State of the Art and Challenges},
  author = {Di Meglio, Alberto and Jansen, Karl and Tavernelli, Ivano and Alexandrou, Constantia and Arunachalam, Srinivasan and Bauer, Christian W. and Borras, Kerstin and Carrazza, Stefano and Crippa, Arianna and Croft, Vincent and de Putter, Roland and Delgado, Andrea and Dunjko, Vedran and Egger, Daniel J. and Fern\'andez-Combarro, Elias and Fuchs, Elina and Funcke, Lena and Gonz\'alez-Cuadra, Daniel and Grossi, Michele and Halimeh, Jad C. and Holmes, Zo\"e and K\"uhn, Stefan and Lacroix, Denis and Lewis, Randy and Lucchesi, Donatella and Martinez, Miriam Lucio and Meloni, Federico and Mezzacapo, Antonio and Montangero, Simone and Nagano, Lento and Pascuzzi, Vincent R. and Radescu, Voica and Ortega, Enrique Rico and Roggero, Alessandro and Schuhmacher, Julian and Seixas, Joao and Silvi, Pietro and Spentzouris, Panagiotis and Tacchino, Francesco and Temme, Kristan and Terashi, Koji and Tura, Jordi and T\"uys\"uz, Cenk and Vallecorsa, Sofia and Wiese, Uwe-Jens and Yoo, Shinjae and Zhang, Jinglei},
  journal = {PRX Quantum},
  volume = {5},
  issue = {3},
  pages = {037001},
  numpages = {49},
  year = {2024},
  month = {Aug},
  publisher = {American Physical Society},
  doi = {10.1103/PRXQuantum.5.037001},
  url = {https://link.aps.org/doi/10.1103/PRXQuantum.5.037001}
}

\end{document}